\documentclass[aps,prx,10pt,twocolumn,superscriptaddress,notitlepage,floatfix]{revtex4-2}

\usepackage[utf8]{inputenc}
\usepackage{graphicx}
\graphicspath{{../}}

\pdfoutput=1 

\usepackage{listings}

\usepackage[colorlinks=true, hyperindex, breaklinks, linkcolor=blue, urlcolor=blue, citecolor=blue]{hyperref}

\usepackage{dsfont}
\usepackage{amsmath}
\usepackage{amssymb,bm}
\usepackage{amsthm}

\usepackage{amsfonts}
\usepackage{amstext}
\usepackage{algorithm}
\usepackage{algorithmic}
\usepackage{array, makecell}
\usepackage{boldline}
\usepackage[capitalise]{cleveref}
\usepackage{graphicx}
\usepackage[caption=false]{subfig}
\usepackage[normalem]{ulem}
\usepackage{cancel}
\usepackage{xcolor}

\newcommand{\nocontentsline}[3]{}
\newcommand{\tocless}[2]{\bgroup\let\addcontentsline=\nocontentsline#1{#2}\egroup}

\makeatletter 
    
\renewcommand\onecolumngrid{
\do@columngrid{one}{\@ne}%
\def\set@footnotewidth{\onecolumngrid}
\def\footnoterule{\kern-6pt\hrule width 1.5in\kern6pt}%
}

\renewcommand\twocolumngrid{
        \def\footnoterule{
        \dimen@\skip\footins\divide\dimen@\thr@@
        \kern-\dimen@\hrule width.5in\kern\dimen@}
        \do@columngrid{mlt}{\tw@}
}%

\makeatother 

\begin{document}

\onecolumngrid
\title{Hardware-efficient quantum error correction via concatenated bosonic qubits}


\author{Harald Putterman} 
\email{putterma@amazon.com}
\author{Kyungjoo Noh} 
\author{Connor T. Hann} 
\author{Gregory S. MacCabe} 
\author{Shahriar Aghaeimeibodi} 
\author{Rishi N. Patel} 
\author{Menyoung Lee} 
\author{William M. Jones} 
\author{Hesam Moradinejad} 
\author{Roberto Rodriguez} 
\author{Neha Mahuli} 
\author{Jefferson Rose} 
\author{John Clai Owens} 
\author{Harry Levine} 
\author{Emma Rosenfeld} 
\altaffiliation[Current affiliation: ]{Google Research}
\author{Philip Reinhold} 
\author{Lorenzo Moncelsi}

\author{Joshua Ari Alcid}
\author{Nasser Alidoust} 
\author{Patricio Arrangoiz-Arriola} 
\author{James Barnett}
\author{Przemyslaw Bienias} 
\author{Hugh A. Carson} 
\author{Cliff Chen} 
\author{Li Chen} 
\author{Harutiun Chinkezian} 
\author{Eric M. Chisholm}
\author{Ming-Han Chou} 
\affiliation{AWS Center for Quantum Computing, Pasadena, CA 91125, USA}
% ...

\date{\today}

\begin{abstract}

In order to solve problems of practical importance~\cite{Gidney2021,dalzell2023}, quantum computers will likely need to incorporate quantum error correction, where a logical qubit is redundantly encoded in many noisy physical qubits~\cite{Shor1995,kitaev1997quantum, Knill1998}. The large physical-qubit overhead typically associated with error correction motivates the search for more hardware-efficient approaches~\cite{Cochrane1999Macroscopically,Aliferis2008,Fukui2017,Tuckett2018_ultrahigh,Guillaud2019Repetition,Puri2020_bias,Guillaud2021_error,Darmawan2021,BonillaAtaides2021_XZZX,Chamberland2022,Regent2023highperformance,Gouzien2023,Ruiz2024}. Here, using a microfabricated superconducting quantum circuit~\cite{Blais2004,Blais2021}, we realize a logical qubit memory formed from the concatenation of encoded bosonic cat qubits with an outer repetition code of distance $d=5$~\cite{Guillaud2019Repetition}. The bosonic cat qubits are passively protected against bit flips using a stabilizing circuit~\cite{mirrahimi2014,leghtas2015,touzard2018, Lescanne2020, singlecat2024}.  Cat-qubit phase-flip errors are corrected by the repetition code which uses ancilla transmons for syndrome measurement.  We realize a noise-biased CX gate which ensures bit-flip error suppression is maintained during error correction.  We study the performance and scaling of the logical qubit memory, finding that the phase-flip correcting repetition code operates below threshold, with logical phase-flip error decreasing with code distance from $d=3$ to $d=5$. Concurrently, the logical bit-flip error is suppressed with increasing cat-qubit mean photon number. The minimum measured logical error per cycle is on average $1.75(2)\%$ for the distance-3 code sections, and $1.65(3)\%$ for the longer distance-5 code, demonstrating the effectiveness of bit-flip error suppression throughout the error correction cycle. These results, where the intrinsic error suppression of the bosonic encodings allows us to use a hardware-efficient outer error correcting code, indicate that concatenated bosonic codes are a compelling paradigm for reaching fault-tolerant quantum computation.

\end{abstract}

\maketitle
\vspace*{10\baselineskip}

\newpage
\twocolumngrid
\tocless\section{Introduction}

For quantum computers to solve problems in materials design, quantum chemistry, and cryptography, where known speed-ups relative to classical computations are attainable, currently proposed algorithms require trillions of qubit gate operations to be applied in an error-free manner~\cite{Gidney2021,dalzell2023}. Despite impressive progress over the last few decades in reducing qubit error rates at the physical hardware level, the state-of-the-art remains some nine orders of magnitude away from these requirements.  One path towards closing the error-rate gap is through quantum error correction (QEC)~\cite{Shor1995,kitaev1997quantum, Knill1998}.  Similar to classical error correction used in communications~\cite{Shannon1948} and data storage~\cite{Hamming1950}, QEC can realize an exponential reduction in errors through the redundant encoding of information across many noisy physical qubits.

Recently, QEC experiments have been performed in various hardware platforms, including superconducting quantum circuits~\cite{Krinner2022,Zhao2022,Acharya2023,Sundaresan2023,Acharya2024}, trapped ions~\cite{Egan2021,RyanAnderson2021}, and neutral atoms~\cite{Bluvstein2024}. Some of these experiments are approaching~\cite{Acharya2023}, or have surpassed~\cite{Acharya2024}, the threshold where scaling of the error correcting code size leads to exponential improvements in the logical qubit error rate.  In these experiments, the qubits are realized using a simple encoding into two levels of a physical element, leaving them susceptible to environmental noise that can cause both bit- and phase-flip errors.  Correcting for both types of errors requires QEC codes such as the surface code~\cite{Krinner2022,Zhao2022,Acharya2023,Sundaresan2023}, which have a relatively high overhead penalty~\cite{Gidney2021}. 

A complementary QEC paradigm is to use a layered approach to noise protection, in which one starts from a qubit encoding that natively protects against different noise channels and suppresses errors.  One example is bosonic qubits, where qubit states are encoded in the infinite-dimensional Hilbert space of a bosonic mode (a quantum harmonic oscillator)~\cite{Cochrane1999Macroscopically,Gottesman2001,Jeong2002Efficient}, and the extra dimensionality of the mode provides redundancies that can be utilized to perform bosonic QEC. Experiments taking advantage of this redundancy at the single bosonic mode level to suppress errors have been performed using cat codes~\cite{leghtas2015,ofek2016,touzard2018,Lescanne2020,Berdou2023,MarquetAutoparametric2024,Reglade2024}, binominal codes~\cite{Ni2023}, and GKP codes~\cite{Fluhmann2019,Campagne-Ibarcq2020,Sivak2023}.  At the same time, various proposals have been put forward to further scale bosonic QEC by concatenating it with an outer code across multiple bosonic modes~\cite{Cochrane1999Macroscopically,Fukui2017,Tuckett2018_ultrahigh,Guillaud2019Repetition,Puri2020_bias,Guillaud2021_error,Darmawan2021,BonillaAtaides2021_XZZX,Chamberland2022,Regent2023highperformance,Gouzien2023,xu2023faulttolerant,Ruiz2024}, leveraging the protection offered in each bosonic mode to reduce the overall resource overhead for QEC. 

In this work, we demonstrate a scalable, hardware-efficient logical qubit memory built from a linear array of bosonic modes using a variant of the repetition cat code proposal in Ref~\cite{Guillaud2019Repetition}. In particular, we stabilize noise-biased cat qubits in individual bosonic modes. Bit-flip errors of the cat qubits are natively suppressed at the physical level, and the remaining phase-flip errors are corrected by an outer repetition code over a linear array of modes.  The use of a repetition code enables low overhead due to its large error rate threshold and linear scaling of code distance with physical qubit number~\cite{Guillaud2019Repetition,Guillaud2021_error,Chamberland2022}. In what follows, we describe a microfabricated superconducting quantum circuit which realizes a distance $d=5$ repetition cat code logical qubit memory, present a noise-biased CX gate for implementing error syndrome measurements with ancilla transmons, and study the logical qubit error-correction performance.

\vspace{4ex}
\tocless\section{Quantum device realizing a distance-5 repetition code of cat qubits}

\begin{figure*}[t!]
    \centering
    \includegraphics[width=\textwidth]{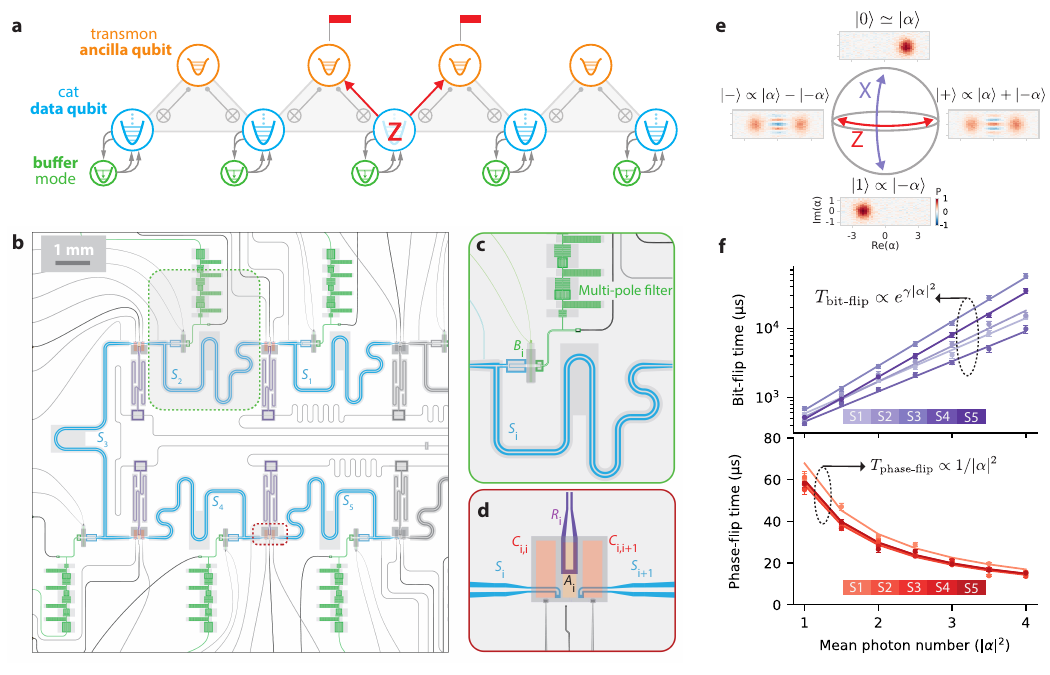}
    \caption{\textbf{Repetition code of bosonic qubits.} (a) Schematic diagram of the repetition code device. Data qubits $\text{S1}, \cdots, \text{S5}$ (blue) are encoded into the Hilbert space of a quantum harmonic oscillator.  Two photons from each oscillator are exchanged with a photon in a damped buffer mode $\text{B1}, \cdots, \text{B5}$ (green).  The ancilla qubits $\text{A1}, \cdots, \text{A4}$ (orange) are transmon qubits which detect $Z$ errors on the data qubits by measuring repetition-code stabilizers. Here we show a $Z$ error on a data qubit being detected by two ancilla qubits. (b) Circuit layout of the repetition code device.  The five bosonic modes ($S_i$) are implemented as coplanar waveguide resonators.  Each resonator is connected to a buffer mode ($B_i$). Buffer modes are damped through a multi-pole filter. Ancilla transmons ($A_i$) are connected to the storage modes by tunable couplers ($C_{i,j}$). The device is formed from two chips bump-bonded together with linear elements (e.g., storage modes, readout resonators, and filters) on the bottom chip and nonlinear elements (e.g., ancillas, couplers, and buffers) on the top chip. Zoomed-in circuit sections showing (c) a storage-buffer subsystem and (d) an ancilla transmon coupled to its neighboring storage modes by tunable couplers. (e) Cat qubit encoding in a bosonic mode. We show experimentally measured Wigner functions of the four basis states of a cat qubit with arrows representing $X$ and $Z$ errors. (f) Bit-flip and phase-flip times of the five cat qubits in our device under simultaneous two-photon dissipation. Error bars incorporate sampling variance and fit uncertainty. } 
    \label{fig:device}
\end{figure*}

A schematic of the repetition code device we use and the corresponding superconducting circuit layout are shown in \cref{fig:device}(a) and Figs.~\ref{fig:device}(b)-(d), respectively. The distance $d=5$ repetition code consists of five bosonic modes that host the data qubits (blue), along with four ancilla qubits (orange). The bosonic modes, also referred to as storage modes, are coplanar waveguide resonators with frequencies in the range $5.24-5.41$~GHz.  The storage modes have an average $T_{1}$ ($T_{2}$) time of over $60$~$\mu$s ($80$~$\mu$s). The ancilla qubits are fixed-frequency transmons with frequencies in the range $5.20-5.32$~GHz, narrowly detuned from their neighboring storage-mode frequencies. The ancilla qubits are coupled to the storage modes through a tunable-transmon coupler~\cite{Yan2018Tunable,Sung2021CZ}. By applying a flux pulse on a tunable coupler we can turn on a dispersive interaction between an ancilla transmon and a storage mode. This dispersive interaction is used to realize a controlled-$X$ operation (CX gate), with the ancilla transmon as the control and the data qubit as the target. Using the CX gates, we measure the repetition code stabilizers $\hat{X}_{i}\hat{X}_{i+1}$ (gray triangles), equivalent to measuring the joint photon-number parity of two neighboring storage modes. Each ancilla qubit is dispersively measured using a readout resonator coupled to a dedicated Purcell filter~\cite{Walter2020} and reset by applying microwave tones that remove the qubit excitations through the readout resonator~\cite{Magnard2018}. See \cref{app:device_fab,app:frequency_targeting_procedure_for_chi_matching,app:control_lines_and_fridge_setup,app:component_calibrations_and_parameters} for further details on device fabrication, experimental set-up, and component parameters.

As mentioned, each data qubit in our system is a cat qubit encoded in a storage mode~\cite{mirrahimi2014,leghtas2015,touzard2018}. The basis states of a cat qubit are shown in \cref{fig:device}(e) along with their experimental Wigner tomograms~\cite{lutterbach1997}. The $|0\rangle$ and $|1\rangle$ computational basis states are approximately the $|\alpha\rangle$ and $|{-}\alpha\rangle$ coherent states, respectively, with a mean photon number of $|\alpha|^{2}$. The complementary basis states are exactly the even and odd cat states $|\pm\rangle\propto |\alpha\rangle\pm|{-}\alpha\rangle$. Thus a bit-flip ($X$) error is a $180$-degree rotation in the phase space mapping $|\alpha\rangle \leftrightarrow |{-}\alpha\rangle$, and a phase-flip ($Z$) error corresponds to a parity flip between the even and odd cat states.  Owing to the phase-space separation of the $|\pm\alpha\rangle$ coherent states, bit-flip error rates can be exponentially suppressed with cat size $|\alpha|^2$~\cite{Lescanne2020,Berdou2023,MarquetAutoparametric2024,Reglade2024,singlecat2024}.  In contrast, phase-flip errors, which are caused by single-photon loss and heating, have a rate which increases linearly with $|\alpha|^2$.  

\begin{figure*}[t!]
    \centering
    \includegraphics[width=\textwidth]{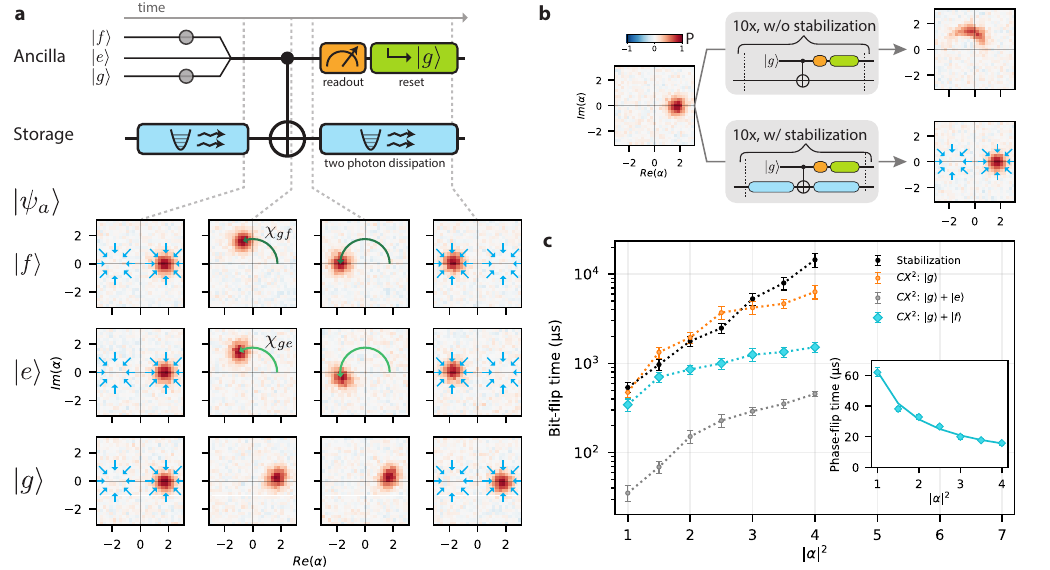}
    \caption{\textbf{Noise-biased $\text{CX}$ gate between a transmon and a cat qubit.} (a) CX gate sequence. At the start of the sequence the ancilla is initialized and cat-qubit stabilization (blue arrows) is on.  The stabilization is then turned off and the CX gate is applied between the ancilla and cat qubit. After the CX gate, the stabilization is turned back on and the ancilla qubit is readout and reset to $|g\rangle$.  Through the experimentally measured Wigner functions, we show the evolution of the storage state during the sequence for each of the ancilla states $|g\rangle$, $|e\rangle$, and $|f\rangle$, with an initial storage-mode state $|\alpha\rangle$. (b) Storage-mode Wigner tomograms before and after 10 applications of the CX gate sequence with the ancilla in $|g\rangle$ and storage initialized in $|\alpha\rangle$.  The sequence is applied with and without stabilization to show the importance of stabilization in preventing error accumulation. (c) Characterization of the CX gate.  We apply repeated CX$^2$ cycles (see main text) with a 3us cycle duration, and plot the measured bit-flip time of the cat qubit as a function of cat-qubit photon-number, $|\alpha|^2$, for different ancilla states. Inset shows the measured phase-flip times when the ancilla is initialized to $|g\rangle+|f\rangle$. Error bars incorporate sampling variance and fit uncertainty.}
    \label{fig:architecture}
\end{figure*}

In this work we stabilize cat qubits using two-photon dissipation which ensures that the cat-qubit amplitude $\alpha$, and thus the noise bias, are maintained over time ~\cite{mirrahimi2014,leghtas2015,touzard2018}. To realize the two-photon dissipation, we couple each storage mode to a lossy buffer mode (green) which is implemented using a version of the asymmetrically-threaded SQUID element~\cite{Lescanne2020,singlecat2024}.  We apply a flux pump to the buffer which converts pairs of photons in the storage into one photon in the buffer and vice versa. The buffer mode is heavily damped in order to dissipate this photon and realize the two-photon loss on the storage mode. The buffer mode is also linearly driven to produce a coherent two-photon drive on the storage mode to complement the two-photon loss, stabilizing the storage to the $|\pm\alpha\rangle$ manifold. The loss spectrum of each buffer is colored through a 4-pole metamaterial bandpass filter~\cite{Mirhosseini2018} such that the lifetime of the storage mode is not degraded by the strong buffer loss channel. Moreover, the buffer-mode parameters are carefully chosen to minimize other parasitic buffer-induced nonlinearities on the storage mode. This enables long cat-qubit bit-flip times even under pulsed cat-qubit stabilization, a crucial operation in our architecture, where two-photon dissipation is turned off for a significant fraction of a cycle. For more details on our cat-qubit realization we refer the readers to \cref{app:buffer_model} and Ref.~\cite{singlecat2024}. 

In \cref{fig:device}(f) we show the bit-flip and phase-flip times of all five data cat qubits when they are being simultaneously stabilized by two-photon dissipation. The bit-flip times of our cat qubits increase exponentially with the mean photon number $|\alpha|^{2}$, while the phase-flip times degrade as $1/|\alpha|^2$ as expected. The phase-flip times correspond to effective storage lifetimes under two-photon dissipation, $T_{1,\text{eff}}$, in the range $57-68$~$\mu$s.  A particularly important feature of our cat qubits is that a large noise bias is achieved even with small values of $|\alpha|^2$. Concretely, at $|\alpha|^2=2$, we achieve greater than $1~\text{ms}$ bit-flip times and $27 -33~\mathrm{\mu s}$ phase-flip times. This constitutes a sizable ($>30$) noise bias and at the same time a long phase-flip time in comparison to an error correction cycle time ($2-3$~$\mu$s). 

\vspace{4ex}
\tocless\section{Noise-biased CX gates in the repetition cat code}

For optimum performance of the repetition cat code, we must ensure that the large noise bias of the cat qubits is retained under all operations used in repetition code syndrome measurements. The key operation which enables syndrome measurements is the CX gate between a data cat qubit and an ancilla qubit with computational states $|0_a\rangle$ and $|1_a\rangle$. The CX gate can propagate errors in the ancilla qubit to the data cat qubit. Therefore, we need to realize a noise-biased CX gate which minimizes undesired bit flips on the target cat qubit caused by ancilla errors. 

While cat qubits can be used as ancilla qubits to implement noise-biased CX gates, these gates can induce significant control errors and require a complex drive scheme~\cite{Guillaud2019Repetition,Puri2020_bias,Chamberland2022}.  To avoid this issue, we use fixed-frequency transmons as ancilla qubits whose lowest three energy eigenstates are denoted by $|g\rangle, |e\rangle, |f\rangle$.  We realize the CX gate with a storage-ancilla dispersive coupling which can be thought of as an ancilla-state-dependent rotation of the storage-mode state in phase space~\cite{Schuster2007,Leghtas2013,sun2014}. Specifically, the CX gate is realized by having the data cat qubit rotate by $180$ degrees conditional on the ancilla being in $|1_a\rangle$. However, if the ancilla qubit is simply encoded in the $|g\rangle/|e\rangle$ manifold of a transmon, $|e\rangle\rightarrow|g\rangle$ decay events of the transmon during a CX gate can dephase the data cat qubit and induce bit flips. 

To circumvent this challenge, we follow Refs.~\cite{Rosenblum2018,Reinhold2020, Ma2020} and encode an ancilla qubit in the states $|0_{A}\rangle = |g\rangle$ and $|1_{A}\rangle = |f\rangle$, and engineer an approximately ``$\chi$-matched'' dispersive interaction between the ancilla and the storage mode in the form of  $\hat{H}=\chi_{ge} \hat{a}^\dagger \hat{a} |e\rangle\langle e| + \chi_{gf} \hat{a}^\dagger \hat{a} |f\rangle\langle f|$ with $\chi_{ge} \simeq \chi_{gf}$ (here $\hat{a}$ and $\hat{a}^\dagger$ are the annihilation and creation operators of the storage mode).  With the $\chi$-matching, even if the ancilla decays from $|f\rangle$ to $|e\rangle$, the storage mode will continue to rotate at a similar rate, and thus additional bit-flip errors on the data cat qubit are suppressed. This ensures that the noise bias of a CX gate is robust against the first-order ancilla decay events. Only higher-order or suppressed ancilla error mechanisms, such as two sequential decay events or heating will cause bit-flip errors on the data cat qubit.  

There are several additional important features of our CX gate. First, a tunable coupler mediates the dispersive coupling between a storage mode and an ancilla. This allows us to turn on the dispersive coupling when a CX gate is applied while maintaining high extinction when it is off.  Second, unlike previous demonstrations where the $\chi$-matching condition is achieved through a strong off-resonant drive on a transmon~\cite{Rosenblum2018,Reinhold2020}, we realize a natively $\chi$-matched dispersive interaction without strong drives by targeting carefully chosen frequency detunings between storage and ancilla (see \cref{app:device_hamiltonian,app:frequency_targeting_procedure_for_chi_matching} and the references therein). Lastly, the $\chi$-matching condition does not need to be satisfied exactly because small mis-rotations during the CX gate due to a mismatched $\chi$ ratio (e.g., $0.8 \lesssim \chi_{ge}/\chi_{gf} \lesssim 1.2$) can be corrected by subsequent two-photon dissipation for cat-qubit stabilization (see \cref{app:chi_matching_requirements}). 

In \cref{fig:architecture}(a) we show a control sequence involving a CX gate similar to the one used for an error correction syndrome measurement. We illustrate the robustness of our CX gate to ancilla decay by measuring the action of the gate on a storage mode prepared in a $|\alpha\rangle$ coherent state with the ancilla prepared in the $|g\rangle$, $|e\rangle$, or $|f\rangle$ states.  Before the CX gate begins we turn off the cat-qubit stabilization to allow the storage mode to rotate freely.  Then we activate the CX gate by applying a flux pulse on the tunable coupler.  As shown by the Wigner tomograms, the storage mode does not rotate when the ancilla is in $|g\rangle$, while it rotates by approximately $180$ degrees over the course of a CX gate when the ancilla is in $|e\rangle$ or $|f\rangle$.  Due to the imperfect $\chi$-matching the storage mode has slightly overrotated when the ancilla is in $|e\rangle$.  In addition, miscalibrations, self-Kerr nonlinearities, and decoherence can cause misrotations and distortion of the storage-mode states.  Notably, all these imperfections can be corrected with high probability when the two-photon dissipation is turned back on after the CX gate, as demonstrated in the last column of the Wigner tomograms. To further highlight the importance of applying the two-photon dissipation, in \cref{fig:architecture}(b) we show the results of $10$ repetitions of the CX gate cycle with and without the pulsed cat-qubit stabilization. When the two-photon dissipation is not applied, errors accumulate over multiple rounds causing significant distortion in the final storage mode state. When two-photon dissipation is applied every cycle, the storage mode stays well confined to the ideal target coherent state.  

We quantify the performance of our CX gate, including its noise bias, in a way that is representative of how we will use it during error correction. To do so, we repeatedly apply the same pulse sequence as in \cref{fig:architecture}(a), but with one difference: the single CX is replaced with the equivalent of two CX gates (a $\text{CX}^{2}$ gate).  This ensures that, similar to a stabilizer measurement, our cycle error rate is first-order insensitive to ancilla state preparation errors. 

In \cref{fig:architecture}(c) we show bit-flip times measured during repeated CX$^2$ cycles for a representative interaction between ancilla $\text{A1}$ and storage $\text{S1}$. Each cycle has a length of $3~\mathrm{\mu s}$ (where the $\text{CX}^2$ gate length is $504~\mathrm{ns}$). We measure the bit-flip times with the ancilla in state $|g\rangle+|f\rangle$, as would be used for syndrome extraction. As control experiments, we repeat the same procedure with the ancilla in $|g\rangle$ and $|g\rangle+|e\rangle$ as well. The black curve serves as a reference showing the exponential increase of the bit-flip times as a function of the mean photon number $|\alpha|^{2}$ in the case where the two-photon dissipation is continually applied (as in \cref{fig:device}(d)). When the gates are applied with the ancilla in $|g\rangle$, bit-flip times exceeding $5~\text{ms}$ are achieved. The small degradation relative to the reference performance at $|\alpha|^2\gtrsim 3$ is due to heating events in the ancilla or coupler during the $\text{CX}^{2}$ gate. With the ancilla in $|g\rangle+|e\rangle$, bit-flip times are severely limited to well under $1~\text{ms}$ due to the storage dephasing caused by the first-order $|e\rangle\rightarrow |g\rangle$ decay errors of the ancilla.  With the initial ancilla state $|g\rangle+|f\rangle$, we recover bit-flip times over $1$~ms at $|\alpha|^{2} \ge 3$ due to the insensitivity to single decay events of the ancilla afforded by $\chi$-matching (here $\chi_{ge}/\chi_{gf} \approx 1.1$; see \cref{app:chi_matching_requirements}). In the inset of \cref{fig:architecture}(c), we also report the corresponding phase-flip times with the ancilla in the state $|g\rangle+|f\rangle$. An effective storage lifetime $T_{1,\text{eff}}$ of $63\pm 2$~$\mu$s is inferred, showing no substantial difference from $T_{1,\text{eff}}=68\pm 2$~$\mu$s measured in \cref{fig:device} in the absence of CX$^2$ gate application.  In terms of error probabilities per cycle, the bit-flip and phase-flip errors per cycle are respectively $(3.5 \pm 0.4)\times 10^{-3}$ and $(9.6 \pm 0.4)\times 10^{-2}$ at $|\alpha|^{2}=2$, corresponding to a noise bias $>25$.

\vspace{4ex}
\tocless\section{Correcting phase-flip errors with the repetition code}

\begin{figure*}[t!]
    \centering
    \includegraphics[width=\textwidth]{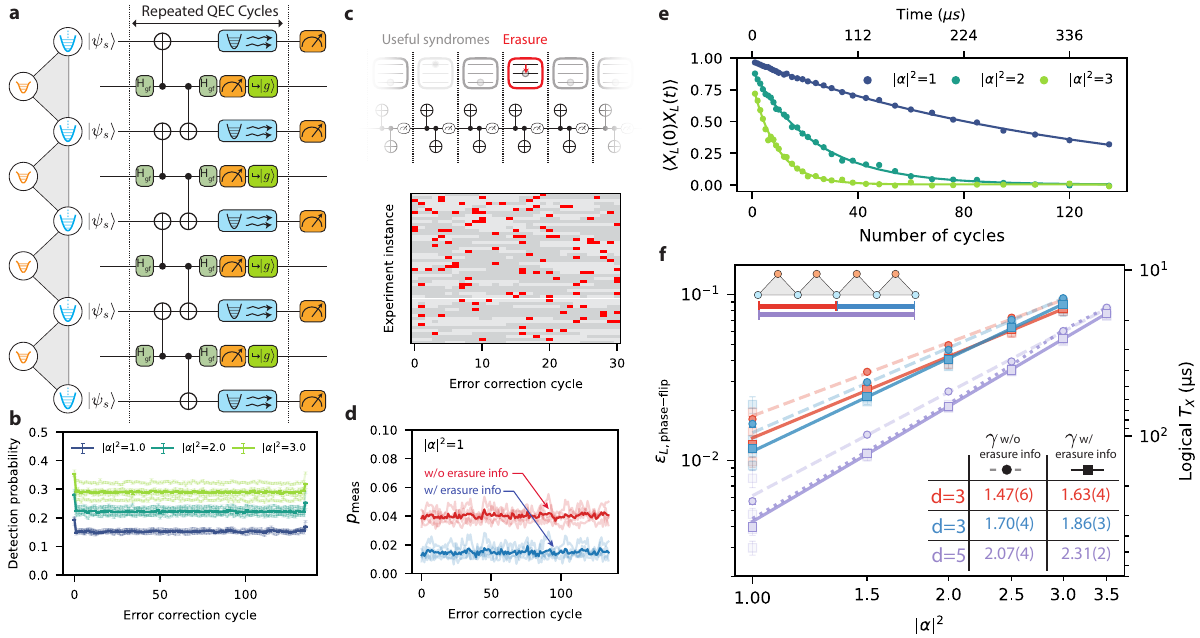}
     \caption{\textbf{Detecting and correcting phase-flip errors with the repetition code.} (a) Error correction circuit, showing repeated error correction cycles with duration $2.8~\mathrm{\mu s}$. (b) Detection probabilities for the measured stabilizers versus error correction cycle for three different cat qubit photon numbers, $|\alpha|^2=1,2,3$. Bold traces correspond to the average over the individual stabilizer traces.  Error bars represent standard error of the mean. (c) Depiction of erasures occurring in the repetition code experiment due to ancilla decay from $|f\rangle$ to $|e\rangle$.  We show shots of the experiment from a representative stabilizer where $|f\rangle$ is light gray, $|g\rangle$ is dark gray, and the erasure state $|e\rangle$ is shown in red. (d) Effective measurement error before and after accounting for the erasure for $|\alpha|^2=1$. Bold traces correspond to averaging over the stabilizers. (e) Example fits of the decay of the $X$ logical operator for the distance-5 repetition code for different photon numbers. (f) Error corrected logical phase-flip probability per cycle ($\epsilon_{L, \text{phase-flip}} = p_Z+p_Y$) and logical $X$ lifetime ($T_X$) as a function of $|\alpha|^2$ for the different repetition code sections.  Data and fits are shown with (squares and solid curves) and without (circles and dashed curves) inclusion of erasure information.  The fits are to the power law $(|\alpha|^2)^\gamma$ for $|\alpha|^2 \geq 1.5$.  Faded points indicate fits binned by the number of even cat states $|+\rangle$ in the initial state and serve to indicate the spread from asymmetric error rates at low photon number (see \cref{app:distribution_of_initial_states_in_logical_X_lifetime_experiments} for more details). The dotted purple curve shows the simulated logical phase-flip probability for the distance-5 section.  Error bars incorporate sampling variance and fit uncertainty.  }
    \label{fig:logical_X}
\end{figure*}

Equipped with the noise-biased CX gates, we now demonstrate the ability to correct the dominant phase-flip errors using a repetition code. Phase-flip errors are detected by repeatedly measuring the repetition code’s stabilizer generators, $\hat{X}_i \hat{X}_{i+1}$ (for $i = 1,\cdots,d-1$). As shown in \cref{fig:logical_X}(a), each measurement of a stabilizer generator, referred to as a syndrome measurement, comprises initialization of the ancilla $\text{A}_{i}$, two CX gates between $\text{A}_{i}$ and its adjacent data qubits $\text{S}_{i}$ and $\text{S}_{i+1}$, and finally measurement and reset of the ancilla. During the measurement and reset, we turn on the dissipative stabilization on all the cat qubits. Each syndrome measurement cycle has a conservatively chosen duration of $2.8$~$\mu$s, with CX gate lengths across the device ranging from $292-552$~ns.

After running an experiment with many error correction cycles we decode the syndrome measurements using minimum-weight perfect matching (MWPM). As the first step in this decoding process, we compare the outcomes of consecutive syndrome measurements. Consecutive measurement outcomes that differ indicate an error. We refer to the comparisons of consecutive syndrome measurements as detectors, and differing consecutive measurements as detection events~\cite{Kelly2015, Chen2021}. 

In \cref{fig:logical_X}(b) we plot the probability of detection events over time for each ancilla, and for different values of the cat mean photon number $|\alpha|^2$. These probabilities increase in proportion with $|\alpha|^2$, reflecting the fact that the phase-flip error rates of the cat qubits scale with photon number. Notably, the detection probabilities in our system are approximately constant over time. We attribute the constant detection probabilities here to the dissipative stabilization of the cat qubits, which naturally prevents the accumulation of leakage out of the cat qubit subspace without requiring additional protocols for active leakage suppression~\cite{Miao2023,lacroix2023}. 

Further improvements in error decoding can be achieved by making use of the fact that  $|f\rangle\rightarrow |e\rangle$ ancilla transmon decay errors constitute detectable erasure errors~\cite{Bennett1997, Grassl1997, Levine2024, Koottandavida2024}. Specifically, while the $\chi$-matching ensures that decay to $|e\rangle$ is unlikely to cause a bit-flip error, the decay has a high probability ($\sim 50\%$) to cause a syndrome measurement error.  As shown in \cref{fig:logical_X}(c), a $T_1$-decay error from $|f\rangle \to |e\rangle$ can be understood as an erasure because it takes the ancilla outside of its computational subspace $|g\rangle$/$|f\rangle$.  We detect these erasures using a three-state transmon readout that separately resolves $|g\rangle$, $|e\rangle$, and $|f\rangle$. The heatmap shows the occurrence of erasure events (indicated in red) interspersed among valid syndromes in the data (gray shades).  When an erasure occurs, the corresponding syndrome measurement provides no information about errors in the data qubits. We account for these erased syndromes in decoding by constructing detectors only using the non-erased syndromes (see \cref{app:decoding_with_erasure_information} for more details). As shown in \cref{fig:logical_X}(d), doing so effectively reduces the syndrome measurement error probability by over a factor of two for $|\alpha|^2=1$. 

We characterize the repetition code's ability to correct cat qubit phase-flip errors by measuring the decay time of the repetition code logical operator $\hat{X}_L=\hat{X}_1$. To do so, we initialize the system into a logical $X$ state by measuring the parity of each cat qubit's steady state under stabilization, which randomly prepares the logical qubit into one of the $2^{d}$ possible product cat states (e.g. $|+\rangle|-\rangle|-\rangle|+\rangle|+\rangle$). Next, we perform a variable number of syndrome measurement cycles.  Finally we extract $\hat{X}_L$ by measuring the parity of each storage-mode state.  Corrections from the MWPM decoding are applied in software. We fit $\langle \hat{X}_L(t=0)\hat{X}_L(t=t) \rangle$ to a decaying exponential and define the decay time constant, $T_X$, as the logical $X$ lifetime. The averaging of $\langle \hat{X}_L(t=0)\hat{X}_L(t=t) \rangle$ is based on the distribution from which the product cat states are sampled, which notably is nonuniform especially at low $|\alpha|^2$ (see \cref{app:distribution_of_initial_states_in_logical_X_lifetime_experiments} for further discussion). Example exponential fits are shown in \cref{fig:logical_X}(e) for various cat mean photon numbers $|\alpha|^2$ for the distance-5 code. From $T_X$ we compute the logical phase-flip error per cycle as $\epsilon_{L, \text{phase-flip}}={T_{\mathrm{cycle}}}/{(2 T_X)}$ (equivalent to $p_Z + p_Y$).

The cat qubit architecture gives us the ability to study the performance of the error correcting code in situ, since we can tune the data qubit phase-flip error rate by varying $|\alpha|^2$. In \cref{fig:logical_X}(f) we plot the measured $\epsilon_{L, \text{phase-flip}}$ versus $|\alpha|^2$ for the distance-5 repetition code and the two minimally overlapping distance-3 repetition codes contained within it. As the photon number increases, we see the expected increase in the logical error probability as the likelihood of a higher-order error not being caught by the error correction increases.  Across the measured range of $|\alpha|^2$, we find that the distance-5 code outperforms the distance-3 subsections. This indicates that the physical phase-flip error rates of our system are below the repetition code's error threshold for the entire range of photon numbers we consider. Note also that there is a sizable reduction in the logical phase-flip rate when the erasure information is incorporated (e.g., by $\sim 20\%$ at $|\alpha|^{2}=1.5$ for $d=5$), a result of the reduced effective measurement error probabilities achieved via the erasure conversion.  

More quantitatively, the logical phase-flip rate is expected to scale as $(|\alpha|^{2})^{\gamma}$~\cite{Dennis2002, Fowler2012}, where $|\alpha|^2$ is a proxy for the cat qubit phase-flip error rate.  When the erasure information is incorporated, we estimate from fits to the measured logical phase-flip probability versus $|\alpha|^2$, scaling exponents of $\gamma = 1.63 \pm 0.04$ and $\gamma=1.86\pm 0.03$ in the two $d=3$ subsections, and $\gamma = 2.31 \pm 0.02$ in the full $d=5$ section. The increase in scaling exponent from $d=3$ to $d=5$ shows that the increased code distance is providing greater resiliency to phase-flip errors. Although the measured values of $\gamma$ are lower than the ideal values, $\gamma=(d+1)/2$, they are consistent with simulations (shown in \cref{fig:logical_X}(f) as a dotted purple curve for the distance-5 code) based on a simple model that incorporates the measured probabilities of cat phase flips, ancilla erasures, and syndrome measurement error. We attribute the scaling behavior to the close proximity of our data qubit phase-flip error rates to the code threshold---a regime where the idealized scaling is not generally expected to hold (see \cref{app:explanation_logical_phase_scaling}). 

\vspace{4ex}
\tocless\section{Maintaining long bit-flip times in a repetition cat code}

Having demonstrated the ability to correct the dominant phase-flip errors of cat qubits via a repetition code, we now proceed to characterize the logical bit-flip rates. Unlike the logical phase flips which are corrected using the repetition code syndrome measurements, logical bit flips are passively suppressed at the level of the individual cat qubit encodings.  As a result achieving long logical bit-flip times is challenging because any single cat-qubit bit-flip event in any part of the repetition code directly causes a logical bit-flip error. Moreover simultaneous syndrome extraction across the entire chain of the repetition code can cause various types of crosstalk. Here, we present strategies for overcoming these challenges and demonstrate that long logical bit-flip times can be maintained during the syndrome extraction of the repetition code in our device.

To achieve a low logical bit-flip error rate, large noise bias must be maintained on every single cat qubit while syndrome measurements are performed.  This requires all the storage-ancilla interactions to have a sufficiently $\chi$-matched interaction, which we achieve by accurately targeting storage-ancilla detunings across the entire device (see \cref{app:frequency_targeting_procedure_for_chi_matching}). Additionally, we carefully tune the syndrome measurements to avoid parasitic effects that can induce cat-qubit bit-flip errors, including measurement-induced state transitions that can excite the couplers and ancillas~\cite{Sank2016},  spurious two-level systems (TLSs) in the ancillas~\cite{Klimov2018}, non-adiabatic errors from the CX gate, and undesired nonlinear buffer resonances. In addition, moving from isolated $\text{CX}$ gate tune-up to simultaneous stabilizer measurements, the $\text{CX}$ gate fidelities can degrade due to crosstalk. To counter this we perform in situ calibration of storage and ancilla phases associated with the $\text{CX}$ gates (see \cref{app:insitu_calibration}).   We also find that due to frequency collisions, some readout resonators can be unintentionally excited by buffer flux pumps. We mitigate this crosstalk mechanism though active compensation.  

\begin{figure}[t!]
    \centering
    \includegraphics[width=\columnwidth]{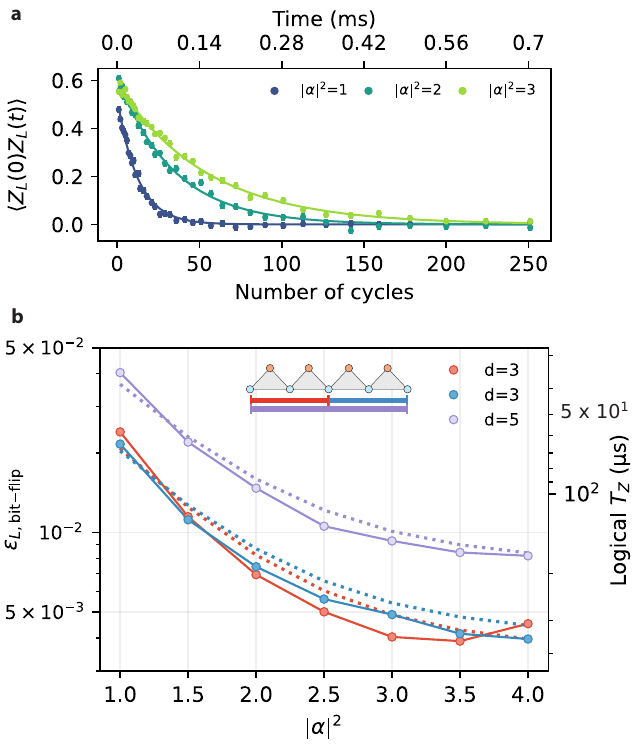}
    \caption{\textbf{Characterizing logical bit-flip error rates.} (a) Fitted decay curve of the logical $Z$ operator for several storage mean photon numbers for the distance-5 code. Error bars represent sampling variance.  (b) Logical bit-flip probability per cycle ($\epsilon_{L,\text{bit-flip}} = p_X+p_Y$) and logical $Z$ lifetime ($T_Z$) as a function of $|\alpha|^2$ for the two distance-3 repetition-code sections and the distance-5 section. Solid lines correspond to data and dotted lines correspond to a phenomenological model. Error bars, capturing sampling variance and fit uncertainty, are smaller than the markers.  }
    \label{fig:logical_Z}
\end{figure}

With the calibration carefully tuned to avoid additional bit-flip mechanisms at the repetition-code level, we move onto characterizing the logical bit-flip probabilities by measuring the decay time, $T_{Z}$, of the logical $Z$ operator $\hat{Z}_{L} = \hat{Z}_{1}\hat{Z}_{2}\cdots\hat{Z}_{d}$. To do so, we first initialize the data cat qubits in a tensor-product of coherent states (e.g., $|\alpha\rangle^{\otimes d}$ and $|{-}\alpha\rangle^{\otimes d}$). Then, we repeatedly apply a variable number of syndrome extraction cycles. Finally, we perform single-shot $Z$-basis measurements (see \cref{app:storage_z_measurement}) on all of the data cat qubits to measure $\hat{Z}_{L}$. Note that after the first round of stabilizer measurements, the data qubits are projected into an eigenstate of the logical $Z$ operator up to random phase flips that do not affect the logical $Z$ measurement. The logical $Z$ lifetime, $T_{Z}$, is then obtained by fitting the decay curve of $\langle \hat{Z}_L(t=0)\hat{Z}_L(t=t) \rangle$ to an exponential.  Example fits are shown in \cref{fig:logical_Z}(a).  From $T_Z$ we compute the logical bit-flip error per cycle as $\epsilon_{L, \text{bit-flip}}={T_{\mathrm{cycle}}}/{(2 T_Z)}$ (equivalent to $p_X + p_Y$).

In \cref{fig:logical_Z}(b) we show $\epsilon_{L, \text{bit-flip}}$ as a function of $|\alpha|^2$ for the distance-5 (purple) and two distance-3 (red and blue) sections.  At a low cat mean photon number of $|\alpha|^2=1$, the logical bit-flip error per cycle are approximately $2\%$ for the two distance-3 sections and $4\%$ for the distance-5 section.  As the cat mean photon number increases to $|\alpha|^{2}=4$, the logical bit-flip error per cycle drop to below $0.5\%$ for the two distance-3 sections and below $1\%$ for the distance-5 section, due to the increased level of bit-flip protection from the cat qubits.  Note that $\epsilon_{L, \text{bit-flip}}$ combines together the bit-flip error contributions from all the cat qubits and $\text{CX}$ gates. Since the distance-5 section has more bit-flip error locations than the distance-3 sections, it has higher logical bit-flip probability. Nevertheless, the large noise bias maintained throughout the error correction cycle allows us to achieve sub-$1\%$ logical bit-flip probability even for the distance-5 section which involves $5$ cat qubits and $8$ $\text{CX}$ gates.

Also shown in \cref{fig:logical_Z}(b) is a phenomenological model of the logical bit-flip errors. In this model, the total logical bit-flip error probability is a sum of $|\alpha|^2$-dependent, idling bit-flip error probabilities for each cat qubit, together with additional $|\alpha|^2$-independent bit-flip probabilities for each CX gate. The values of these physical error probabilities are fit from independent $\text{CX}$ and cat-qubit characterization experiments (see \cref{app:logical_bit_flip_projections}). The agreement between the model and measurements indicates that there is no significant degradation in the bit-flip performance of individual cat qubits or CX gates when integrated together into the repetition code. 

\vspace{4ex}
\tocless\section{Overall memory lifetime and error budget}

\begin{figure}[t!]
    \centering
    \includegraphics[width=\columnwidth]{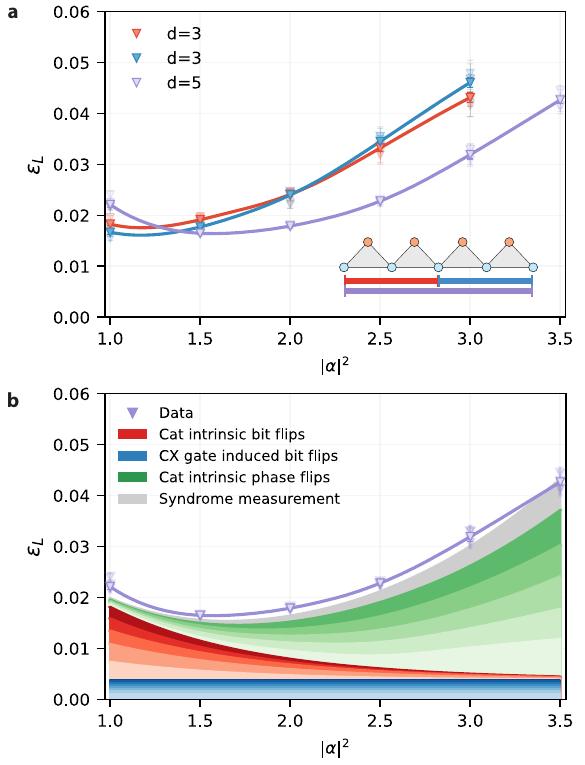}
    \caption{\textbf{Logical qubit memory performance.} (a) Overall logical error per cycle of the repetition cat code, $\epsilon_{L}=(\epsilon_{L, \text{bit-flip}}+\epsilon_{L, \text{phase-flip}})/2 = (p_X + 2p_Y + p_Z)/2$, versus cat qubit mean photon number $|\alpha|^2$.  As in \cref{fig:logical_X}(f), the faded points correspond to fits to groupings of the data by the number of even cat states in the initial state.  The lines are guides to the eye, computed by interpolating both the logical phase-flip and bit-flip probabilities, $\epsilon_{L,\text{phase-flip}}$ and $\epsilon_{L,\text{bit-flip}}$, respectively. (b) Error budget for the distance-5 repetition code using erasure information.  Different shades of color correspond to the different per cat (or per CX gate) contribution to the error budget. Error bar incorporate sampling variance and fit uncertainty.}
    \label{fig:total_lifetime}
\end{figure}

Combining together the logical bit-flip and phase-flip probabilities, we show in \cref{fig:total_lifetime}(a) the overall logical error per cycle \cite{Krinner2022,Acharya2023}, $\epsilon_{L}= (\epsilon_{L,\text{phase-flip}} + \epsilon_{L,\text{bit-flip}} ) / 2 = (p_X + 2p_Y + p_Z )/ 2$, for the repetition cat codes.  Since $\epsilon_{L,\text{phase-flip}}$ increases with $|\alpha|^{2}$, while $\epsilon_{L,\text{bit-flip}}$ decreases with $|\alpha|^{2}$, $\epsilon_{L}$ is minimized at a certain value of $|\alpha|^{2}$. We find the measured logical error probabilities of the distance-3 sections are minimized near $|\alpha|^2=1$, while that of the distance-5 code is minimized at a higher photon number near $|\alpha|^2 = 1.5$. This is because the shorter codes provide less protection against phase-flip errors, but simultaneously have fewer locations for physical bit-flip errors to lead to logical bit-flip errors. Thus, as the mean photon number increases to $|\alpha|^2=1.5$ or higher, the performance of a distance-3 section quickly becomes limited by phase-flip errors which are not sufficiently suppressed by a short repetition code. In contrast, the distance-5 code has better protection from phase-flip errors. This enables the distance-5 code to operate at higher values of $|\alpha|^2$ and benefit from the larger noise bias of the cat qubits. The best measured performance for the distance-5 section is $\epsilon_{L}=1.65\% \pm 0.03 \%$ at $|\alpha|^2=1.5$. This is comparable to the best observed performance for the distance-3 sections which are $\epsilon_{L}=1.83\% \pm 0.03 \%$ and $\epsilon_{L}=1.67\% \pm 0.04 \%$ at $|\alpha|^2=1$. 

For each value of $|\alpha|^2\geq 1.5$, the logical error rate of the distance-5 code is lower than that of the distance-3 sections. Without noise bias, logical error rates would only increase with code distance, since the decrease in logical phase-flip error rate would be outweighed by the corresponding increase in logical bit-flip error rate.  However, with the large noise bias of the cat qubits and the CX gates, the logical phase-flip error contribution dominates for mean photon numbers above $|\alpha|^2 = 1.5$, and thus we benefit from using a larger distance repetition code.

Shown in \cref{fig:total_lifetime}(b), we use models of the logical bit-flip and phase-flip errors to construct an error budget for the distance-5 repetition cat code (see \cref{app:error_budget} for details). The error budget is broken into four error mechanisms: cat intrinsic bit-flip errors (red), CX gate induced bit-flip errors (blue), cat intrinsic phase-flip errors capturing idling and CX gate phase-flips (green), and syndrome measurement errors (gray). The first two contribute to the logical bit-flip rate while the latter two contribute to the logical phase-flip rate.  The bit-flip mechanisms dominate at small $|\alpha|^2$, and the phase-flip mechanisms dominate at large $|\alpha|^2$. The minimum logical error rate is achieved at $|\alpha|^2\simeq 1.5$ where the bit-flip and phase-flip contributions are comparable. Notably, at this optimal value of $|\alpha|^2\simeq 1.5$, the cat intrinsic  bit-flip and phase-flip errors are the dominant contributors. Thus the minimum logical error rate of our repetition cat code is limited primarily by the individual cat-qubit errors, rather than by additional CX gate induced bit-flip errors or syndrome measurement errors caused by the ancilla transmons.  

\vspace{4ex}
\tocless\section{Conclusion and Outlook}

In this work, we have performed error correction using a concatenated bosonic code, where bit-flip errors are suppressed with a bosonic cat code, and residual phase-flip errors are corrected with a repetition code. This experiment serves as a promising first step in taking advantage of bosonic qubits, and additionally noise bias, to improve the hardware efficiency of quantum error correction.  Furthermore, having constructed our logical qubit memory using planar microfabrication processes, this work highlights the potential scalability of the concatenated bosonic qubit architecture.

The logical error in our current device is dominated by intrinsic cat bit-flip and phase-flip errors (see \cref{fig:total_lifetime}(b)), but there are several strategies to reduce these errors in the near term. To reduce both types of error, the cycle time could be reduced by a factor of $\sim 2$ by removing  padding in the pulse sequence and achieving more uniform CX gate lengths across the device. Furthermore cat-qubit bit-flip rates can be reduced (see \cref{app:buffer_model,app:two_photon_dissipation_calibration}) by using optimized circuit parameters to achieve higher performing two-photon dissipation as demonstrated in Ref.~\cite{singlecat2024}. We project that with these improvements, an overall logical error per cycle approaching $0.5\%$ (limited by transmon errors) is achievable with a distance-5 code even without improvements in the component coherence times. 

The use of ancillary transmons for syndrome measurements is critical to our experiment, enabling coherence-limited, noise-biased CX gates.  While the current performance of our experiment is not limited by the cat bit-flip errors induced by ancilla transmon double decay and heating, these mechanisms would ultimately place a lower bound on the logical error probability of our repetition cat code. This performance floor can be lowered in the near term by improving the coherence of the ancilla transmons. A scalable, long-term approach to overcome the remaining limitation, while still enjoying the practical benefits of the ancilla transmon, is to concatenate cat qubits into rectangular surface codes~\cite{Chamberland2022} or XZZX surface codes~\cite{BonillaAtaides2021_XZZX} which are tailored to noise-biased qubits. We analyze this approach in Ref.~\cite{theorypaper2024}, and find that significant hardware-efficiency improvements are possible relative to the case without biased noise. 

An alternative approach to overcome the transmon-induced limitations is to utilize cat qubits as the ancillas as well as the data qubits. This was proposed early on in Ref. ~\cite{Guillaud2019Repetition}, but existing proposals for cat-cat CX gates are hampered by large control errors~\cite{Guillaud2019Repetition,Guillaud2021_error,Chamberland2022}.
Searching for ways to implement syndrome measurements with a large noise bias but without undesired control errors~\cite{Cohen2017,Xu2023,Gautier2023} thus represents an important direction for future research. Indeed, if the performance of gates were limited only by the cats' intrinsic bit-flip and phase-flip rates, sizable reductions in logical-memory overhead would be possible with realistic device parameters. For example, in Ref.~\cite{singlecat2024} we show a cat qubit with bit-flip times approaching $1~\mathrm{s}$ at $|\alpha|^2=5$ that correspond to a bit-flip error per cycle of $10^{-6}$ assuming $1~\mathrm{\mu s}$ error-correction cycles. With improvement of the storage lifetime to $\sim300~\mathrm{\mu s}$~\cite{Place2021}, we project (see \cref{app:optimistic_overhead_estimates}) that an overall logical error per cycle below $10^{-5}$ could be achieved with a repetition code comprising only $d=11$ cat qubits. Further, with $\gtrsim 100~\mathrm{s}$ bit-flip times~\cite{Berdou2023}, and ms-scale storage $T_1$~\cite{Reagor2016}, algorithmically-relevant logical error per cycle of $\sim10^{-8}$ could be achieved with similar overhead. While these examples assuming coherence-limited gates are idealized hypotheticals, they nevertheless highlight the potential of cat qubits to enable hardware-efficient logical qubits.

\vspace{4ex}
\tocless\section{acknowledgments}
\label{sec:ack}

We thank the staff from across the AWS Center for Quantum Computing that enabled this project. We also thank Fiona Harrison, Harry Atwater, David Tirrell, and Tom Rosenbaum at Caltech, and Simone Severini, Bill Vass, James Hamilton, Nafea Bshara, and Peter DeSantis at AWS, for their involvement and support of the research activities at the AWS Center for Quantum Computing. 

\vspace{4ex}
\tocless\section{Author contributions}
The transmon-ancilla architecture was developed by H.P., K.N., and C.H. The CX gate implementation was developed by H.P. K.N., and C.H. The device parameter specification was led by H.P. and K.N.  Circuit-level modeling of the device was led by K.N. The design of the device was led by S.A., with earlier versions led by M.L. The fabrication was led by G.M. and M.M., with key process modules developed by M.J., H.M., R.R., N.M., and J.R. The fridge and instrumentation setup was specified by H.P., R.P., J.O., and L.M. The repetition code experiment calibration was developed by H.P with inputs from R.P., J.O, H.L., E.R. and P.R. H.P. developed the repetition code experiment protocols and performed the experiment. Analysis of the data was performed by H.P. with input from K.N. and C.H. C.H. and H.P. implemented the decoding using erasure information. C.H. performed the performance simulations and logical error budgeting. The tuning procedure to achieve the required frequency targeting was developed by K.N., H.P., S.A., M.J., M.M., and G.M. The project was managed by M.M., and overseen by F.B. and O.P.  The bulk of the manuscript was written by H.P., K.N., and C.H., with O.P., M.M., G.M., C.R., J.P., Y.Y, H.L., S.A., and F.B. reviewing and editing the manuscript. All other authors contributed to developing technical infrastructure such as fab modules, control hardware, cryogenic hardware, software, calibration modules, design tools, and simulation packages used for the experiment and its analysis.

\vspace{4ex}
\tocless\section{Data availability}
Data is available from the authors upon reasonable request.

\vspace{4ex}
\tocless\section{Competing interests}
The authors declare no competing interests.

\bibliographystyle{naturemag}
\let\oldaddcontentsline\addcontentsline
\renewcommand{\addcontentsline}[3]{}
\bibliography{rep_cat}
\let\addcontentsline\oldaddcontentsline

\clearpage
\appendix

\onecolumngrid
\begin{center}
{\large \textbf{Supplemental information for "Hardware-efficient quantum error correction via concatenated bosonic qubits"} \par}
\end{center}
\twocolumngrid
\tableofcontents

\section{Device fabrication}
\label{app:device_fab}

The repetition code device consists of two silicon dies fabricated separately on high-resistivity silicon then flip-chip bonded together \cite{Foxen2018, Das2018}, as described in Ref.~\cite{singlecat2024}. The $16 \times 14~\text{mm}^2$ aluminum-based ``qubit'' die contains Al/AlOx/Al Josephson junctions to form nonlinear circuit elements in the ancilla transmons, couplers, and buffers. The second, $20 \times 20~\text{mm}^2$ tantalum-based die is patterned to form storage resonators and other linear elements such as the readout resonators, Purcell filters for the readout resonators, and filters for the buffers. Indium bumps are deposited on both dies to facilitate flip-chip bonding and electrical connection between the dies, as well as an under-bump metallization layer on the aluminum-based die. Bond-stop spacers are used to improve the uniformity of the gap between dies. After flip-chip bonding of the paired dies, signal line pads at the periphery of the tantalum-based die are wirebonded to a PCB with aluminum wirebonds.

We have separated the fabrication of constituent ``qubit'' and tantalum-based dies into distinct process flows tailored for their individual requirements. This separation has several benefits. First, this allows us to fabricate high-coherence storage modes in thin-film tantalum~\cite{Place2021,Crowley2023} without process integration constraints imposed by fabricating Al/AlOx/Al Josephson junctions on the same wafer. Second, we are free to test and select die pairs which have favorable frequency alignment throughout the unit cells of the repetition code, particularly between the ancillas on the qubit die and storage resonators on the tantalum-based die. By using the estimated Josephson junction energies measured at room-temperature to inform the circuit layout for the tantalum-based dies, we can partially account for both systematic and stochastic junction energy variations in the flip-chip integrated device, as detailed in \cref{app:frequency_targeting_procedure_for_chi_matching}. Third, systematic imperfections in the fabricated Josephson junction energies on the qubit die relative to their designed parameters (e.g. due to spatial gradients across a wafer) can generally be reduced by keeping the die size smaller. The latter two benefits are especially important in scaling up from unit-cell devices as in Ref.~\cite{singlecat2024} to repetition codes.

\section{Control lines and fridge setup}
\label{app:control_lines_and_fridge_setup}

\begin{figure*}[t!]
    \centering
    \includegraphics[width=1.85\columnwidth]{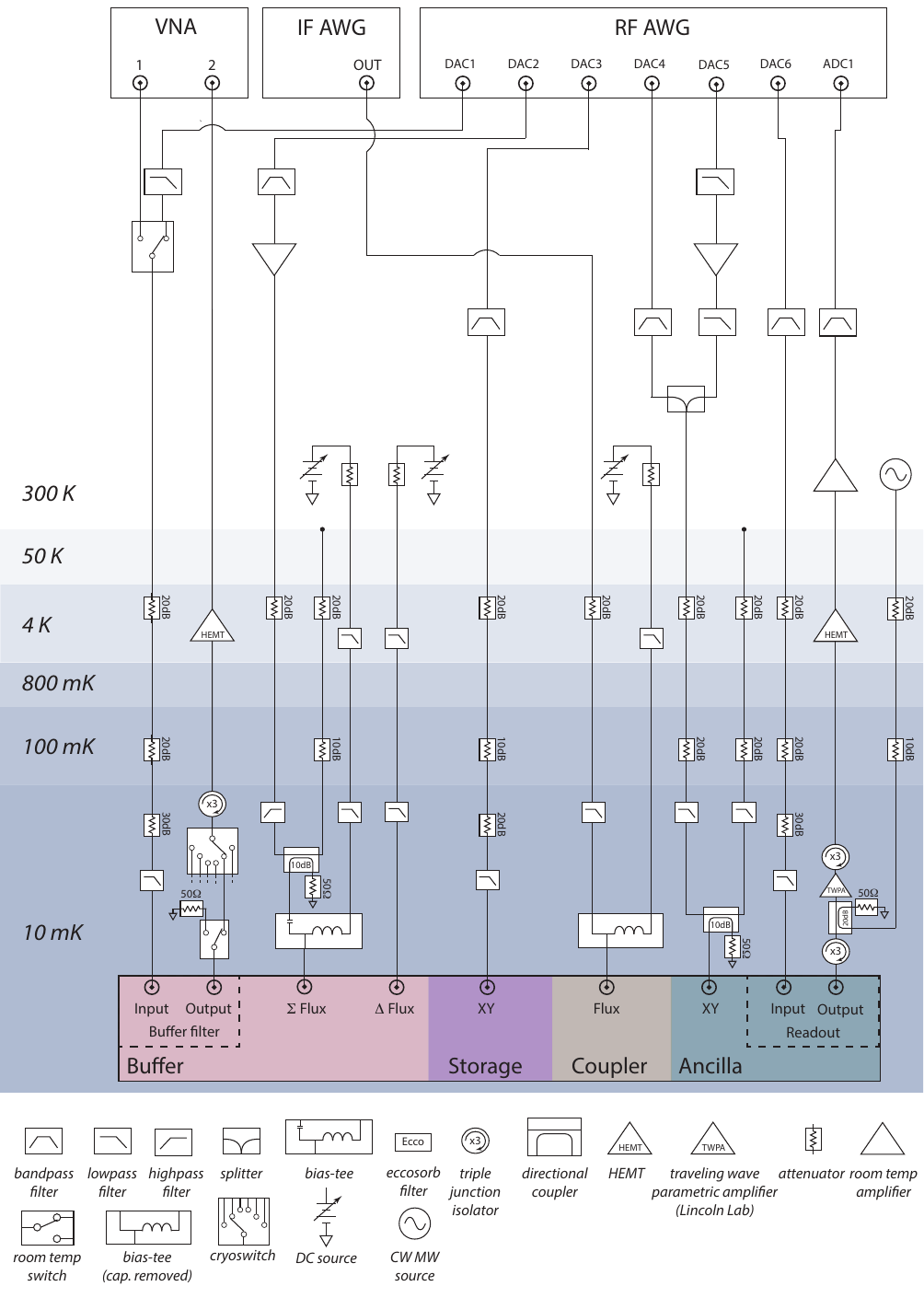}
    \caption{\textbf{Wiring diagram.}  Room temperature and fridge setup for one unit cell.  Eccosorb filters, room temperature switches for toggling between characterization of different components, and additional lines used for crosstalk compensation (see \cref{app:crosstalk}) are not included in the diagram.  }
    \label{app_fig:wiring_diagram}
\end{figure*}

The buffer is flux-biased by $\Sigma$ and $\Delta$ control lines which drive flux symmetrically and antisymetrically in the buffers flux loops~\cite{Lescanne2020}.  To drive the buffer mode and set the steady-state for the two-photon dissipation, we drive an input line capacitively-coupled to the first unit cell of the filter array.  The buffer output is a transmission line coupled to the last element of the buffer mode filter.  This input line and output line can also be used for calibration of the buffer mode by spectroscopically measuring the buffer mode.  When not performing readout of the buffer, the buffer output is terminated to a 50 Ohm environment.  The flux pump to realize the two-photon dissipation is applied on the $\Sigma$ flux bias line.

The storage mode can be displaced with a charge-coupled drive line.  The coupler bias is controlled through a flux line which addresses its SQUID loop.  The ancilla has an XY control line and can be read out through its readout resonator coupled to a transmission line.  The ancillas on the chip are grouped so that two or three ancillas are read out using the same readout line in a multiplexed manner~\cite{Jeffrey2014, Walter2020}.

The dilution fridge and room temperature setup used for one unit cell of the device is shown in \cref{app_fig:wiring_diagram}.  This setup is duplicated across all of the unit cells in the repetition code. Note that compared to Ref.~\cite{singlecat2024} the wiring diagram is almost identical. The main differences are the buffer output line configuration which uses a cryoswitch to allow multiple buffers to be read out with one readout line and some minor differences in room temperature filtering.

We temperature cycle the dilution fridge to shift spurious two-level systems (TLSs) \cite{Klimov2018} if they are at problematic frequencies (see \cref{app:noise_mechanisms}).  The need for this could be minimized in the future by moving to tunable-frequency ancilla qubits.

\section{Basis and error rate convention}
\label{app:basis_and_error_rate_convention}

Similarly as in Refs.~\cite{Guillaud2019Repetition,Chamberland2022}, we use the basis convention where the complementary basis states $|\pm\rangle$ of the cat qubit are given by the even and odd cat states, i.e., 
\begin{align}
    |\pm\rangle \equiv |C^{\pm}_{\alpha}\rangle = \frac{|+\alpha\rangle \pm |{-}\alpha\rangle}{\sqrt{2(1 \pm e^{-2|\alpha|^{2}})}} . 
\end{align}
Thus, the computational basis states $|0\rangle = (|+\rangle + |-\rangle) / \sqrt{2}$ and $|1\rangle = (|+\rangle - |-\rangle) / \sqrt{2}$ are approximately given by the coherent states, i.e., $|0\rangle \simeq |+\alpha\rangle$ and $|1\rangle \simeq |{-}\alpha\rangle$ in the limit of $e^{-2|\alpha|^{2}} \ll 1$. In this basis convention, a Pauli $X$ operator $\hat{X}$ of the cat qubit is simply given by the photon number parity operator $\exp[i\pi\hat{n}]$ where $\hat{n}=\hat{a}^{\dagger}\hat{a}$ is the photon-number operator. As a result, an $X$-basis readout of a cat qubit can be straightforwardly implemented by a photon-number parity measurement. 
    
Since the computational basis states are given by two coherent states $|\pm\alpha\rangle$, reading out a cat qubit in the $Z$ basis requires the ability to distinguish the two coherent states. One way to achieve this is to perform displaced-parity measurements and average the measurement outcomes over many shots (see, e.g., Ref.~\cite{singlecat2024}). However, this approach does not realize the $Z$-basis readout in a single-shot manner which is necessary for characterizing one of the logical operators of the repetition cat code. Thus in this work, we instead use displaced-vacuum population measurements \cite{theorypaper2024} to perform a single-shot $Z$-basis readout of a cat qubit (see \cref{app:storage_z_measurement} for more details).    

In our basis convention, the cat qubits have noise bias towards phase-flip (Z) errors which are caused by parity flips due to photon losses. Thus as an outer error-correcting code, we use a distance-$d$ repetition code whose stabilizer generators are given by $\hat{X}_{i}\hat{X}_{i+1}$ for $1 \le i \le d-1$. With this choice, the stabilizers of the repetition code anti-commute with the $Z$ errors and thus can extract non-trivial error syndromes of the dominant $Z$ errors of the data cat qubits. 

The logical $Z$ operator of the repetition code is given by $\hat{Z}_{L}=\hat{Z}_{1}\hat{Z}_{2}\cdots \hat{Z}_{d}$. Then the correctability of the $Z$ errors is illustrated by the large irreducible Hamming weight of the logical $Z$ operator. In contrast, any single-qubit Pauli $X$ operator acts as a logical $X$ operator of the repetition code, i.e., $\hat{X}_{L} = \hat{X}_{i}$ for any $1\le i \le d$. For example, the action of $\hat{X}_{1}$ on the repetition code states is equivalent to that of $\hat{X}_{2}$ since these two operators only differ by the multiplication of a stabilizer $\hat{X}_{1}\hat{X}_{2}$ which acts trivially in the code space. As a result, a bit-flip (X) error on any single qubit can directly cause a logical $X$ error in the repetition code. Thus it is crucial that the bit-flip errors on the data cat qubits are suppressed at the physical level. 

To characterize the logical lifetimes and the error rates of the repetition cat code, we measure $\langle \hat{X}_{L}(t=0)\hat{X}_{L}(t=t)\rangle$ and $\langle \hat{Z}_{L}(t=0)\hat{Z}_{L}(t=t)\rangle$ using the pulse sequences in \cref{app_fig:X_basis_experiment} and \cref{app_fig:Z_basis_experiment}, respectively. Then we fit $\langle \hat{X}_{L}(t=0)\hat{X}_{L}(t=t)\rangle$ to an exponential decay curve with a constant offset, i.e., $\langle \hat{X}_{L}(t=0)\hat{X}_{L}(t=t)\rangle = A\exp[-t/T_{X}] + B$ and define the logical $X$ lifetime as the decay time constant $T_{X}$. Note that the constant offset is added to account for the non-zero asymptotic value of $\langle \hat{X}_{L}(t=0)\hat{X}_{L}(t=t)\rangle$ in the long time limit, which is expected in the small $|\alpha|^{2}$ regime due to the asymmetric phase-flip rates between the even and odd cat states (see \cref{app:distribution_of_initial_states_in_logical_X_lifetime_experiments} for more details). For the logical $Z$ lifetime, we fit $\langle \hat{Z}_{L}(t=0)\hat{Z}_{L}(t=t)\rangle$ to an exponential decay curve without a constant offset, i.e., $\langle \hat{Z}_{L}(t=0)\hat{Z}_{L}(t=t)\rangle = A\exp[-t/T_{Z}]$ and define the extracted decay time constant $T_{Z}$ to be the logical $Z$ lifetime. A constant offset is not included in this case because the bit-flip rates are not state-dependent and thus $\langle \hat{Z}_{L}(t=0)\hat{Z}_{L}(t=t)\rangle$ is expected to vanish in the long time limit. Based on these logical $X$ and $Z$ lifetimes $T_{X}$ and $T_{Z}$, we define the logical phase-flip and bit-flip error per cycle as
\begin{align}
    \epsilon_{L,\text{phase-flip}} &= p_{Z} + p_{Y} \equiv \frac{T_{\text{cycle}}}{2T_{X}}, 
    \nonumber\\
    \epsilon_{L,\text{bit-flip}} &= p_{X} + p_{Y} \equiv \frac{T_{\text{cycle}}}{2T_{Z}}, 
\end{align}
where $T_{\text{cycle}}$ is the error-correction cycle time which is given by $T_{\text{cycle}} = 2.8~\mathrm{\mu s}$ in our experiments. Note that the factor of $2$ in the denominator emerges from the fact that two subsequent errors cancel out in a simple Lindblad noise model with Pauli jump operators~\cite{singlecat2024, Krinner2022,Acharya2023,Sundaresan2023,Acharya2024}. Following Refs.~\cite{Krinner2022,Acharya2023,Sundaresan2023,Acharya2024}, we define an overall logical error per cycle as 
\begin{align}
    \epsilon_{L} \equiv \frac{1}{2}(\epsilon_{L,\text{phase-flip}} + \epsilon_{L,\text{bit-flip}}) . 
\end{align}
Lastly we remark that for the single-cat bit-flip and phase-flip times, we follow the same convention (including the factor of $2$) as in Ref.~\cite{singlecat2024}.

\section{Device Hamiltonian}
\label{app:device_hamiltonian}

Here we provide more details on the Hamiltonian of a unit cell of our device (see e.g., \cref{fig:device}(c) and \cref{fig:architecture}(a)). In particular, we focus on the aspects that are crucial for realizing a noise-biased CX gate between a data cat qubit and an ancilla transmon qubit. The desired effective Hamiltonian for realizing the CX gate takes the form 
\begin{align}
    \hat{H} &= \Big{(}g_{2}(t)(\hat{a}^{2} - \alpha^{2})\hat{b}^{\dagger} + \text{H.c.}\Big{)} 
    \nonumber\\
    &\quad + \chi_{ge}(\Phi_{x,c}(t))\hat{a}^{\dagger}\hat{a} |e\rangle\langle e| + \chi_{gf}(\Phi_{x,c}(t))\hat{a}^{\dagger}\hat{a} |f\rangle\langle f|. 
\end{align}
As will be made clear shortly, however, $g_{2}(t)$ and $\Phi_{x,c}(t)$ are never turned on simultanesouly in our experiments. Here, $\hat{a}$ and $\hat{b}$ are the annihilation operators of the storage and the buffer, respectively. $g_{2}(t)$ is the strength of the three-wave mixing (3WM) interaction between them. Combined with the buffer loss $\kappa_{b}\mathcal{D}[\hat{b}]$, the storage-buffer interaction $g_{2}(t)(\hat{a}^{2} - \alpha^{2})\hat{b}^{\dagger} + \text{H.c.}$ realizes a two-photon dissipation of the form $\kappa_{2}(t)\mathcal{D}[\hat{a}^{2}-\alpha^{2}]$ which stabilizes a cat qubit with a mean photon number of approximately $|\alpha|^{2}$. Note that similarly as in Ref.~\cite{singlecat2024}, we use the terminology ``pure two-photon dissipation'' to refer to the dynamics due to $\kappa_{2}\mathcal{D}[\hat{a}^{2}]$ (i.e., two-photon dissipation with $\alpha=0$). 

Note that $|e\rangle$ and $|f\rangle$ are the first and the second excited states of the ancilla transmon. Thus, $\chi_{ge}(\Phi_{x,c}(t))$ and $\chi_{gf}(\Phi_{x,c}(t))$ represent the dispersive shift of the storage mode frequency as the ancilla transmon is excited from the ground state $|g\rangle$ to an excited state $|e\rangle$ and $|f\rangle$, respectively. During a CX gate, the ancilla transmon should ideally be in the $|g\rangle/|f\rangle$ manifold. However in practice the ancilla may have a non-zero $|e\rangle$ state population for various reasons such as state-preparation errors and an $|f\rangle\rightarrow |e\rangle$ decay process.  

The dispersive coupling strengths $\chi_{ge}(\Phi_{x,c}(t))$ and $\chi_{gf}(\Phi_{x,c}(t))$ depend on the external flux $\Phi_{x,c}$ of a tunable coupler. Each coupler in our system is implemented by a tunable-frequency transmon. Ideally when the coupler is in an ``off'' position ($\Phi_{x,c}=0$ maximizing the coupler frequency), both $\chi_{ge}$ and $\chi_{gf}$ should vanish. Moreover when the coupler is in the ``on'' position ($\Phi_{x,c} = \Phi_{0}/2$, or the half flux quantum, minimizing the coupler frequency), we ideally need to achieve the ``$\chi$-matching'' condition $\chi_{ge} = \chi_{gf}$ such that the bit-flip rates of a cat qubit are first-order insensitive to the processes that put the ancilla transmon in the $|e\rangle$ state when it is supposed to be in the $|f\rangle$ state. 

In our implementation of the $\text{CX}$ gate, the two-photon dissipation and the flux pulse are never turned on simultaneously. That is, the two-photon dissipation is turned on only when the coupler is in the ``off'' position (i.e., $\Phi_{x,c}(t)=0$ when $|g_{2}(t)|>0$) and not when a coupler flux pulse is applied (i.e., $g_{2}(t)=0$ when $|\Phi_{x,c}(t)|>0$).  This way, the cat qubit in the storage mode can freely rotate when a coupler flux pulse activates the storage-ancilla dispersive coupling.  

In this section, we first provide more details on the buffer modes in our device including nonidealities in some unit cells in \cref{app:buffer_model}. Then moving on to the $\text{CX}$ gates we discuss in \cref{app:CX_Kerr_oscillator_model} how the $\chi$-matching condition can be natively achieved for the $\text{CX}$ gates through frequency targeting based on a simple Kerr oscillator model. Finally in \cref{app:CX_circuit_quantization_model}, we provide a circuit-quantization-level model of a CX gate in our device and discuss various practical aspects of the storage-ancilla interaction for implementing the CX gate.

\subsection{Buffer mode for implementing two-photon dissipation}
\label{app:buffer_model}

The buffer mode in our device is implemented as an ATS~\cite{Lescanne2020} with additional inductors~\cite{singlecat2024} in series with the two side junctions. To protect the storage modes from the strong engineered decay on the buffer modes (e.g., with $\kappa_{b} / 2\pi \sim 10~\text{MHz}$), we use a 4-pole metamaterial bandpass filter for each buffer. The bandwidth of the filter is approximately $1\text{GHz}$ and the transmission drops sharply outside of the filter passband. This ensures that the lifetimes of the storage modes are not limited by the buffer's loss channel as in Ref.~\cite{singlecat2024}. 

In practice, various buffer-induced nonlinearities such as the storage self Kerr $K_{s}$ and storage-buffer cross Kerr $\chi_{sb}$ can negatively affect the performance of two-photon dissipation and degrade the bit-flip times of a cat qubit. Following the method detailed in Ref.~\cite{singlecat2024}, we have carefully optimized the circuit parameters of the buffer mode (especially its average side junction energy and the serial inductance) such that the self Kerr of the buffer mode is minimized at the saddle points (where the buffer frequency is first-order insensitive to the flux deviations). As a result, the buffer-induced storage self Kerr and the storage-buffer cross Kerr are predicted to be small in most unit cells of our devices (i.e., $|K_{s}/2\pi| < 2~\text{kHz}$ and $|\chi_{sb}/2\pi| < 120~\text{kHz}$) compared to the 3WM interaction strength $g_{2}/2\pi$ which range from $300~\text{kHz}$ to $450~\text{kHz}$. In one of the unit cells which exhibits the worst bit-flip performance (i.e., S4), the buffer-induced nonlinearities are predicted to be large ($K_{s}/2\pi = 2.7~\text{kHz}$ and $\chi_{sb}/2\pi = 390~\text{kHz}$ compared to $g_{2}/2\pi \sim 400~\text{kHz}$) due to mis-targeting of the buffer junction energies. These predictions are made using a detailed circuit-quantization-level model of our device, where the model parameters are systematically tuned up to match the measured buffer spectrum in the same way as in Ref.~\cite{singlecat2024}.   

The desired 3WM mixing interaction between the storage and the buffer (i.e., $g_{2}\hat{a}^{2}\hat{b}^{\dagger} + \text{H.c.}$) is realized by pumping the sigma flux of the buffer mode~\cite{Lescanne2020} with a pump frequency of $\omega_{p} = 2\bar{\omega}_{s}-\bar{\omega}_{b}$. Here, $\bar{\omega}_{s}$ and $\bar{\omega}_{b}$ are the Stark-shifted frequencies of the storage and the buffer under the buffer pump. As shown in Ref.~\cite{singlecat2024}, this buffer pump can generally lead to undesired side-band transitions. In the parameter regime of our device, the most relevant side-band transition takes the form of $\hat{a}^{\dagger}\hat{b}^{\dagger 3} + \text{H.c.}$ due to the resonance condition $2\omega_{p} \sim \bar{\omega}_{s} + 3\bar{\omega}_{b}$. In some of the unit cells of our device, the undesired process $\hat{a}^{\dagger}\hat{b}^{\dagger 3} + \text{H.c.}$ is not sufficiently detuned from the desired 3WM process $\hat{a}^{2}\hat{b}^{\dagger} + \text{H.c}$ and leads to additional parity-flip rates of our storage mode under the buffer pump (see for example \cref{app_fig:coherences_and_frequencies,app_fig:storage_coherence_measurements}). 

\subsection{Simple Kerr oscillator model of the CX gate and $\chi$-matching}
\label{app:CX_Kerr_oscillator_model}

To understand the basic aspects of the CX gate and $\chi$-matching, we first provide here a toy model of the storage-ancilla subsystem based on a simple Kerr oscillator model of a transmon. The key conclusion of this analysis is that an ideal $\chi$-matching condition is achieved approximately when the storage frequency is higher than the ancilla frequency by the magnitude of the ancilla transmon's anharmonicity (assuming the qubit has a negative anharmonicity).  

We consider the Hamiltonian of the form 
\begin{align}
    \hat{H} = \omega_{s,0}\hat{a}^{\dagger}\hat{a} + \omega_{q,0}\hat{q}^{\dagger}\hat{q} + \frac{K_{q}}{2}\hat{q}^{\dagger 2}\hat{q}^{2} + g(\hat{a}\hat{q}^{\dagger} + \text{H.c.}).  
\end{align}
Here, $\hat{a}$ and $\hat{q}$ are the annihilation operators of the storage mode and the ancilla transmon qubit. Note that $K_{q}$ is the self Kerr (or the anharmonicity) of the qubit in isolation without the coupling term $g(\hat{a}\hat{q}^{\dagger} + \text{H.c.})$, i.e., $\omega_{ef} - \omega_{ge} = K_{q}$. For a transmon, the self Kerr $K_{q}$ is negative and typically given by $K_{q} / 2\pi \simeq -200~\text{MHz}$. 

We treat the coupling term $g(\hat{a}\hat{q}^{\dagger} + \text{H.c.})$ perturbatively. Then, we find the following storage frequencies conditioned on the qubit being in $|g\rangle, |e\rangle, |f\rangle$ (see, e.g., Eq. (41) of Ref.~\cite{Blais2021})
\begin{align}
    \omega_{s, q:|g\rangle} &\simeq \omega_{s,0} +  \frac{g^{2}}{\Delta_{0}} , 
    \nonumber\\
    \omega_{s, q:|e\rangle} &\simeq \omega_{s,0} +  \frac{2g^{2}}{\Delta_{0} - K_{q} }  - \frac{g^{2}}{ \Delta_{0} }, 
    \nonumber\\
    \omega_{s, q:|f\rangle} &\simeq \omega_{s,0} + \frac{3 g^{2}}{ \Delta_{0} - 2K_{q}  }  - \frac{2 g^{2}}{\Delta_{0} - K_{q} } 
\end{align}
to the second order in $g^{2}$, where $\Delta_{0} \equiv \omega_{s,0} - \omega_{q,0}$. Note that these conditional storage frequencies are independent of the storage photon number $n$ to the second order in $g^{2}$ but this is generally not the case if the higher order terms are included (e.g., due to the qubit-induced self Kerr of the storage mode).   

The above conditional storage frequencies subsequently yield 
\begin{align}
    \chi_{ge} \equiv \omega_{s, q:|e\rangle} - \omega_{s, q:|g\rangle} &\simeq \frac{2g^{2}K_{q}}{\Delta_{0}(\Delta_{0} - K_{q})} 
    \nonumber\\
    \chi_{gf} \equiv \omega_{s, q:|f\rangle} - \omega_{s, q:|g\rangle} &\simeq \frac{2g^{2}K_{q}(2\Delta_{0}-K_{q})}{\Delta_{0}(\Delta_{0} - K_{q})(\Delta_{0}-2K_{q})}  
\end{align}
to the second order in $g^{2}$. Note that the ratio between $\chi_{ge}$ and $\chi_{gf}$ is given by
\begin{align}
    \frac{\chi_{ge}}{\chi_{gf}} &= \frac{\Delta_{0} - 2K_{q}}{2\Delta_{0} - K_{q}} . \label{app_eq:naive_prediction_on_the_chi_ratio}
\end{align}
Thus the $\chi$-matching condition $\chi_{ge} = \chi_{gf}$ is achieved when $\Delta_{0} - 2K_{q} = 2\Delta_{0} - K_{q}$ which is equivalent to $\Delta_{0} = -K_{q}$ or more explicitly 
\begin{align}
    \omega_{s,0} &= \omega_{q,0} - K_{q}.  
\end{align}
Hence for a transmon with a negative anharmonicity $K_{q}<0$, the above perturbative analysis suggests that an ideal $\chi$-matching condition is achieved when the storage frequency is higher than the ancilla frequency by the magnitude of the ancilla transmon's anharmonicity (e.g., approximately $200~\text{MHz}$)~\cite{Reinhold2019thesis}.  

\subsection{Full circuit-quantization-level numerical model of the CX gate and $\chi$-matching}
\label{app:CX_circuit_quantization_model}

The Kerr oscillator model gives a useful intuition on how the $\chi$-matching condition can be natively achieved through frequency targeting. In practice, various realistic aspects of our device (e.g., presence of the couplers) are not accounted for in the simple Kerr oscillator mode. Thus, we use a more detailed circuit-quantization-level model to accurately predict the $\chi_{ge}/\chi_{gf}$ ratio. Specifically, besides the storage, ancilla, and the coupler that directly participate in the $\text{CX}$ gate, our model also includes other elements that are adjacent to the storage and the ancilla. For example, the buffer mode and additional modes associated with the buffer's serial inductors are included in the model since they affect the properties of the storage mode. Similarly, even though the readout resonator and another coupler in the same unit cell (for realizing the $\text{CX}$ gate in a different direction) do not directly participate in the $\text{CX}$ gate, they are included in the model as they affect the properties of the ancilla. To make the diagonalization of such a large system tractable, we use the DMRG-X method~\cite{Khemani2016,ITensor2022} (i.e., density matrix renormalization group method for excited states). We ensure that the results converge with respect to the local dimension and the bond dimension within the accuracy of interest. 

Recall that in our device, we use a tunable-frequency transmon as a coupler that mediates a tunable dispersive coupling between a storage mode and an ancilla transmon. The basic principles of this tunable dispersive coupling are described in detail in Ref.~\cite{singlecat2024}. In this previous work, the tunable dispersive coupling was used simply as a way to characterize the state of a storage mode at the end of a pulse sequence. However in this work, the tunable dispersive coupling has an important additional role of mediating an entangling gate (i.e., a noise-biased CX gate) between a data cat qubit and an ancilla transmon. Since these CX gates are repeatedly applied in the bulk of a repetition cat code error-correction sequence, precise engineering of the tunable dispersive coupling is crucial for the performance of our repetition cat code. Here, we complement Ref.~\cite{singlecat2024} by presenting additional details of the tunable dispersive coupling related to the implementation of a noise-biased CX gate.

In our noise-biased CX gate, the computational basis states of an ancilla transmon are given by $|g\rangle$ and $|f\rangle$ where the latter may decay to the $|e\rangle$ state. Thus it is essential to track the properties of the system when the transmon is in $|f\rangle$ as well as in $|g\rangle$ and $|e\rangle$. We present various properties of a CX gate predicted by our circuit-quantization-level model of the device. The parameters of the model are chosen to reproduce the observed frequency spectrum on one of the CX gates in our device (i.e., CX gate between A2 and S3). With these parameters, as the coupler flux goes from $\Phi_{x,c}=0$ to $\Phi_{x,c} = \Phi_{0}/2$, the frequency of the coupler tunes from $9.0~\text{GHz}$ to $6.6~\text{GHz}$, approaching those of the storage and the ancilla.    

\begin{figure}[t!]
    \centering
    \includegraphics[width=1\columnwidth]{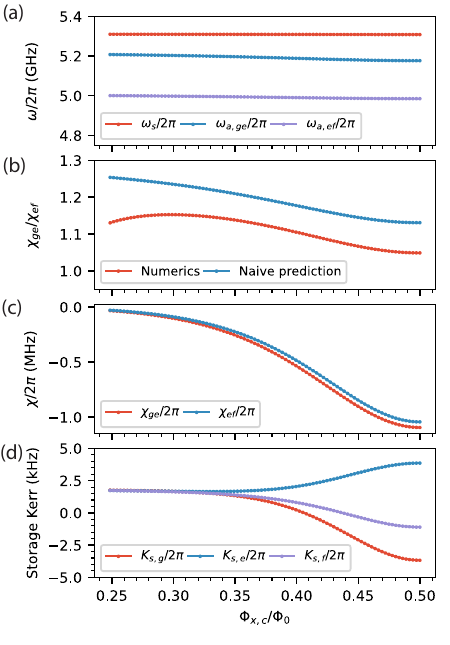}
    \caption{\textbf{Full circuit-quantization-level numerical model of a CX gate and $\chi$-matching.} (a) Frequencies of the storage, ancilla, and the tunable coupler as a function of the coupler flux $\Phi_{x,c}$ in the unit of the flux quantum $\Phi_{0}$. (b) The ratio $\chi_{ge}/\chi_{gf}$ as a function of the coupler flux. A crude prediction based on the perturbative expression \cref{app_eq:naive_prediction_on_the_chi_ratio} is added as a reference (blue curve). The changes in the $\chi_{ge}/\chi_{gf}$ ratio are mostly due to the increasing storage-ancilla detuning as the coupler approaches its minimum frequency position at $\Phi_{x,c}/\Phi_{0} = 0.5$. (c) Storage-ancilla $\chi_{ge}$ and $\chi_{gf}$ as a function of the coupler flux. (d) Self Kerr of the storage mode conditioned on the ancilla being in $|g\rangle,|e\rangle,|f\rangle$ as a function of the coupler flux.   }
    \label{app_fig:CX_DMRGX_numerics}
\end{figure}

In \cref{app_fig:CX_DMRGX_numerics}(a), we show the frequencies of the storage mode and the ancilla transmon as a function of the coupler flux $\Phi_{x,c}$ which are predicted by our model. Similarly as in Ref.~\cite{singlecat2024}, the coupler-ancilla coupling is designed to be much stronger than the coupler-storage coupling such that the storage mode does not inherit large non-linearities from the coupler. As a result, the ancilla frequency is more significantly shifted than the storage frequency due to the level repulsion with the coupler as the coupler approaches its minimum frequency and gets closer to the storage and the ancilla. This then results in an increase in the storage-ancilla detuning as the coupler approaches $\Phi_{x,c} = \Phi_{0}/2$ which is where the CX gate is operated at. 

The dependence of the storage-ancilla detuning on the coupler flux is important because the $\chi_{ge}/\chi_{gf}$ ratio varies as the detuning changes. In particular, we aim to ensure that the $\chi$-matching condition $\chi_{ge} / \chi_{gf} = 1$ is achieved at the operating point of the CX gate $\Phi_{x,c} = \Phi_{0}/2$. Thus the idling frequencies at $\Phi_{x,c}=0$ need to be chosen carefully to account for the frequency shifts due to the coupler as the coupler approaches the operating point. In \cref{app_fig:CX_DMRGX_numerics}(b), we show how the $\chi_{ge}/\chi_{gf}$ ratio varies as a function of the coupler flux $\Phi_{x,c}$. For reference, we also feature a naive prediction of the $\chi_{ge}/\chi_{gf}$ ratio based on the dressed storage-ancilla detuning and the ancilla anharmonicity and using \cref{app_eq:naive_prediction_on_the_chi_ratio}. Note that the $\chi_{ge} / \chi_{gf}$ ratio decreases and gets closer to the ideal value of $1$ as the coupler approaches $\Phi_{x,c} = \Phi_{0}/2$. This is because the storage-ancilla detuning is increased due to the asymmetric repulsion with the coupler as discussed above. This behavior can also be explained qualitatively by the expression in \cref{app_eq:naive_prediction_on_the_chi_ratio}. Despite the qualitative agreement, the naive prediction based on the Kerr oscillator model does not agree quantitatively with the one from the full circuit-quantization-level model. Note that we rely on the latter model in designing our system since it agrees well with the experimental data for several key metrics such as the $\chi_{ge}/\chi_{gf}$ ratio. For example for the $\text{A2}\leftrightarrow \text{S3}$ interaction considered here, the $\chi_{ge}/\chi_{gf}$ ratio is measured to be $1.04$ at the coupler's minimum frequency position (see \cref{app:storage_ancilla_interaction_characterization}). The circuit-quantization-level model predicts $\chi_{ge}/\chi_{gf} = 1.05$ at the same operating point while the naive perturbative expression predicts $\chi_{ge}/\chi_{gf} = 1.13$. This level of discrepancy can be important, especially on the edge of the $\chi$-matching requirement $0.8\lesssim \chi_{ge}/\chi_{gf} \lesssim 1.2$.   

With the chosen model parameters, the storage-ancilla dispersive shifts $\chi_{ge}$ and $\chi_{gf}$ are given by around $-1~\text{MHz}$ in an approximately $\chi$-matched manner as shown in \cref{app_fig:CX_DMRGX_numerics}(c). This level of the magnitude of the dispersive shift approximately corresponds to a CX gate length of $\sim500\text{ns}$. Furthermore, \cref{app_fig:CX_DMRGX_numerics}(d) shows that the magnitude of the storage self Kerr remains well below $5~\text{kHz}$ at the operating point of the CX gate for all relevant transmon states $|g\rangle,|e\rangle,|f\rangle$. Importantly, these Kerr values are at least two orders of magnitude smaller than the dispersive coupling strength. Thus the storage-ancilla interaction remains sufficiently dispersive even with many photons in the storage mode (e.g., $|K_{s}\alpha|^{2} \ll |\chi_{ge}|,|\chi_{gf}|$ for an average photon number $|\alpha|^{2}$ on the order of $10$). This property is crucial for hosting a cat qubit with a large number of photons in the storage mode.  

Lastly, we remark that the storage self Kerr behaves smoothly as a function of the coupler flux. This indicates that our system is free of undesired resonances (e.g., of the kind discussed in Ref.~\cite{singlecat2024}). While the undesired resonances are absent in our system, some unintended transitions can still occur due to a flux pulse applied to the coupler. In particular across all the unit cells in our device, the storage-ancilla detuning ranges from $40~\text{MHz}$ to $150~\text{MHz}$ when the coupler is in the ``off'' position. If the flux pulse is not adiabatically ramped up and down (relative to the energy scale of the storage-ancilla detuning), it can contain a significant spectral density at the storage-ancilla detuning and induce an ``iSWAP''-like excitation exchange between the storage and the ancilla. Such an excitation exchange then causes additional heating and decay of the ancilla and results in a significant degradation of the bit-flip times of our cat qubits. 

\section{Frequency targeting procedure for achieving the $\chi$-matching condition}
\label{app:frequency_targeting_procedure_for_chi_matching}

\begin{figure}[t!]
    \centering
    \includegraphics[width=\columnwidth]{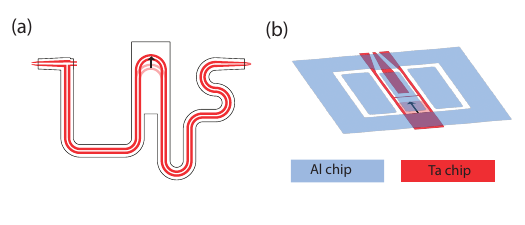}
    \caption{\textbf{Tuning the tantanlum-based die for frequency targeting.} (a) Example of decreasing the storage mode frequency by increasing the length of the storage resonator on the tantalum-based die. The black outlines are the Al chip cutout. (b) Example of decreasing the ancilla mode frequency by increasing the degree of metallization in the tantalum-based die. The extra metallization in the tantalum-based die is achieved by reducing the length of the associated cutout. This then increases the ancilla's total capacitance, leading to reduced anharmonicity and the frequency of the ancilla.}
    \label{app_fig:bottom_chip_tuning}
\end{figure}

As described in \cref{app:device_hamiltonian}, native $\chi$-matching imposes a stringent requirement on targeting of the storage-ancilla detuning. Targeting of the storage frequency is relatively straightforward since the storage modes are simple CPW resonators. However, targeting of the ancilla frequency is challenging due to systematic and stochastic variations in the Josephson junction energies of the ancilla junctions. We address this challenge by systematically optimizing the design parameters of the tantalum-based die based on the estimated junction energies of the ancilla junctions in the aluminum-based die. 

At room temperature, we directly measure the resistances of all the ancilla junctions in the aluminum-based die that will eventually form a final device. We then estimate the cryogenic ancilla junction energies based on the outcome of the room-temperature resistance measurement. In particular, we rely on an empirical relationship between the room-temperature junction resistances and the cryogenic junction energies which is established based on prior tests.

Provided with the estimated ancilla junction energies, we optimize various design parameters of the tantalum-based die. First as shown in \cref{app_fig:bottom_chip_tuning}(a), we tune the length of the storage resonators (and thus their frequencies) such that the storage modes have an optimal detuning with respect to the expected ancilla frequencies. This can correct for a global mistargeting of the ancilla frequencies relative to the desired targets. Second as shown in \cref{app_fig:bottom_chip_tuning}(b), we individually tune the total capacitance of the ancilla transmons by varying the length of the associated cutouts in the tantalum-based die. This then results in predictable changes in the anharmonicity and frequency of the ancilla transmons. These changes can be used to correct for individual random variations in the ancilla frequencies relative to the desired targets. To make the design optimization systematic, we use the full circuit-quantization-level model of the CX gate presented in \cref{app:CX_circuit_quantization_model}. In particular, we aim to minimize the metric $|\chi_{ge}/\chi_{gf} - 1|$ at a value of the coupler flux which yields $\chi_{gf} / 2\pi = -1.35~\text{MHz}$ such that the $\chi_{ge}/\chi_{gf}$ ratios are optimized for a representative CX gate length of approximately $500~\text{ns}$.

Note that we have carefully designed both the aluminum-based die and the tantalum-based die such that they can flexibly accommodate a wide range of storage frequencies and ancilla capacitances. We further remark that for this design tuning procedure to succeed in practice, we need an accurate relationship between the room-temperature junction resistance and the cryogenic junction energy. In particular, after the ancilla junction resistance measurement, the aluminum-based die has to wait until its tailor-designed tantalum-die is fabricated. Thus, a good model of the evolution (i.e., ``aging'') of the junction resistances is also required to ensure that the two dies remain as an optimal pair at a later point when they are ready to be flip-chip bonded. Lastly, the flip-chip gap should be consistently given by the expected value across the entire device because it affects the capacitances (and thus also frequencies) of the storages and the ancillas.

The frequency targeting of the two specific ancillas A1 and A2 illustrate the importance of the frequency tuning procedure described here. The anharmonicities of A1 and A2 were tuned by $5\%$ resulting in the reduction of the frequencies of these ancillas by approximately $100~\text{MHz}$. As a result, these ancillas achieved good measured $\chi$-matching ratios of $1.03 < \chi_{ge}/\chi_{gf}< 1.12$ which they otherwise would not have if it were not for the frequency tuning procedure.

\section{Component calibrations and parameters}
\label{app:component_calibrations_and_parameters}

\subsection{Mode frequencies and coherences}
\label{app:mode_frequencies_and_coherences}

\begin{figure*}[t!]
    \centering
    \includegraphics[width=1.8\columnwidth]{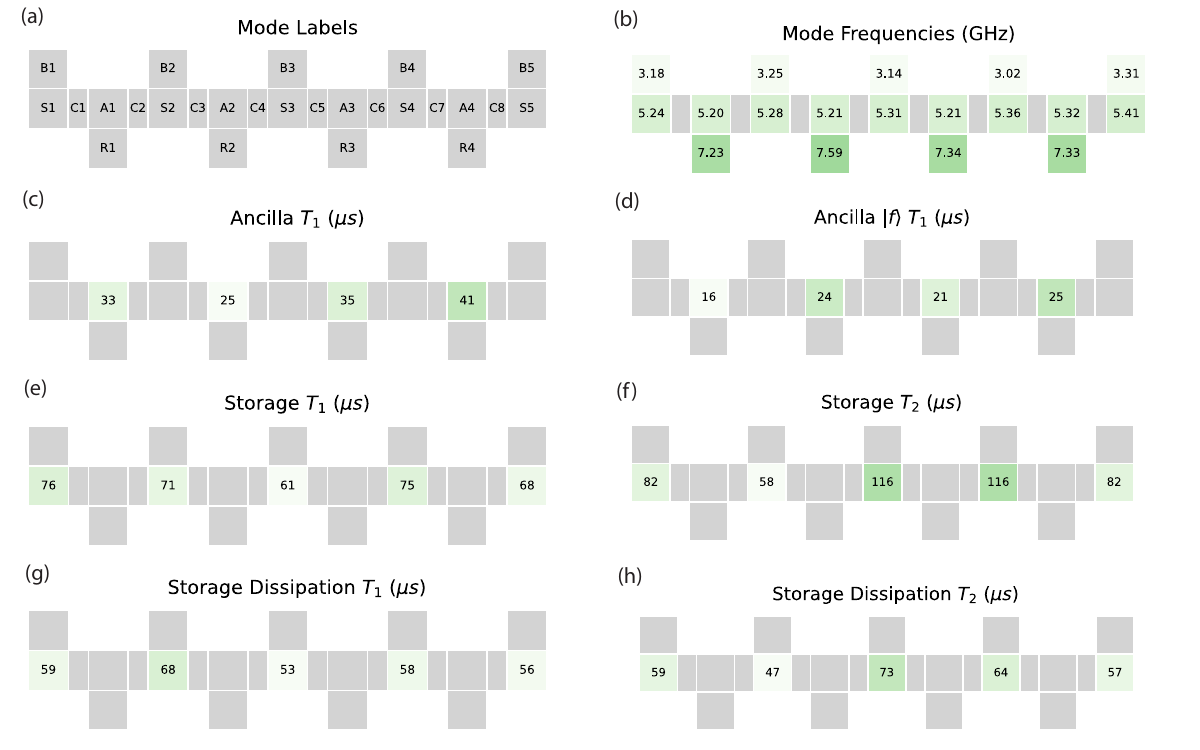}
    \caption{\textbf{Mode frequencies and coherences.}  (a) Labels for the storages, buffers, couplers, ancillas, and readout resonators in the circuit.  (b) Frequencies of the modes. (c)(d) $T_1$ of the ancilla ge and ef transitions.  These measurements are taken with the couplers in the off position and should be viewed only as approximate since the ancilla coherences can fluctuate over time and vary when moving the coupler from the ``off'' to ``on'' position (see \cref{app:cx_ancilla_T1} and \cref{app:noise_mechanisms}).  (e)(f) Storage $T_1$ and $T_2$ without two-photon dissipation being applied. (g)(h) Storage $T_1$ and $T_2$ with two-photon dissipation being applied. }
    \label{app_fig:coherences_and_frequencies}
\end{figure*}

The couplers in our device idle at their maximum frequency positions. The maximum frequencies of the couplers are above $8.5~\text{GHz}$ and their minimum frequencies range from $6.25~\text{GHz}$ to $6.75~\text{GHz}$. Each buffer mode is biased to one of its two saddle points. When choosing which one of the two saddle points each buffer should operate on, we take into consideration the impact of problematic pump-induced resonances discussed in \cref{app:buffer_model} and Ref.~\cite{singlecat2024}. The frequencies of all the important modes of our device are shown in \cref{app_fig:coherences_and_frequencies}(b).  Note that the reported ancilla and storage frequencies are measured with all the couplers at their idle (maximum frequency) positions.  

In \cref{app_fig:coherences_and_frequencies} we also show the $|e\rangle\rightarrow |g\rangle$ and $|f\rangle\rightarrow |e\rangle$ decay times of the ancilla transmons ($T_{1,e\rightarrow g}$ and $T_{1,f\rightarrow e}$) as well as the storage $T_1$ and $T_2$ with or without a pure two-photon dissipation being applied.  The storage mode lifetimes and dephasing times are reported by averaging over the $T_{1}$ and $T_{2}$ data collected during the repetition code experiment run. The ancilla decay times $T_{1,e\rightarrow g}$ and $T_{1,f\rightarrow e}$ are from the measurements taken at the time of the $\text{CX}^2$ gate performance characterization.  Since the ancilla decay and dephasing times can fluctuate over time the reported values in \cref{app_fig:coherences_and_frequencies} should only be viewed as a representative snapshot. Note also that the ancilla decay times can vary as a function of the coupler flux (see \cref{app_fig:noise_affecting_performance}).

\subsection{Storage $T_1$ and $T_2$ measurements}
\label{app:storage_t1_t2_measurements}

\begin{figure*}[t!]
    \centering
    \includegraphics[width=1.8\columnwidth]{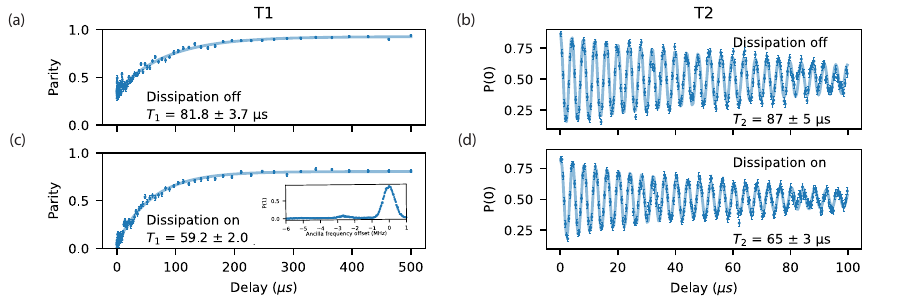}
    \caption{\textbf{Storage $T_1$ and $T_2$ measurements.} Example storage (a) $T_{1}$ and (b) $T_{2}$ measurements without the pure two-photon dissipation applied during the delay for one of the storage modes S1 in our device.  (c) $T_{1}$ and (d) $T_{2}$ measurements with the pure two-photon dissipation turned on during the delay for the same storage mode S1. We observe degradation in the storage $T_1$ and $T_2$ with pure two-photon dissipation applied during the delay. In addition the steady-state parity of the storage mode is shifted when the pure two-photon dissipation is turned on.  The inset shows the number splitting after applying two-photon dissipation for $100~\mathrm{\mu s}$ after starting from the vaccuum. The non-zero steady state parity can be attributed to buffer pump induced resonances between the storage and buffer which cause undesired heating of the storage mode (occurring in this unit cell as well as some other unit cells in our device; see \cref{app:buffer_model} and Ref.~\cite{singlecat2024} for more details).}
    \label{app_fig:storage_coherence_measurements}
\end{figure*} 

\begin{figure*}[t!]
    \centering
    \includegraphics[width=2\columnwidth]{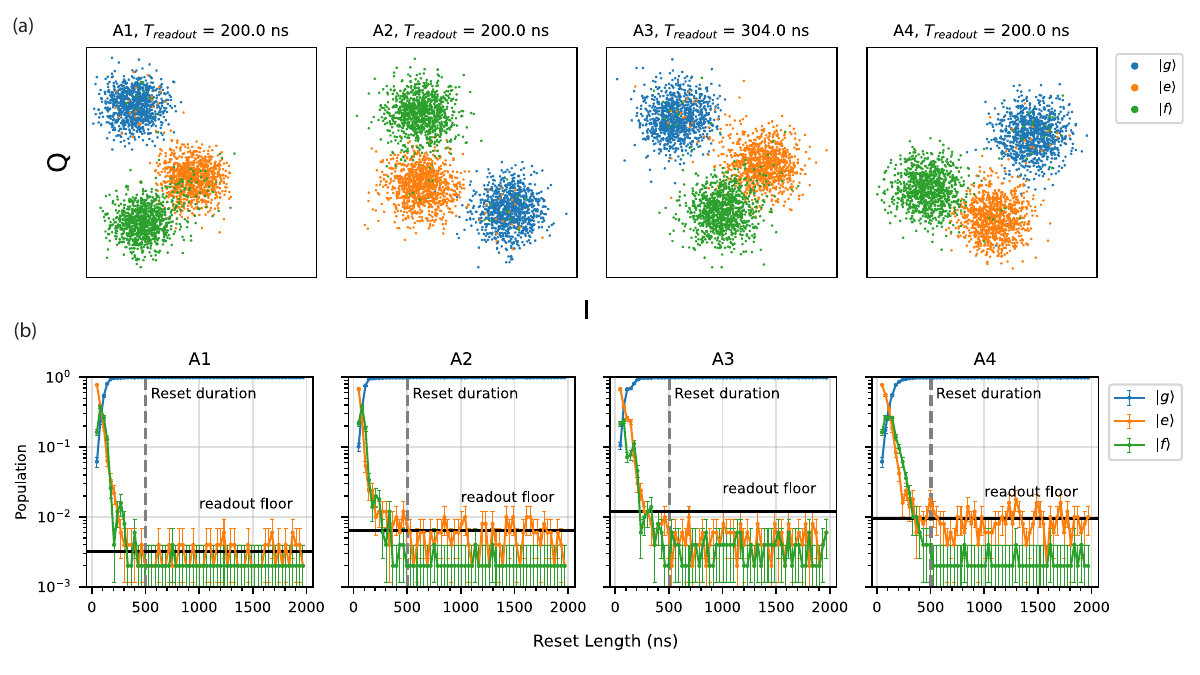}
    \caption{\textbf{Ancilla transmon readout and reset.}  (a) Readout clouds for the ancilla states $|g\rangle$, $|e\rangle$, and $|f\rangle$ for all the ancillas, demonstrating good distinguishability of the three ancilla states. (b) $|g\rangle$, $|e\rangle$, and $|f\rangle$ populations of the ancilla under the reset pulses as a function of time. In these experiments, each ancilla is initialized in the $|e\rangle$ state. The combined $|e\rangle$ and $|f\rangle$ populations reach the readout floor in approximately $500~\text{ns}$ or less.  The readout floor corresponds to the probability of measuring not-$|g\rangle$ when $|g\rangle$ is prepared.  }
    \label{app_fig:readout_and_reset}
\end{figure*}

We measure the $T_1$ and $T_2$ of the storage modes both with and without two-photon dissipation being applied.  We follow a similar procedure to~\cite{singlecat2024}. To measure the storage $T_{1}$ and $T_{2}$ with the pure two-photon dissipation turned on, we displace the storage mode into a coherent state with a large coherent state amplitude of $|\alpha|=2$. Then we apply pure two-photon dissipation for $6~\mathrm{\mu s}$ to prepare the state $(|\hat{n}=0\rangle + |\hat{n}=1\rangle) / \sqrt{2}$ (up to relative phase and state preparation errors) and subsequently apply a variable-length delay with the pure two-photon dissipation turned on. When we measure the storage $T_{1}$ and $T_{2}$ without the pure two-photon dissipation turned on during the delay, we mostly follow the same approach above with the differences that the initial coherent state amplitude is now given by $|\alpha|=0.75$ and the pure two-photon dissipation is turned off during the variable-length delay. We use a smaller coherent state amplitude for the case without the pure two-photon dissipation during the delay to ensure the storage mode is well confined within the $|\hat{n}=0\rangle/|\hat{n}=1\rangle$ manifold after the state preparation. After the delay period we perform photon-number parity measurement to measure the storage $T_{1}$ while for the storage $T_{2}$ measurement, we displace the storage mode by $\pm \ln(\sqrt{1/2}) \simeq \pm 0.83$ and apply a vacuum-selective pulse on the ancilla.

Note that in practice other mechanisms besides single photon loss and dephasing can impact the storage $T_{1}$ and $T_{2}$ as measured with the methods described above. For example our parity-based storage $T_1$ measurements are also sensitive to heating of the storage mode which affects the photon number parity of the storage mode. In particular as discussed in Ref.~\cite{singlecat2024} and \cref{app:buffer_model}, some unit cells in our device are subject to the undesired buffer-pump-induced resonances of the form $\hat{a}^{\dagger}\hat{b}^{\dagger 3} + \text{H.c.}$ which can induce heating of the storage mode under the two-photon dissipation.

\cref{app_fig:storage_coherence_measurements} shows an example of these $T_{1}$ and $T_{2}$ measurements performed on one of the storage modes S1. We observe degradation of both the $T_1$ and $T_2$ of the storage mode when the pure two-photon dissipation is applied during the delay. Additionally, we observe that the steady-state parity of the storage mode under the pure two-photon dissipation is significantly shifted compared to when the pure two-photon dissipation is not applied during the delay. We attribute the shift in steady state parity and decreased lifetime under two-photon dissipation to the undesired heating of the storage induced by the buffer pump due to the mechanism mentioned above. This is further evidenced by the small population of the excited storage state shown in the inset of \cref{app_fig:storage_coherence_measurements}(c) which presents the number splitting spectrum of the storage after two-photon dissipation is applied for $100~\mathrm{\mu s}$ to the initial vacuum state.  However, as demonstrated in Ref.~\cite{singlecat2024}, the issue of the buffer-pump-induced heating of the storage mode can be remedied by carefully arranging the storage and buffer frequencies to avoid these undesired resonances. Thus, we expect that these undesired effects will not limit the dissipative cat qubit architectures in the future.

The storage $T_1$ and $T_2$ reported in \cref{app_fig:coherences_and_frequencies} are obtained by averaging over multiple storage $T_1$ and $T_2$ datasets collected during the repetition code experiment run as outlined in \cref{app:characterization_procedure}.

\subsection{Ancilla transmon readout and reset}
\label{app:readout_and_reset}

\begin{figure*}[t!]
    \centering
    \includegraphics[width=2.07\columnwidth]{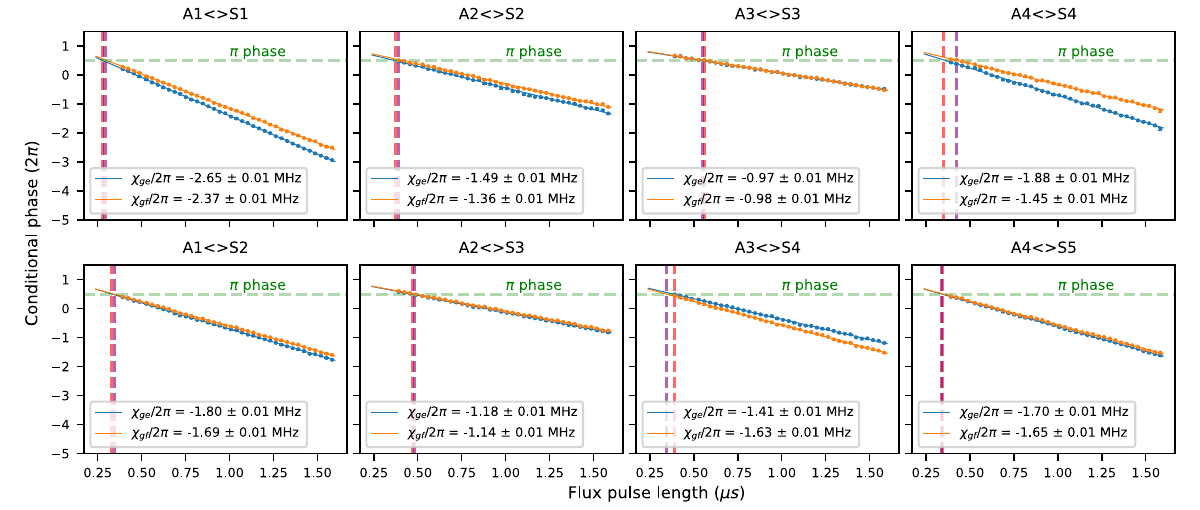}
    \caption{\textbf{Storage conditional phase measurements.} Results of the storage conditional phase measurements for all $8$ storage-ancilla pairs (corresponding to $8$ couplers) are shown. In these measurements, each coupler is flux pulsed to the ``on'' position which is chosen to be its minimum frequency position. The green horizontal line indicates the target conditional phase of $\pi$ required to achieve a CX gate. The vertical lines indicate the gate lengths for the CX gates in the $|g\rangle/|e\rangle$ (purple) and $|g\rangle/|f\rangle$ (red) manifold used in the repetition-code error-correction experiment. The storage-ancilla dispersive coupling strengths $\chi_{ge}$ and $\chi_{gf}$ are determined by fitting the conditional phase as a function of the flux pulse length. In most cases $\chi_{ge}$ is well matched to $\chi_{gf}$ within the $20\%$ relative precision, i.e., $|\chi_{ge} - \chi_{gf}|/|\chi_{gf}| < 0.2$. The S4-A4 interaction has a particularly large mismatch between $\chi_{ge}$ and $\chi_{gf}$ due to a small detuning between the storage and ancilla when the respective coupler in the ``on'' position.     }
    \label{app_fig:conditional_phase}
\end{figure*}

Each ancilla transmon is read out by driving the readout resonator to which it is dispersively coupled~\cite{Blais2004}. The readout resonator is inductively coupled to a feed line through a single-pole $\lambda/4$ Purcell filter to allow for fast readout without degrading the lifetime of the ancilla transmon. To allow for fast readout, the readout line has a traveling wave parametric amplifier (TWPA) from MIT Lincoln Laboratory~\cite{Macklin2015}.  We use a readout integration length which is $80~\text{ns}$ longer than the readout pulse length. Then a ringdown time of $64~\text{ns}$ is added to evacuate the excitations in the readout resonator. These readout time scales are chosen conservatively and could be significantly reduced since the decay rates of the readout resonators all exceed $\kappa_{r}/2\pi > 5~\text{MHz}$ in our device. The readout drive amplitude and the pulse length are tuned to avoid measurement-induced state transitions~\cite{Sank2016} which, in our system, may excite the ancilla and coupler to a highly excited state (e.g., not reset by the reset tones discussed below). We use the readout pulse duration of $200~\text{ns}$ for the first three ancillas (A1, A2, and A3) and $304~\text{ns}$ for the last ancilla (A4). The readout ``clouds'' for the three ancilla states $|g\rangle$, $|e\rangle$, and $|f\rangle$ are shown in \cref{app_fig:readout_and_reset} for all four ancilla transmons.   The state-preparation and measurement (SPAM) error for distinguishing $|g\rangle$ and $|e\rangle$ is under $2\%$ on all ancillas with a mean of $1.5\%$.

Reset of each ancilla transmon is initiated after the readout ringdown is completed. Each transmon is reset by driving the $|f0\rangle \leftrightarrow |g 1\rangle$ transition between the ancilla and the readout resonator while also driving the $|e\rangle\leftrightarrow |f\rangle$ transition of the ancilla at the same time~\cite{Magnard2018}. With this reset strategy, both the $|e\rangle$ and $|f\rangle$ states of the ancilla can be unconditionally reset to the ground state $|g\rangle$.  Duration of these reset pulses is chosen to be $504~\text{ns}$. This reset duration is long enough for the combined $|e\rangle$ and $|f\rangle$ populations to reach the readout floor which is given by the probability of measuring not-$|g\rangle$ when $|g\rangle$ is prepared.  We have found that repeatedly applying the ancilla reset pulses can cause elevated thermal populations which can subsequently degrade the noise bias of the $\text{CX}$ gates. Thus, we have configured the ancilla transmon XY lines carefully to mitigate this problem, as shown in \cref{app_fig:wiring_diagram}. Similarly, as in the case of the readout, the reset duration is chosen conservatively and further optimizations could speed up the ancilla reset. 

\subsection{Storage-ancilla dispersive couplings}
\label{app:storage_ancilla_interaction_characterization}

\begin{figure*}[t!]
    \centering
    \includegraphics[width=2\columnwidth]{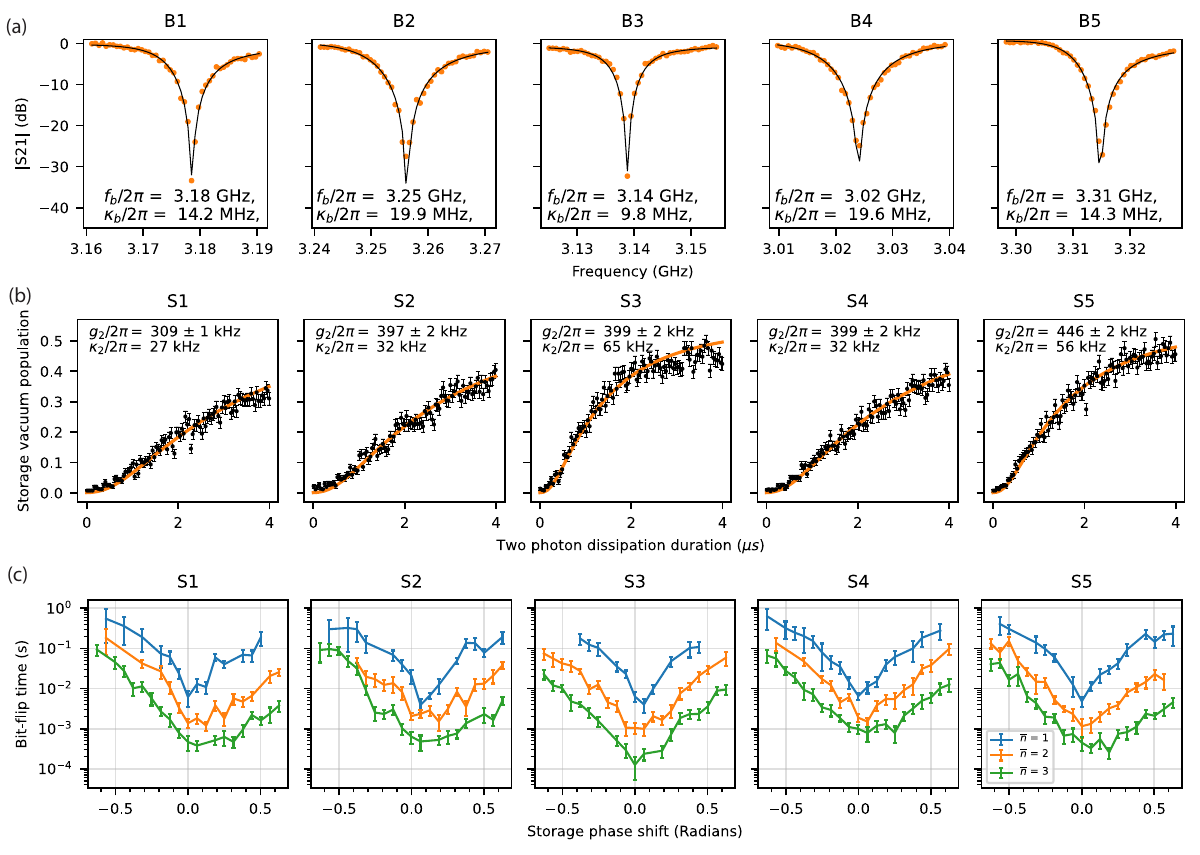}
    \caption{\textbf{Buffer parameters.}  (a) $|S_{21}|$ and fit for each of the buffer modes.  $\kappa_b$ (buffer loaded $\kappa$) and $f_b$ (buffer frequency) are reported based on the fits.  (b) Data to characterize the storage mode two-photon dissipation strength.  The storage mode is initialized to the coherent state $|\alpha\rangle$ and pure two-photon dissipation is applied for a variable duration.  We measure the vacuum state population of the storage mode.  The data is fit against a master equation model to determine $g_2$.  (c) Data to characterize the asymmetry of the two-photon dissipation.  We perform a pulsed stabilization experiment where we intentionally apply a phase between the storage mode state and the dissipative stabilization.  The asymmetry of the dissipative map can be ascribed to the calibration of the $\Delta_b$ detuning for the stabilization and has implications on CX gate fidelity particularly on interactions with significant $\chi$ mismatch. }
    \label{app_fig:buffer_details}
\end{figure*}

In this section, we characterize the storage-ancilla interactions at their ``on'' positions. There are $8$ storage-ancilla pairs, corresponding to $8$ couplers. In our experiments, the ``on'' position of each interaction is realized by pulsing the associated coupler to its minimum frequency position, where the storage-ancilla dispersive coupling strength is maximized. 

An initial characterization of the storage-ancilla dispersive coupling strength is performed using a storage conditional phase measurement which is described in detail in Ref.~\cite{singlecat2024}. In this experiment, we determine the storage phase as a function of coupler flux pulse length with the ancilla in the $|g\rangle$, $|e\rangle$, and $|f\rangle$ states.  By comparing the storage phase evolution for different ancilla states we can determine several important parameters of our device. First by analyzing the storage phase with the ancilla in $|g\rangle$ we can determine an initial value of the storage's interaction frequency (i.e., storage frequency at an ``on'' position; see \cref{app:cx_calibration} for how this is further refined later). Second, the slope of the storage phase evolution with the ancilla in $|e\rangle$ (or $|f\rangle$) relative to the ancilla in $|g\rangle$ provides us the value of the storage-ancilla $\chi_{ge}$ (or $\chi_{gf}$). Lastly, an initial value of the CX gate length in the $|g\rangle/|e\rangle$ (or $|g\rangle/|f\rangle$) manifold is determined by finding where a relative storage phase of $\pi$ is reached between the cases where the ancilla is in $|g\rangle$ versus $|e\rangle$ (or $|f\rangle$). Note that the $\text{CX}$ gate lengths are refined later in \cref{app:cx_calibration}. 

The results of the storage conditional phase measurements are shown in \cref{app_fig:conditional_phase} for all $8$ interactions. The $\text{CX}$ gate length in the $|g\rangle/|e\rangle$ (or $|g\rangle/|f\rangle$) manifold is indicated by the red (or purple) vertical line in the plots. Ideally all the interactions need to be perfectly $\chi$-matched (i.e. $\chi_{ge}/\chi_{gf}=1$. Due to the imperfections in the frequency targeting (even after the tuning strategy in \cref{app:frequency_targeting_procedure_for_chi_matching} is employed), the realized $\chi_{ge}/\chi_{gf}$ ratios range from $0.85$ to $1.3$. Except for the $\text{S4}\leftrightarrow \text{A4}$ interaction, all remaining $7$ interactions have a $\chi$-mismatch of $\chi_{ge}$ and $\chi_{gf}$ less than $20\%$, i.e., $|\chi_{ge} - \chi_{gf}| / |\chi_{gf}| < 0.2$. However as illustrated in \cref{fig:architecture} and discussed in the main text, the over- or under-rotation due to a small relative $\chi$ mismatch can be tolerated in the repetition-code error-correction experiments thanks to the cat qubit stabilization via the two-photon dissipation process (see ~\cref{app:chi_matching_requirements}).

\subsection{Two-photon dissipation calibration}
\label{app:two_photon_dissipation_calibration}

The buffer modes are parked on one of their two saddle points. At a saddle point, the frequency of the buffer is first-order insensitive to the fluctuation in both its sigma and delta fluxes. For all five buffers, the chosen saddle points are located within the buffer's filter passband. Thus all the buffers have radiative loss rates of $\sim 10~\text{MHz}$ or higher at the chosen saddle points as shown in \cref{app_fig:buffer_details}(a). These high buffer loss rates are necessary for satisfying the adiabatic elimination condition for the two-photon dissipation. Despite the large buffer loss rates, the storage-mode lifetimes are not limited by the buffer loss channels since the buffer loss environment is colored via a 4-pole metamaterial bandpass filter.

Tuning up the two-photon dissipation requires calibrating two frequencies, $\omega_p$ and $\omega_d$. These frequencies correspond to the sigma-flux pump frequency and the buffer drive frequency. For a generic set of pump and drive frequencies, the two-photon exchange Hamiltonian between the storage ($\hat{a}$) and the buffer ($\hat{b}$) is given by 
\begin{align}
    \hat{H}=g_2\hat{a}^2\hat{b}^{\dagger} e^{i\Delta_p t} -g_2 \alpha^2\hat{b}^{\dagger} e^{-i\Delta_d t}+ \text{H.c.}, 
\end{align}
in the rotating frame with respect to the Stark-shifted storage and buffer frequencies $\bar{\omega}_{s}$ and $\bar{\omega}_{b}$. Here $\Delta_p=\omega_p-(2\bar{\omega}_s-\bar{\omega}_b)$ is the detuning of the pump that realizes the 3WM condition and $\Delta_d=\omega_d-\bar{\omega}_b$ is the detuning of the buffer drive.  When calibrating the two-photon dissipation it is particularly important that storage detuning, given by $\Delta\equiv (\omega_p+\omega_d)/2-\bar{\omega}_s = (\Delta_p+\Delta_d)/2$~\cite{Lescanne2020}, is minimized.  Miscalibration of $(\omega_p+\omega_d)/2$ (i.e., a non-zero residual storage detuning $\Delta$) can degrade the bit-flip times of a cat qubit by a multiplicative factor. In contrast to the storage detuning which needs to be carefully minimized, the bit-flip times of a cat qubit are not as sensitive to the conjugate detuning $\Delta_b\equiv (\Delta_p-\Delta_d)/2$ which we refer to as the buffer detuning.  Note also that while the buffer detuning $\Delta_{b}$ does not significantly affect the bit-flip times of a cat qubit in isolation, it can have a more noticeable impact in the context of performing a noise-biased CX gate with an ancilla transmon (see \cref{app:dissipative_map_asymmetry}).  

In our experiment, the buffer detuning $\Delta_{b}$ is calibrated to a precision of $\sim 1~\text{MHz}$ while the storage detuning $\Delta$ is calibrated to a precision of order $1~\text{kHz}$ or less. To characterize the Stark-shifted storage frequency to a high accuracy we use a Ramsey inteferometry experiment (of the kind used for the storage $T_{2}$ measurement as in \cref{app:storage_t1_t2_measurements}) with the pure two-photon dissipation turned on during the delay. Then the pump frequency is updated to match the measured Stark-shifted storage frequency $\bar{\omega}_{s}$. These calibrations are repeated often to account for the drift in the storage frequency and update the pump frequency. See \cref{app:characterization_procedure} for more details.    

We take into consideration the effects of undesired buffer-pump-induced resonances (discussed in Ref.~\cite{singlecat2024} and \cref{app:buffer_model}) when choosing at which one of the two saddle points each buffer should operate. However, in some unit cells of our device, both saddle points are susceptible to the buffer-pump-induced resonances. In these unit cells, we pump the sigma flux of the buffer more weakly so that the negative effects of the pump-induced resonances can be mitigated. However, this leads to a weaker 3WM interaction strength $g_{2}$ and correspondingly a weaker two-photon dissipation strength $\kappa_{2}$. This is one of the reasons why cat qubits in some unit cells (e.g., S6) have poorer bit-flip performance (including less favorable exponent in the exponential scaling of the bit-flip times in $|\alpha|^{2}$) compared to the ones in other unit cells. Through a different choice of the storage and buffer frequencies as in Ref.~\cite{singlecat2024}, this issue can be avoided in the future.

Parameters relating to the buffer and two-photon dissipation calibration are shown in \cref{app_fig:buffer_details}.  In \cref{app_fig:buffer_details} (b) we characterize the two-photon dissipation strength by initializing the storage mode into a coherent state $|\alpha|^2=7$ and applying pure two-photon dissipation ($\kappa_{2}D[\hat{a}^2]$).   We measure the vacuum population of the storage mode as a function of time.  The population in the vacuum state rises to $0.5$ as the two-photon dissipation takes the storage mode state into the $|\hat{n}=0\rangle/|\hat{n}=1\rangle$ manifold.  By fitting this decay to a numerical model we determine the $g_2/2\pi$ of the storage modes which range between $300~\text{kHz}$ and $450~\text{kHz}$.  The two-photon dissipation rates of the storage modes vary between $27~\text{kHz}$ and $65~\text{kHz}$. For more details on this procedure see Ref.~\cite{singlecat2024}.

\subsection{Displacement and dissipation amplitude calibrations}

\begin{figure*}[t!]
    \centering
    \includegraphics[width=2\columnwidth]{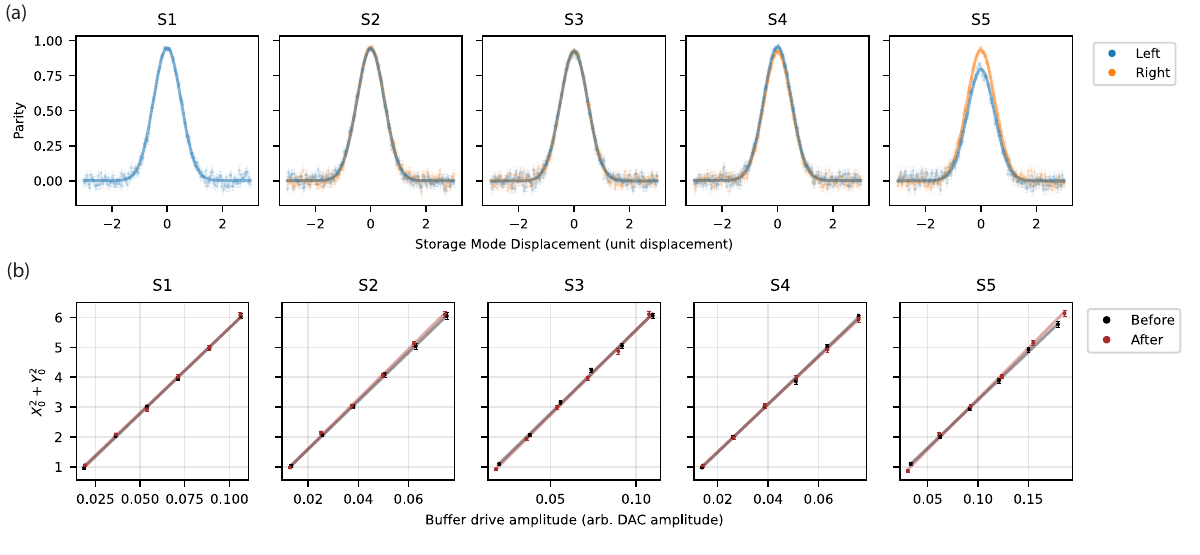}
    \caption{\textbf{Storage displacement and dissipation amplitude calibrations.} (a) Storage mode displacement calibration.  We calibrate the storage displacement by measuring the displaced parity of the vacuum state.  In blue we show the measurements using the ancilla on the left side of an ancilla mode and in orange the measurements using the right side of an ancilla mode.  The left measurement of the last storage has lower contrast due to imperfect tracking of the ancilla phase.  (b) Dissipation amplitude calibration.  We calibrate the mapping of buffer drive amplitude to storage mean photon number by measuring Wigner tomograms as a function of buffer drive amplitude.  From the Wigner tomograms we determine the storage mean photon number which we fit to a linear model as a function of the buffer drive amplitude.  We show the calibrations taken in the week before the repetition code experiment which were used for the data of the main text.  We also show a calibration taken a few days after the repetition code experiment to give a sense of how the calibration can drift.  }
    \label{app_fig:storage_alpha_calibrations}
\end{figure*}

In this section we discuss how we calibrate the storage-mode displacement as well as the relationship between the buffer drive amplitude and the mean photon number of a cat qubit. We follow the same method presented in Ref.~\cite{singlecat2024}. 

Calibration of the storage mode displacement is done by measuring the displaced parity of the storage mode vacuum state. Specifically, we calibrate the displacement by comparing to the expected functional form $e^{-2|\alpha|^2}$ where $\alpha$ is the displacement amplitude. In \cref{app_fig:storage_alpha_calibrations} we show the displacement calibration for all five storage modes. On all but the first storage mode S1, we show two calibration results where one is obtained by using the ancilla on the left-hand side of the storage (e.g., A1 for S2) and the other is obtained by using the ancilla on the right-hand side (e.g., A2 for S2). The calibrated storage-mode displacements are consistent within the $2\%$ relative accuracy regardless of whether the ancilla on the left-hand side versus the right-hand side is used. In the repetition code experiments, we use the displacement calibration obtained via the ancilla on the right-hand side of each storage.

We calibrate the buffer drive amplitude by measuring the storage mode steady-state Wigner function for various buffer drive amplitudes.  To reach a storage-mode steady state we start with the storage mode in the vacuum and apply two-photon dissipation for $200~\mathrm{\mu s}$.  In the calibration we target measuring storage mode mean photon numbers ranging from $1$ to $6$. The storage mode Wigner function is fit to a pair of diametrically opposed 2D Gaussians from which we determine the steady state storage average photon number $|\alpha|^2$.  Note that at $|\alpha|^2=1$ the photon-number parity does not exactly vanish, indicating that the sum-of-two-Gaussians model is not accurate in the $|\alpha|\rightarrow 0$ limit.  Nonetheless, in simulations we find that this effect causes only a few-percent-level error on the fitted photon number near $|\alpha|^2=1$. We fit the $|\alpha|^2$ versus buffer drive amplitude relationship to a linear model to complete the calibration. In subsequent experiments we use this linear model to determine which buffer drive amplitude to use to achieve a target $|\alpha|^2$ of a cat qubit.  In \cref{app_fig:storage_alpha_calibrations} we show the result of this calibration for all five storage modes. The black lines show the calibrations used for the repetition code experiment which were taken in the days before.  To give a sense of how this calibration can fluctuate over time, we show calibrations taken a couple days after the repetition code experiment run (brown lines). The calibrations before and after are consistent to within $1\%\sim 8\%$ relative accuracy at $|\alpha|^2=1$ across all five storage modes. At $|\alpha|^2=2$ the calibrations are consistent to within the $1\% \sim 3\%$ relative accuracy. 

\subsection{$\text{CX}$ (and $\text{CX}^2$) gate calibration procedure}
\label{app:cx_calibration}

\begin{figure*}[t!]
    \centering
    \includegraphics[width=\textwidth]{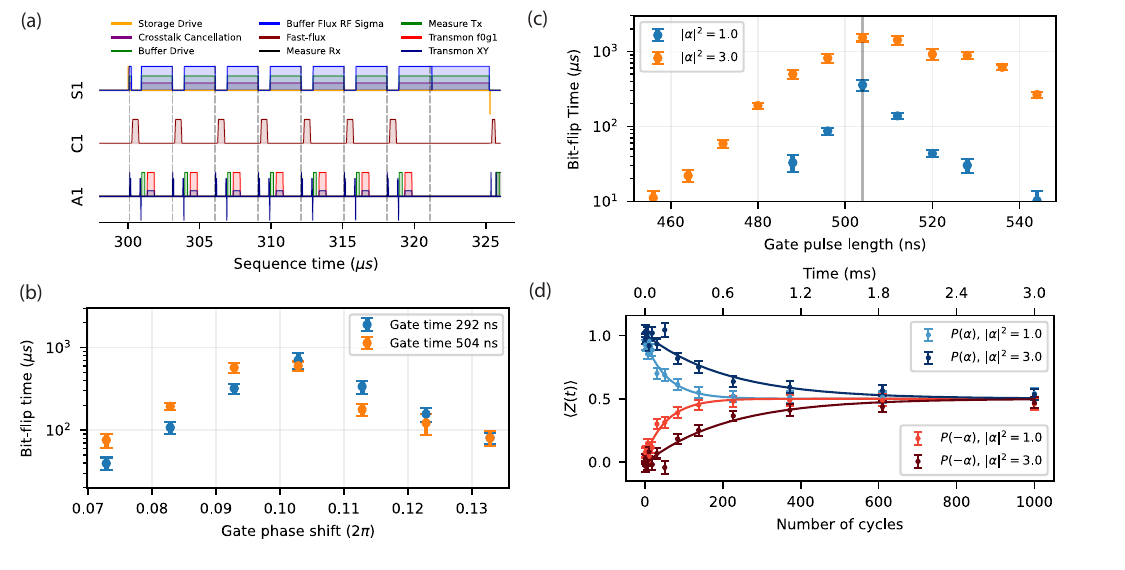}
    \caption{\textbf{$\text{CX}^2$ gate calibration procedure.} (a) Pulse sequence for characterizing the cat-qubit bit-flip times under a $\text{CX}^2$ gate. (b) Bit-flip time of a cat qubit with $|\alpha|^{2}$ as a function of the storage phase shift for two different gate lengths. In this experiment the ancilla is in its ground state $|g\rangle$. The storage interaction frequency and the phase correction are fine tuned based on a linear relationship between the optimal storage phase shift and the gate length. (c) Bit-flip time of a cat qubit under a $\text{CX}^2$ gate as a function of the $\text{CX}^2$ gate length. In this experiment, the ancilla prepared in the $|f\rangle$ state and is reset every cycle. (d) Example exponential fits for determining the bit-flip times under a $\text{CX}^2$ gate. These fits correspond to the data in \cref{fig:architecture} with the ancilla prepared in the state $|g\rangle+|f\rangle$.  }
    \label{app_fig:cx_calibration}
\end{figure*}

The $\text{CX}$ gates in our experiment are parameterized by the coupler flux pulse amplitude, flux pulse shape parameters (e.g., ramp times), storage interaction frequency, storage phase correction, ancilla interaction frequency, ancilla phase correction, and flux pulse length (i.e., gate length). The $\text{CX}^2$ gates are calibrated in an analogous way to how the $\text{CX}$ gates are calibrated, with one difference that the pulse length is calibrated to achieve the full $2\pi$ conditional rotation instead of the $\pi$ conditional rotation. In this section we explain these parameters and provide details on how they are calibrated. Note that a subset of the $\text{CX}$ and $\text{CX}^2$ gate parameters (e.g., coupler flux pulse amplitude, flux pulse length, and storage interaction frequency) was initially determined in \cref{app:storage_ancilla_interaction_characterization}. Here we refine the gate parameters by directly optimizing bit-flip rates.  

The pulse sequence for measuring the $\text{CX}$ and $\text{CX}^2$ gate bit-flip rates is shown in \cref{app_fig:cx_calibration}(a). In this pulse sequence, which is representative of syndrome extraction of a repetition-code stabilizer, the storage mode is initially prepared in the $|+\alpha\rangle$ coherent state. Then, we repeatedly apply the cycles of ancilla state preparation, $\text{CX}$ (or $\text{CX}^{2}$) gate, ancilla state unpreparation (inverse of the state preparation), and ancilla readout and reset. When the coupler flux pulse is not applied, two-photon dissipation with some padding is applied to the storage mode to bring it back to the cat-qubit manifold spanned by the $|\pm\alpha\rangle$ coherent states. Each cycle in this pulse sequence takes $3~\mathrm{\mu s}$ and these cycles are applied for a variable number of rounds. Finally at the end of the pulse sequence, we perform displaced parity measurements of the storage mode with displacement amplitudes of $\pm\alpha$ to determine if a bit-flip error (i.e., $|+\alpha\rangle \rightarrow |{-}\alpha\rangle$) has occurred on the cat qubit.

Our $\text{CX}$ (or $\text{CX}^{2}$) gates use flat-top Gaussian waveforms. That is, the flux pulse uses Gaussian shoulders for ramping up and down and is otherwise held at a constant amplitude. The first parameter we set for our $\text{CX}$ (or $\text{CX}^2$) gates is the coupler flux pulse amplitude.  For all the $\text{CX}$ gates used in our repetition-code experiments, we pulse the respective couplers from their maximum frequency position to the minimum frequency position. 

The parameters of the ramps in our coupler flux pulse need to be chosen carefully to avoid unwanted excitation exchange between the storage and the ancilla discussed in \cref{app:CX_circuit_quantization_model}. Specifically spectral component of the flux pulse at the storage-ancilla detuning can convert storage photons into ancilla photons. On interactions where the storage-ancilla detuning is smaller (especially $\text{A1}\leftrightarrow\text{S1}$ and $\text{A4}\leftrightarrow\text{S4}$), these resonance processes are less detuned and thus we use longer flux pulse ramp times ($64~\text{ns}$ for these two interactions) to avoid the undesired excitation exchanges. 

Under the coupler flux pulse, the frequency of storage is shifted by the coupler approaching the storage mode. The storage phase during the coupler flux pulse is tracked by the storage interaction frequency and a storage phase correction which accounts for offsets due to the flux pulse transients.  At the chosen flux pulse amplitude and the pulse shape, we refine both the storage interaction frequency under the flux pulse and calibrate the storage phase correction to optimize bit-flip times. In particular, we apply a simplified version of the pulse sequence in \cref{app_fig:cx_calibration}(a) where only the coupler flux pulse and pulsed two-photon dissipation are performed in each cycle (i.e., omitting the ancilla state preparation, unpreparation, readout, and reset). Then as shown in \cref{app_fig:cx_calibration}(b), we characterize the bit-flip time of a cat qubit with $|\alpha|^{2}=1$ as a function of the storage phase shift for two different gate lengths (i.e., flux pulse lengths), where the presented results are for the $\text{A1}\leftrightarrow \text{S1}$ interaction. For each gate length, we find the optimal storage phase shift that maximizes the bit-flip time of the cat qubit. Then by using a linear relationship between the optimal storage phase shift and the gate length, we extract the storage interaction frequency (from the slope) and obtain the storage phase correction (from the offset). This calibration is repeated often to account for the drift in the storage frequency over time.  

\begin{figure*}[t!]
    \centering
    \includegraphics[width=2\columnwidth]{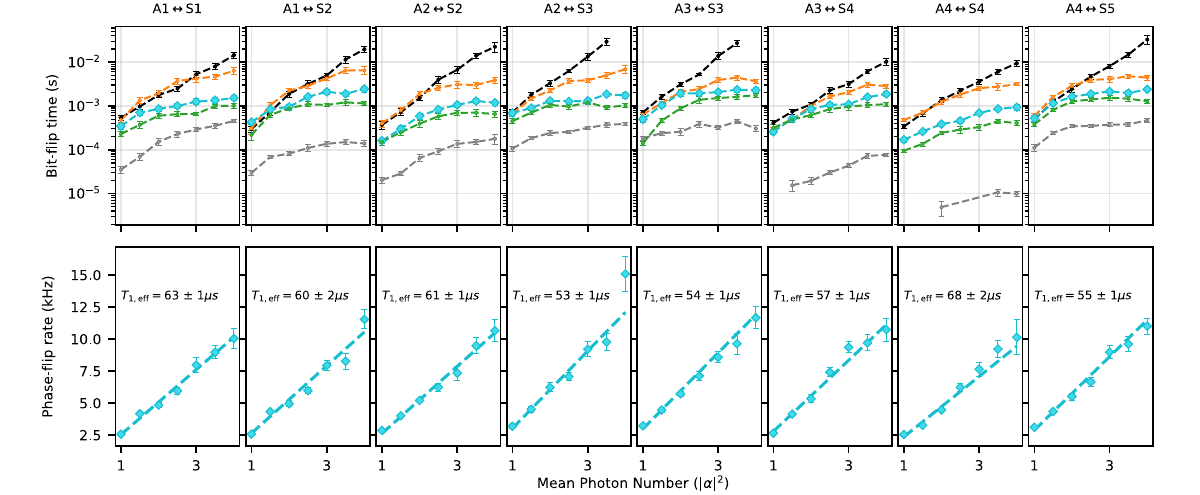}
    \caption{\textbf{Summary of the bit-flip and phase-flip performance of all the $\text{CX}^2$ gates.} Summary of $\text{CX}^2$ cycle performance for all $8$ interactions in the distance-5 repetition code. The curves correspond to a reference static stabilization case (black), the ancilla being prepared to $|g\rangle$ (orange), the ancilla being prepared to $|g\rangle+|e\rangle$ (grey), the ancilla being prepared to $|f\rangle$ (green), and the ancilla being prepared to $|g\rangle+|f\rangle$ (blue). These experiments were performed approximately $5$ days after the repetition code run presented in the main text. The cycle times used in these experiments are $3~\mathrm{\mu s}$.   }
    \label{app_fig:cx_gate_summary}
\end{figure*}

Similar to the storage mode, the ancilla's frequency is also shifted under the coupler flux pulse due to increased hybridization with the coupler. We perform experiments where we track the ancilla phase under the flux pulse to calibrate the ancilla interaction frequency and the ancilla phase correction. These ancilla interaction frequency and phase correction calibrations are not so important for the $\text{CX}^2$ bit-flip time characterization but is important when we want to extract information on storage-mode parities. For example the ancilla phase tracking is important in the repetition code stabilizer measurements \cref{app:X_basis_pulse_sequence} or any time we do storage parity measurements such as the final displaced-parity measurements in \cref{app_fig:cx_calibration}(a). 

\begin{figure*}[t!]
    \centering
    \includegraphics[width=2\columnwidth]{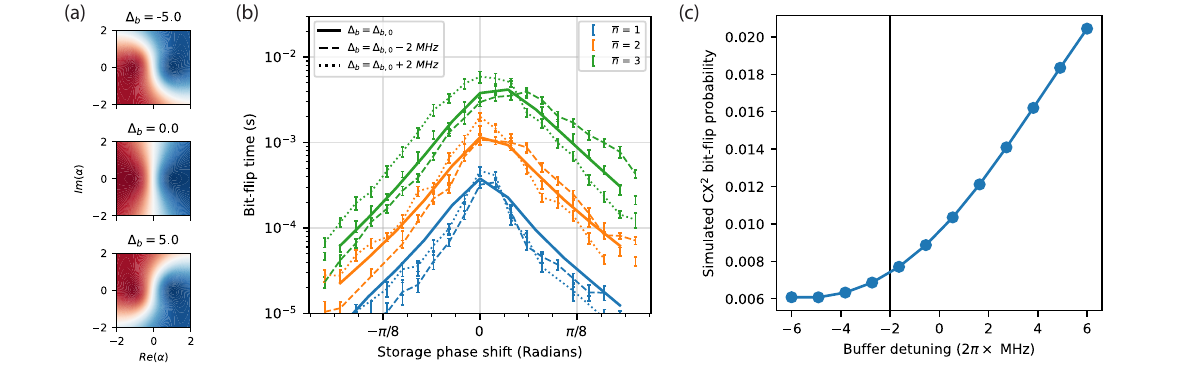}
    \caption{\textbf{Effect of $\Delta_b$ detuning.} (a) Simulated population recovery of two-photon dissipation with different $\Delta_b$ detunings.  The red corresponds to population returning to $|{-}\alpha\rangle$ and the blue corresponds to population returning to $|\alpha\rangle$.  (b) Experimentally measured bit-flip time as a function of storage phase shift for a pulsed stabilization experiment.  We perform a pulsed stabilization experiment where the storage is phase shited every time the two-photon dissipation is turned off.  $\Delta_{b,0}$ corresponds to the nominal value of $\Delta_b$ used in experiments. (c) Simulated bit-flip probability for the $\text{A4}\leftrightarrow \text{S4}$ interaction as a function of buffer detuning ($\Delta_b$).  }
    \label{app_fig:delta_b_effect}
\end{figure*}

Equipped with the above calibrations, the gate length of the $\text{CX}^{2}$ gate, initially determined in \cref{app:storage_ancilla_interaction_characterization}, can be fined tuned by finding the optimal flux pulse length that maximizes the bit-flip time of a cat qubit with the ancilla prepared in the $|f\rangle$ state via the preparation pulses. Specifically, we use a simplified version of \cref{app_fig:cx_calibration}(a) without the ancilla readout. The inclusion of ancilla reset in the pulse sequence is crucial for the $\text{CX}^{2}$ gate length calibration. This is because the target ancilla state is given by $|f\rangle$ and any residual excited state populations of the ancilla can degrade the bit-flip times of the cat qubit, compromising the gate calibration. 

\cref{app_fig:cx_calibration}(c) shows an example result of the $\text{CX}^{2}$ gate length calibration for a specific interaction $\text{A1}\leftrightarrow \text{S1}$. Note that in our calibration, we use cat qubits with $|\alpha|^{2}=1$ when optimizing the $\text{CX}^{2}$ gate length. For comparison we also show in \cref{app_fig:cx_calibration}(c) the bit-flip times of a larger cat qubit with $|\alpha|^2=3$. The achieved bit-flip times at $|\alpha|^{2}=3$ are generally higher and less sensitive to the exact gate length. Notably, the observed bit-flip times are asymmetric in the sign of the timing mismatch (relative to the optimal gate length) for the case with $|\alpha|^{2}=3$. This is related to the asymmetry in how the positive versus negative phase shifts are recovered by the two-photon dissipation (see \cref{app:dissipative_map_asymmetry}). Moreover as discussed in \cref{app:measurement_error_vs_n}, the performance of our architecture may be further improved if all the gates are specifically calibrated for each average photon number $|\alpha|^{2}$, as opposed to exclusively for $|\alpha|^{2}=1$.  

Note that the bit-flip times under a $\text{CX}^{2}$ gate are not sensitive to ancilla state preparation errors to $|g\rangle$ because the storage mode either stays in or returns to the $|+\alpha\rangle$ coherent state regardless of whether the ancilla is in $|g\rangle$ or $|f\rangle$.  In contrast, the bit-flip times under a $\text{CX}$ gate are sensitive to the ancilla state preparation errors since the storage mode ends up in two different coherent states $|\pm\alpha\rangle$ depending on whether the ancilla is in $|g\rangle$ or $|f\rangle$. For this reason, we do not calibrate the $\text{CX}$ gate length using the pulse sequence shown in \cref{app_fig:cx_calibration}(a). Instead, we use stabilizer measurement circuit of the repetition code to tune up the $\text{CX}$ gate lengths (see \cref{app:cx_gate_length_calibration_for_rep_code}).  We also remark that storage phase correction and the ancilla phase correction are further revised in the repetition-code experiments \cref{app:insitu_calibration}.   

A comprehensive summary of the $\text{CX}^{2}$ cycle performance for all $8$ storage-ancilla pairs is given in \cref{app_fig:cx_gate_summary}. Right before collecting the $\text{CX}^{2}$ cycle performance data presented in \cref{app_fig:cx_gate_summary}, we ran the storage phase optimization as in \cref{app_fig:cx_calibration}(b). The $\text{CX}^{2}$ cycle performance data shown in the main text (i.e., \cref{fig:architecture}(c)) is for the specific interaction $\text{A1}\leftrightarrow\text{S1}$ (i.e., first column in \cref{app_fig:cx_gate_summary}). \cref{app_fig:cx_calibration}(d) shows example underlying data and exponential fits for determining the bit-flip times for the same interaction $\text{A1}\leftrightarrow\text{S1}$ with an ancilla initial state of $|g\rangle + |f\rangle$. In these experiments we use the gate length corresponding to a $\text{CX}^2$ gate in the $|g\rangle/|f\rangle$ manifold.  This allows us to see the effect of $\chi$-mismatch by looking at the error rates for the $|g\rangle+|e\rangle$ initial state.  With significant $\chi$-mismatch the error rates at low $|\alpha|^2$ are enhanced for the initial state $|g\rangle+|e\rangle$.

Besides the $\text{A4}\leftrightarrow \text{S4}$ interaction, the bit-flip times under a $\text{CX}^2$ cycle exceed $1~\text{ms}$ with a sufficiently large $|\alpha|^{2}$ for all remaining $7$ interactions in the case when the ancilla is prepared in the $|g\rangle+|f\rangle$ state. The $\text{A4}\leftrightarrow \text{S4}$ interaction is an outlier in terms of the $\text{CX}^2$ bit-flip times. This is understandable given the particularly large $\chi$-mismatch of $\chi_{ge}/\chi_{gf} \simeq 1.3$. For a $2\pi$ rotation this means that an ancilla decay into $|e\rangle$ from $|f\rangle$ can cause misrotation of the storage mode by over $90$ degrees. With such an extreme misrotation, the storage state has a high probability of being recovered incorrectly by the two-photon dissipation, causing bit-flip errors. All $8$ interactions show effective storage mode lifetime $T_{1,\text{eff}}$ over $50~\mathrm{\mu s}$ which is extracted from a linear fit of the phase-flip rates as a function of $|\alpha|^{2}$.  

\subsection{Dissipative map assymmetry}
\label{app:dissipative_map_asymmetry}

We refer to the mapping from an initial density matrix $\hat{\rho}_0$ to the population in the coherent states $|\pm\alpha\rangle$ after two-photon dissipation as the dissipative map. Specifically, we let $P_{D,\pm\alpha}(\hat{\rho}_{0})$ denote the probability of the initial state $\hat{\rho}_{0}$ is mapped to $|\pm\alpha\rangle$ after the two-photon dissipation. The ideal two-photon dissipation with steady states $|\pm\alpha\rangle$, is described by 
\begin{align}
    \frac{d\hat{\rho}(t)}{dt} = D[\hat{a}^2-\alpha^2]\hat{\rho}(t) .  
\end{align}
We consider the case with $\alpha\in \mathbb{R}$.  Then with the ideal two-photon dissipation, if a storage mode is initialized into the state $|\beta\pm i\delta\rangle$ where $\beta,\delta \in \mathbb{R}$, the dissipative map will yield the same population distribution into $|{\pm}\alpha\rangle$ regardless of the sign of $\delta$ for a given $\beta$, i.e., $P_{D,\alpha}( |\beta + i\delta \rangle\langle \beta + i\delta | ) = P_{D,\alpha}( |\beta - i\delta \rangle\langle \beta - i\delta | )$. Similarly in this ideal case, the dissipative map also is symmetric under the sign flip of $\beta$ for a given $\delta$, i.e., $P_{D,\alpha}( |\beta + i\delta \rangle\langle \beta + i\delta | ) = P_{D,-\alpha}( |-\beta + i\delta \rangle\langle -\beta + i\delta | )$ (note the sign flip of $\alpha$ in the subscript).  When these symmetry properties do not hold, we call the dissipative map asymmetric.  

When we perform a $\text{CX}$ (and $\text{CX}^2$) gate between a storage-ancilla pair with near-perfect $\chi$-matching, the symmetry of the two-photon dissipation does not have a big effect on the cat-qubit bit-flip rates since the storage state will rotate close to the target dissipative steady state even if the ancilla decays from $|f\rangle$ to $|e\rangle$ during the gate.  On the other hand if there is a significant $\chi$-mismatch, the asymmetry in the dissipative map can either help or hurt the gate bit-flip rates.  Consider for example the case of performing a $\text{CX}^2$ gate where $\chi_{ge}/\chi_{gf}=1.25$ ($\chi_{ge},\chi_{gf} < 0$). Suppose that the storage is initialized to $|\alpha\rangle$. Then if the ancilla is initialized to $|f\rangle$ and decays to $|e\rangle$ in the very beginning of the gate, the storage mode can overrotate by $\pi/2$ (relative to the target angle of $2\pi$) to the state $|i\alpha\rangle$ instead of $|\alpha\rangle$.  With ideal two-photon dissipation the storage state would be equally recovered to $|\alpha\rangle$ and $|{-}\alpha\rangle$.  This means that ancilla decay events of this form (and state preparation error to $|e\rangle$) have a $50\%$ probability of causing a bit-flip event. If we can add asymmetry to the two-photon dissipation we can make it more likely for the overrotated population to return to the target steady state $|\alpha\rangle$ over the other steady state $|{-}\alpha\rangle$ and reduce the bit-flip probability.  

Here, we show how the dissipative map can be made asymmetric by detuning the pump and drive frequencies of the two-photon dissipation in a specific direction. To do so, we consider the Hamiltonian with detuned 3WM pump and linear drive of the storage-buffer system: 
\begin{align}
    \hat{H}=g_2\hat{a}^2\hat{b}^{\dagger} e^{i\Delta_p t} -g_2\alpha^2\hat{b}^\dagger e^{-i\Delta_d t}+ \text{H.c.} 
\end{align}
Here $\Delta_p$ is the detuning of the pump that realizes the 3WM condition and $\Delta_d$ is the detuning of the linear buffer drive.  We use rotated detunings of the form~\cite{Lescanne2020} 
\begin{align}
    \Delta=(\Delta_p+\Delta_d)/2 , 
    \nonumber\\
    \Delta_b=(\Delta_p-\Delta_d)/2. 
\end{align}
Asymmetric two-photon dissipation is realized by varying $\Delta_b$. With $\Delta=0$, the effective evolution is described by
\begin{align}
    \frac{d\hat{\rho}(t)}{dt}= -i[\hat{H}_{\text{eff}}, \hat{\rho}(t)] +\frac{\kappa_b g_2^2}{\Delta_b^2+(\kappa_b/2)^2} D\left[ \hat{a}^2 - \alpha^2  \right]\hat{\rho}(t), 
\end{align}
with
\begin{align}
    \hat{H}_{\text{eff}}=-\frac{g_2^2\Delta_b}{\Delta_b^2+(\kappa_b/2)^2} (\hat{a}^{\dagger 2}-\alpha^2)(\hat{a}^2-\alpha^2) . 
\end{align}
This shows that the main effects of adding the $\Delta_b$ detuning are to marginally weaken the two-photon dissipation strength (assuming $\kappa_b\gg \Delta_b$) and to introduce a Kerr-cat Hamiltonian~\cite{Puri2017} commensurate with the two-photon dissipation (i.e., with a consistent cat-qubit amplitude $\alpha$).  While the steady states of the two-photon dissipation are unchanged, the use of non-zero $\Delta_b$ can induce significant asymmetry in the dissipative map. We illustrate this in \cref{app_fig:delta_b_effect}(a) by showing the amount of population recovered to $|{\pm}\alpha\rangle$ for an initial coherent state at each point in the phase space. With $\Delta_b=0$ the two-photon dissipation is symmetric while with $\Delta_b/2\pi=\pm 5~\text{MHz}$, significant asymmetry is introduced. The orientation of the asymmetry depends on the sign of $\Delta_b$.   

\begin{figure}[b!]
    \centering
    \includegraphics[width=\columnwidth]{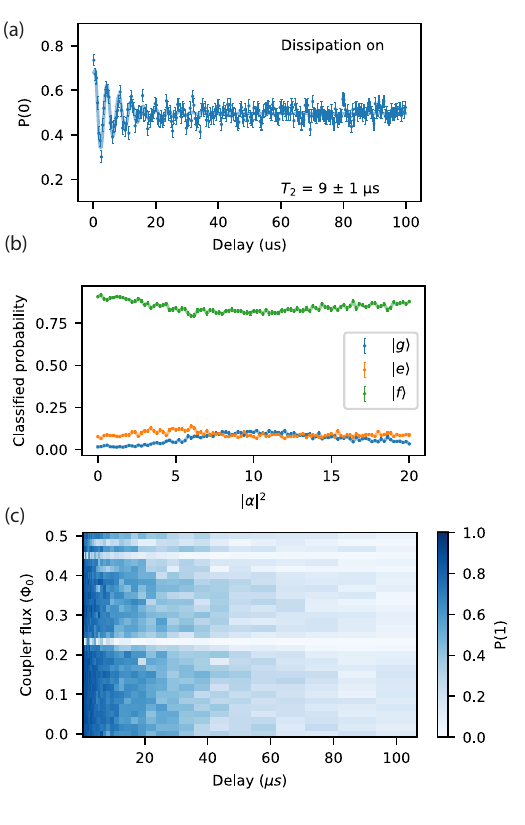}
    \caption{\textbf{Examples of device instability.}  All of these examples are from a different temperature cycle of the chip compared to what was used for the rep code experiment. (a) Example of a storage having a low $T_2$.  In this case the storage is S1.  (b) Example of direct $f\rightarrow g$ decay on the A4,S4 interaction.  We initialize the transmon into $|f\rangle$ and measure the population in $|g\rangle$, $|e\rangle$, and $|f\rangle$ after an 800 ns flux pulse to the on position.  We perform this measurement while exciting the storage mode with variable $|\alpha|^2$.  We observe direct decay of $|f\rangle$ to $|g\rangle$.  (c) Example of the ancilla interacting with TLS as the coupler is tuned from the off to on position on the A3, S3 interaction.  We initialize the ancilla into $|e\rangle$ and apply a coupler flux pulse with variable amplitude and duration.  We observe that there are multiple TLS in the ancilla spectrum.  }
    \label{app_fig:noise_affecting_performance}
\end{figure}

To carefully resolve this asymmetry in experiment we perform a pulsed cat-qubit stabilization experiment where the storage mode is intentionally phase shifted every time the two-photon dissipation is turned off.  By measuring the bit-flip time as a function of the phase shift we can get a sensitive characterization of the asymmetry in the two-photon dissipative map. In \cref{app_fig:delta_b_effect}(b), we show the measured bit-flip times as a function of the storage phase shift for $3$ different photon numbers and $3$ different $\Delta_b$ values. We observe that under the nominal value of $\Delta_{b}$ we use in our experiment ($\Delta_{b,0}$), the dissipative map is asymmetric in favor of recovering positive phase shifts. Decreasing $\Delta_b$ increases the asymmetry in this direction whereas increasing $\Delta_b$ reduces the extend of the asymmetry. Since the dissipative map is approximately symmetric when we use $\Delta_{b} = \Delta_{b,0} +  2\pi \times 2~\text{MHz}$ (dotted curve), we estimate that the $\Delta_{b,0}$ detuning in our experiment is $\Delta_{b,0} / (2\pi) \simeq -2~\text{MHz}$.  

The sign of the phase shift is defined such that a positive phase shift has the same sign as accumulating phase with a lower physical storage frequency.  Thus in our experiment, the orientation of the dissipative map asymmetry is in a favorable direction of correcting over-rotations caused by the $\chi$-mismatch of the form $\chi_{ge}/\chi_{gf}>1$ (in our case $\chi_{ge},\chi_{gf} < 0$). In \cref{app_fig:delta_b_effect}(c) we show the simulated $\text{CX}^2$ bit-flip probabilities for the storage-ancilla pair $\text{A4} \leftrightarrow \text{S4}$ as a function of $\Delta_b$ for $|\alpha|^2=3$. Bit-flip probabilities can be reduced by almost a factor of $2$ by going to negative $\Delta_{b}$ detunings on the scale of $-5~\mathrm{MHz}$. In contrast, positive $\Delta_b$ detuning can result in significantly increased bit-flip probabilities. In the future systematically tuning $\Delta_b$ across the device is an avenue to further optimize the CX gate performance under mismatched $\chi$ ratios. The vertical black line corresponds to $\Delta_{b,0} / 2\pi \approx -2~\text{MHz}$ used in our experiment.

\subsection{Examples of noise affecting repetition code performance}
\label{app:noise_mechanisms}

In \cref{app_fig:noise_affecting_performance} we show a few ways in which device performance can be unstable.  These datasets are not from the time period of the repetition code experiment.  In \cref{app_fig:noise_affecting_performance}(a) we show an example of a storage-mode $T_2$ being degraded.  We intermittently observe that storage $T_2$ may fluctuate which might be due to  spurious two-level systems (TLSs) in buffer or storage.  In \cref{app_fig:noise_affecting_performance}(b) we perform an experiment where the ancilla is initialized to $|f\rangle$, an 800 ns flux pulse is applied, and the ancilla state is read out.  We study the dependence of the ancilla population on the average photon number $|\alpha|^{2}$ of a coherent state in the storage mode.  We observe in this instance that the population of $|f\rangle$ decays directly to $|g\rangle$, bypassing $|e\rangle$. This effect is most pronounced around $|\alpha|^{2}=10$. This could for example be due to a TLS mediating a multi-photon process involving ancilla, coupler, and storage photons.  In \cref{app_fig:noise_affecting_performance}(c) we perform a $T_{1}$ measurement on an ancilla as a function of coupler flux pulse amplitude.  As the coupler moves to its minimum frequency the hybridization of the coupler and ancilla increases and the ancilla frequency is shifted.  Both these effects can result in the ancilla encountering TLSs at specific coupler flux values. In this case we see example of multiple TLSs coupled to the ancilla. These TLSs can cause ancilla population loss and are especially problematic if they are present at the coupler ``on'' position.

\section{Repetition code calibration and pulse sequences}

\subsection{Single-shot Z-basis measurement of a cat qubit}
\label{app:storage_z_measurement}

\begin{figure}[t!]
    \centering
    \includegraphics[width=\columnwidth]{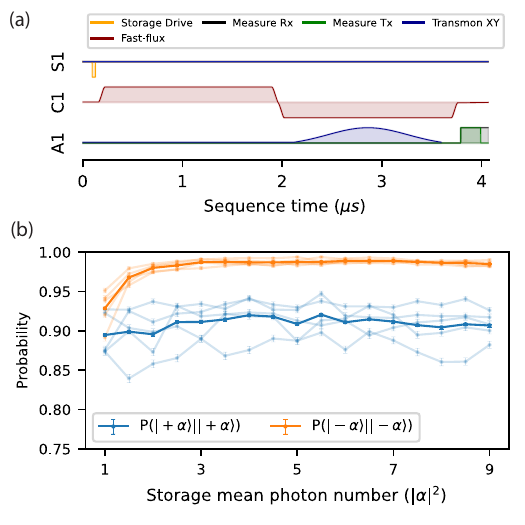}
    \caption{\textbf{Storage mode $Z$-basis measurement.}  (a) Pulse sequence for storage $Z$-basis measurement.  The storage mode is assumed to have started in the cat qubit manifold spanned by the $|\pm\alpha\rangle$ coherent states.  The storage mode is displaced by $-\alpha$. The vacuum population of the storage mode is read out by using a weak vacuum-selective $\pi$ pulse on the ancilla. (b) Characterization of the storage mode $Z$-basis measurement.  The storage mode is prepared into either $|+\alpha\rangle$ or $|{-}\alpha\rangle$ and the two-photon dissipation is applied for $10~\mathrm{\mu s}$. Then finally the storage $Z$-basis measurement is performed. The faded lines indicate the assignment fidelity for each of the storage modes and the solid lines indicate the average performance, averaged over all the storage modes.  }
    \label{app_fig:storage_z_measurement}
\end{figure}

To characterize the repetition code logical $Z$ observable we need to measure $\hat{Z}_{L} = \hat{Z}_{1}\hat{Z}_{2}\cdots\hat{Z}_{d}$. This requires single-shot $Z$-basis measurement of the cat qubits since the outcomes of the $Z$-basis measurements of all the data cat qubits need to be multiplied together in a given shot to evaluate the logical $Z$ operator. As mentioned in \cref{app:basis_and_error_rate_convention}, the $Z$-basis readout scheme based on displaced parity measurements does not suffice for characterizing the repetition code as they require averaging over many shots. Here, we present an alternative $Z$-basis readout scheme that works in a single-shot manner and demonstrate it experimentally. In particular we show that through an adequate symmetrization, the measurement axis of our readout scheme is not tilted from the $Z$ basis under various imperfections (transmon errors and non-orthogonality of the coherent states in the small $|\alpha|^{2}$ regime). 

To perform a single-shot $Z$-basis measurement on a cat qubit, we first displace the storage by either $+\alpha$ or $-\alpha$ such that one of the two computational basis states of a cat qubit is mapped to the vacuum state. Then we measure the vacuum population of the storage mode by using an ancilla transmon. In particular, the vacuum population measurement is enabled by the dispersive coupling between the storage mode and the ancilla. Due to the dispersive interaction, the frequency of the ancilla is shifted when the storage mode is not in the vacuum state. Thus by applying a sufficiently weak drive on the ancilla, we can realize a selective $\pi$ pulse of the ancilla (exciting it from $|g\rangle$ to $|e\rangle$) conditioned on the storage mode being in the vacuum state. Finally by reading out the state of the ancilla in the end, we can determine whether the cat qubit was in the $|0\rangle \simeq |+\alpha\rangle$ state or the $|1\rangle\simeq |{-}\alpha\rangle$ state in a single-shot manner.   

The pulse sequence for the storage $Z$ measurement is shown in \cref{app_fig:storage_z_measurement}(a). First the storage mode is displaced. Then a net-zero flux pulse is applied on the coupler such that the storage-ancilla dispersive coupling is turned on. In the second half of the net-zero flux pulse, a weak $\pi$ pulse is applied to the ancilla, which is resonant only when the storage mode is in the vacuum state. Finally the ancilla is read out. The weak vacuum-selective pulses use Gaussian waveforms with pulse lengths of either $1.6~\mathrm{\mu s}$ or $1.8~\mathrm{\mu s}$ depending on the unit cell.  

In \cref{app_fig:storage_z_measurement}(b) we show the probability of correctly assigning a cat qubit state in the $Z$-basis readout as a function of $|\alpha|^2$. For this $Z$-basis readout data, the storage mode is displaced by $-\alpha$. The faint lines correspond to the $Z$-basis readout fidelities of the five cat qubits in our system. The dark lines represent the average performance. Note that the probability of correctly measuring $|{-}\alpha\rangle$ is higher than for $|+\alpha\rangle$. This is because when the storage mode is in $|+\alpha\rangle$, it is displaced to the vacuum state $|\hat{n}=0\rangle$. This results in the vacuum-selective $\pi$ pulse exciting the ancilla transmon.  Thus in this case, the assignment fidelity is sensitive to transmon decay during the $\pi$ pulse and the transmon readout. In contrast if the storage is in $|{-}\alpha\rangle$, it is displaced to $|-2\alpha\rangle$ and the ancilla is not excited by the vacuum-selective $\pi$ pulse. In this case the $Z$-basis readout is not sensitive to the ancilla decay and the assignment fidelity is higher.  

In practice we symmetrize the $Z$-basis readout by applying the $+\alpha$ displacement in half of the shots and $-\alpha$ displacement in the remaining half of the shots. This has the effect of averaging out the asymmetry in the readout assignment fidelity discussed above. For a distance-$d$ repetition code we apply all $2^d$ combinations of the $\pm\alpha$ displacements with equal probabilities across the shots.

We note that our single-shot $Z$-basis readout scheme works adequately even in the small $|\alpha|^{2}$ limit provided that we symmetrize over the $\pm\alpha$ displacements. To illustrate this point, we characterize the POVM elements of the readout scheme. Let $P^{(+\alpha)}_{0}(\hat{\rho})$ and $P^{(-\alpha)}_{0}(\hat{\rho})$ be the probability of getting the bit-string $0$ for a given initial state $\hat{\rho}$. Here the superscript indicates whether the initial displacement is $+\alpha$ or $-\alpha$. Since the initial $|0\rangle\simeq |+\alpha\rangle$ state is displaced to the $|+2\alpha\rangle$ state under the $+\alpha$ displacement, $P^{(+\alpha)}_{0}(\hat{\rho})$ is defined as the probability of the storage being in a non-vacuum state after the $+\alpha$ displacement. Similarly, $P^{(-\alpha)}_{0}(\hat{\rho})$ is defined as the probability of the storage mode being in the vacuum state after the $-\alpha$ displacement since $|0\rangle\simeq |+\alpha\rangle$ is mapped to the vacuum state $|\hat{n}=0\rangle$ under the $-\alpha$ displacement. Then, $P^{(\pm \alpha)}_{1}$ is simply given by $P^{(\pm \alpha)}_{1} = 1 - P^{(\pm \alpha)}_{0}$. 

Assuming no incoherent errors, we analytically find 
\begin{align}
    &P^{(\pm \alpha)}_{0}(\hat{\rho}) 
    \nonumber\\
    &= \frac{1}{2}(1 \mp e^{-2|\alpha|^{2}}) \langle +| \hat{\rho}|+\rangle  + \frac{1}{2}(1 \pm e^{-2|\alpha|^{2}}) \langle -| \hat{\rho}|-\rangle
    \nonumber\\
    &\quad + \frac{1}{2}\sqrt{1 - e^{-4|\alpha|^{2}}} ( \langle +| \hat{\rho}|-\rangle + \langle -| \hat{\rho}|+\rangle). 
\end{align}
where $|+\rangle$ and $|-\rangle$ are the even and odd cat states. Note that without symmetrizing the $\pm\alpha$ displacements, our readout scheme can differentiate between the even and odd cat states in the small $|\alpha|^{2}$ limit where $e^{-2|\alpha|^{2}}$ is not negligible (i.e., $P^{(\pm \alpha)}_{0}(|+\rangle\langle +| ) \neq P^{(\pm \alpha)}_{0}(|-\rangle\langle -|)$). This is undesirable since it indicates that the measurement basis is tilted from the $Z$ basis. However by sampling both the $\pm\alpha$ displacements with equal probabilities, we can remove this asymmetry. In particular, we have
\begin{align}
    P_{0}(\hat{\rho}) &\equiv \frac{1}{2}\Big{(}P^{(+ \alpha)}_{0}(\hat{\rho}) + P^{(- \alpha)}_{0}(\hat{\rho}) \Big{)}
    \nonumber\\
    &= \frac{1}{2}(\langle +|\hat{\rho}|+\rangle + \langle -|\hat{\rho}|-\rangle) 
    \nonumber\\
    &\quad + \frac{1}{2}\sqrt{1 - e^{-4|\alpha|^{2}}} ( \langle +| \hat{\rho}|-\rangle + \langle -| \hat{\rho}|+\rangle). 
\end{align}
This probability can be succinctly represented as $P_{0}(\hat{\rho}) = \text{Tr}[\hat{\rho}\hat{F}_{0}]$ with a POVM element 
\begin{align}
    \hat{F}_{0} &= \frac{1}{2}(1 + \sqrt{1 - e^{-4|\alpha|^{2}}}) |0 \rangle\langle 0 | 
    \nonumber\\
    &\quad + \frac{1}{2}(1 - \sqrt{1 - e^{-4|\alpha|^{2}}}) |1\rangle\langle 1|, 
\end{align}
where $|0\rangle$ and $|1\rangle$ are the computational basis states. Similarly the POVM element for measuring $1$ is given by
\begin{align}
    \hat{F}_{1} &= \frac{1}{2}(1 - \sqrt{1 - e^{-4|\alpha|^{2}}}) |0 \rangle\langle 0 | 
    \nonumber\\
    &\quad + \frac{1}{2}(1 + \sqrt{1 - e^{-4|\alpha|^{2}}}) |1\rangle\langle 1|. 
\end{align}

In the large $|\alpha|^{2}$ limit where $e^{-4|\alpha|^{2}}$ can be neglected, these POVM operators reduce to the ideal $Z$-basis readout POVM elements $\hat{F}_{0} = |0\rangle\langle 0|$ and $\hat{F}_{1} = |1\rangle\langle 1|$. From the deviation from these ideal POVM elements, we identify that the intrinsic measurement error rate of our $Z$-basis readout scheme is given by $\frac{1}{2}(1 - \sqrt{1 - e^{-4|\alpha|^{2}}})$. Importantly however, the two POVM elements are diagonal in the computational basis at any value of $|\alpha|^{2}$. Thus, while our $Z$-basis readout scheme is subject to an intrinsic assignment error due to the non-orthogonality of the two coherent states $|\pm\alpha\rangle$, its measurement basis is not tilted from the $Z$ basis.   

\begin{figure*}[t!]
    \centering
    \includegraphics[width=2\columnwidth]{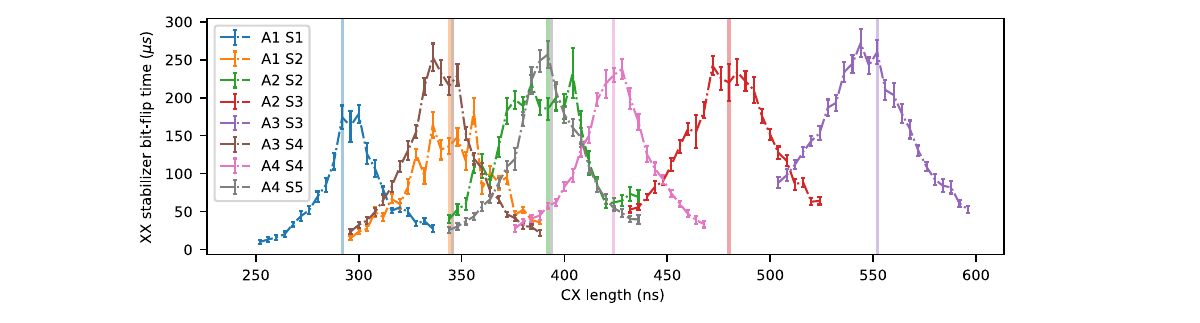}
    \caption{\textbf{$\text{CX}$ gate length calibration and validation summary.} For each $\text{CX}$ gate, we measure the logical bit-flip time of the corresponding $d=2$ repetition code as a function of the $\text{CX}$ gate length. The vertical lines indicate the $\text{CX}$ gate lengths used in the repetition code experiment. Some vertical lines are slightly shifted so they do not overlap with others. The experiment serves as a validation of the gate lengths because the $d=2$ repetition code bit-flip times are maximized near the chosen gate lengths.  }
    \label{app_fig:cx_gate_length_calibration}
\end{figure*}

Finally, we remark that the above symmetrization also eliminates the imbalance due to an asymmetric transmon readout confusion matrix (which could be caused by asymmetries in the intrinsic readout errors as well as the errors in the vacuum-selective ancilla $\pi$ pulse). Let $P^{(t)}_{j\rightarrow k}$ be the probabilities that the transmon was prepared in state $|j\rangle$ and then measured in $|k\rangle$ where $j,k\in \lbrace g,e \rbrace$. Then the probabilities $P^{(\pm \alpha)}_{0}(\hat{\rho})$ are modified as $Q^{(\pm \alpha)}_{0}(\hat{\rho})$ with 
\begin{align}
    Q^{(+\alpha)}_{0}(\hat{\rho}) &= P^{(+\alpha)}_{0}(\hat{\rho}) P^{(t)}_{g\rightarrow g} + (1 - P^{(+\alpha)}_{0}(\hat{\rho}))P^{(t)}_{e\rightarrow g}, 
    \nonumber\\
    Q^{(-\alpha)}_{0}(\hat{\rho}) &= P^{(-\alpha)}_{0}(\hat{\rho}) P^{(t)}_{e\rightarrow e} + (1 - P^{(-\alpha)}_{0}(\hat{\rho}))P^{(t)}_{g\rightarrow e}. 
\end{align}
Then after the symmetrization, we find 
\begin{align}
    Q_{0}(\hat{\rho}) &\equiv \frac{1}{2}\Big{(} Q^{(+\alpha)}_{0}(\hat{\rho}) + Q^{(-\alpha)}_{0}(\hat{\rho}) \Big{)}
    \nonumber\\
    &= \frac{1}{2}(\langle +|\hat{\rho}|+\rangle + \langle -|\hat{\rho}|-\rangle) 
    \nonumber\\
    &\quad + \frac{1}{2}\sqrt{1 - e^{-4|\alpha|^{2}}} \Big{(}1 - \frac{1}{2} (P^{(t)}_{g\rightarrow e} +P^{(t)}_{e\rightarrow g})\Big{)} 
    \nonumber\\
    &\qquad \times ( \langle +| \hat{\rho}|-\rangle + \langle -| \hat{\rho}|+\rangle), 
\end{align}
where we used $P^{(t)}_{g\rightarrow g} + P^{(t)}_{g\rightarrow e} = 1$ and $P^{(t)}_{e\rightarrow g} + P^{(t)}_{e\rightarrow e} = 1$. Thus after the symmetrization over the $\pm\alpha$ displacements, any imbalances due to an asymmetric transmon readout confusion matrix are removed. For example, the even and odd cat states still cannot be distinguished by our single-shot $Z$-basis readout scheme. Note however that the readout contrast is degraded by a factor of $(1 - \frac{1}{2} (P^{(t)}_{g\rightarrow e} + P^{(t)}_{e\rightarrow g}))$ due to the transmon readout assignment errors. As shown in \cref{app_fig:storage_z_measurement}, the $Z$-basis assignment infidelities in our experiments saturate to non-zero value for a sufficiently large average photon number $|\alpha|^{2} \gtrsim 3$. These non-zero saturated infidelities can be attributed to the effective error probabilities $P^{(t)}_{g\rightarrow e}, P^{(t)}_{e\rightarrow g} >0$ which account for the effects due to intrinsic transmon readout assignment infidelities as well as the errors in the vacuum-selective ancilla $\pi$ pulse.   

\subsection{Calibrating the CX gate length for the repetition code}
\label{app:cx_gate_length_calibration_for_rep_code}

Previously in \cref{app:cx_calibration} we described how we fine tuned the $\text{CX}^2$ gate length by directly optimizing the bit-flip times of a single cat qubit. Here we show how we optimize the length of a $\text{CX}$ gate which is used in the repetition code sequence. To fine tune the $\text{CX}$ gate length, we optimize the bit-flip time of a $d=2$ repetition cat code (i.e., bit-flip time under the repeated measurements of a $\hat{X}_{i}\hat{X}_{i+1}$ stabilizer). 

Recall that the bit-flip times under a $\text{CX}^{2}$ gate are insensitive to ancilla state preparation errors. This is because an $\text{X}^{2}$ operation acts trivially on a single cat qubit. Similarly, bit-flip times under a $\hat{X}_{i}\hat{X}_{i+1}$ stabilizer measurement are not sensitive to ancilla state preparation errors because the stabilizer acts trivially on the repetition code states. This is why we use logical bit-flip time of a $\hat{X}_{i}\hat{X}_{i+1}$ to fine tune the length of the $\text{CX}$ gates. In the pulse sequence for measuring the $\hat{X}_{i}\hat{X}_{i+1}$ stabilizer, the associated ancilla transmon (e.g., A1) is prepared in the $|g\rangle + |f\rangle$ state and two $\text{CX}$ gates are applied sequentially between the ancilla and its two neighboring cat qubits (e.g., $\text{A1}\leftrightarrow \text{S1}$ and $\text{A1}\leftrightarrow \text{S2}$). Then, the ancilla is unprepared through the inverse of the state preparation pulse. Finally the ancilla is read out and reset. For the purpose of calibrating the $\text{CX}$ gate length, we use a simplified pulse sequence where the ancilla readout is omitted. However, it is crucial to keep the ancilla reset for similar reasons as in the $\text{CX}^{2}$ gate length calibration. 

To calibrate the length of each $\text{CX}$ gate, we measure the logical bit-flip time of the corresponding $d=2$ repetition code as a function of the $\text{CX}$ gate length of interest. \cref{app_fig:cx_gate_length_calibration} shows the logical bit-flip time of a $d=2$ repetition code as a function of a $\text{CX}$ gate length for all $8$ $\text{CX}$ gates. The $\text{CX}$ gate lengths used in the repetition code experiment are indicated by the vertical lines. These chosen gate lengths are validated by the results in \cref{app_fig:cx_gate_length_calibration} since the logical bit-flip times of all the constituent $d=2$ repetition codes are maximized around the chosen gate lengths. Note that before these experiment, the storage phase corrections were optimized for the nominal $\text{CX}$ gate lengths following a procedure similar to the one described in \cref{app:insitu_calibration}.

\subsection{In situ calibrations of $\text{CX}$ gate for repetition code operation}
\label{app:insitu_calibration}

\begin{figure}[t!]
    \centering
    \includegraphics[width=\columnwidth]{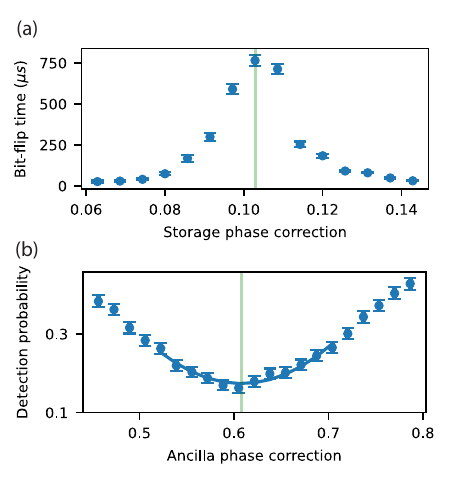}
    \caption{\textbf{Example in situ calibration of the storage and ancilla phase corrections for a repetition code operation.}  (a) Example of optimizing the storage $\text{CX}$ gate phase for a repetition code experiment.  We perform an experiment similar to the repetition code $Z$-basis experiment with the main difference being that the ancilla is in the state $|g\rangle$.  We sweep that phase applied during the $\text{CX}$ gate on the storage mode.  Rather than optimizing the logical bit-flip time we optimize the bit-flip times of the individual storage modes. The vertical green line indicates the optimal point which was selected based on the calibration. (b) Example of optimizing the ancilla $\text{CX}$ phase for a repetition code experiment.  We measure the ancilla detection probability as a function of the phase on the ancilla.  We fit the minimum to a quadratic to find the optimal phase which is indicated by the vertical line.    }
    \label{app_fig:insitu_calibrations}

\end{figure}

Prior to every run of the repetition code experiment we optimize both the storage and ancilla phase corrections. The storage phase corrections affect the logical-bit flip rates while the ancilla phase corrections affect the syndrome measurement error probability of the stabilizers.  Optimization of the storage and ancilla phase correction are each done in parallel across all unit cells.  

To optimize the storage mode phase we perform a slightly simplified repetition code experiment with the ancilla in $|g\rangle$ and readout not applied.  With the ancilla in $|g\rangle$, individual storage $Z$ observables are meaningful quantities since ideally no entanglement between between different storage modes is generated after the first round of the stabilizer measurement.  We in parallel vary the phase of each of the storage modes in a window around their nominal values.  For each storage mode we update the storage mode phase to the value which yields the highest bit-flip time.  We show an example of this calibration for the first storage mode in \cref{app_fig:insitu_calibrations}(a).  We show the data for all of the storage mode phase calibrations from the repetition code run in \cref{app_fig:all_phase_optimizations}(b).  As expected the center storage mode S3 has the calibration which most significantly depends on the repetition code section being used since it can either be on the boundary or in the center.

To calibrate the ancilla phase correction, we perform a repetition code experiment where in parallel we vary the phase correction on each ancilla.  We determine the detection probability for each of the stabilizers as a function of the ancilla phase~\cite{Kelly2015}.  We fit the minimum to a quadratic to determine the optimal value of the ancilla phase.  An example of this is shown in \cref{app_fig:insitu_calibrations}(b).  We show the all of the storage mode phase calibrations from the repetition code run in \cref{app_fig:all_phase_optimizations}(a).

\subsection{Crosstalk cancellation}
\label{app:crosstalk}

\begin{figure}[t!]
    \centering
    \includegraphics[width=\columnwidth]{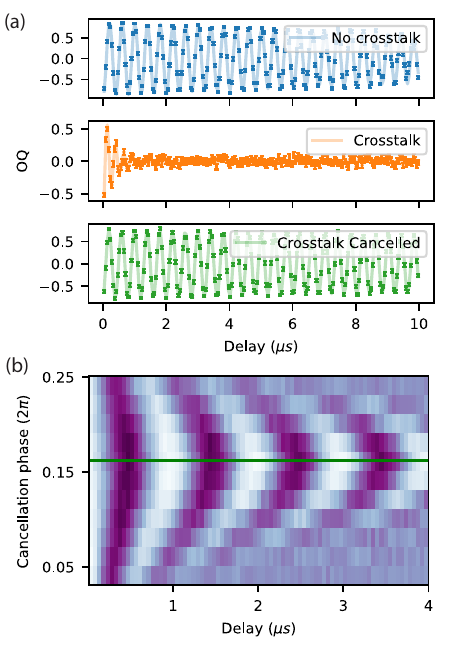}
    \caption{\textbf{Crosstalk from flux pump.}  (a) Ramsey measurements of the ancilla A4 with and without crosstalk from the flux pump on the buffer B1. In blue we show the result of a typical Ramsey measurement on the ancilla A4. In orange we show the same Ramsey measurement on A4 when the buffer flux pump is applied on B1. The crosstalk causes significant excitation of A4's readout resonator R4 which then dephases and causes an average frequency shift of the ancilla A4.  (b) Optimizing the crosstalk cancellation phase.  We perform the ancilla Ramsey measurements in the presence of the buffer flux pump in B1 as a function of the phase of a crosstalk cancellation tone applied on the readout resonator R1.  When the phase is properly chosen we can cancel out the crosstalk from the buffer flux pump.}
    \label{app_fig:crosstalk_calibration}
\end{figure}

Both the readout resonators and the buffer flux pumps in our system have the frequencies ranging from $7~\text{GHz}$ to $8~\text{GHz}$. Thus, the buffer flux pumps for realizing two-photon dissipation may excite the readout resonators due to crosstalk. In our device, we find that the crosstalk to the readout resonator R4 of the ancilla A4 from the flux pumps on the buffers B1 and B2 is significant and problematic. 

The buffer-pump-induced excitation of a readout resonator have several detrimental consequences. First, excitations in the readout resonator cause an average frequency shift and dephasing of the ancilla which can affect the syndrome measurement fidelity of the associated stabilizer. Second, the excitation of the readout resonator can significantly limit the performance of the ancilla reset since the ancilla can be excited by the reset tones if the readout resonator is not in its ground state. Lastly, if the readout resonator is excited, couplers can be excited as they are flux pulsed from their ``off'' to the ``on'' position and pass through the readout resonator.  In particular, the latter two effects (additional heating of the ancillas and couplers) can induce cat-qubit bit flips.  

In \cref{app_fig:crosstalk_calibration}(a), we show an example of  how a buffer flux pump may dephase an ancilla through crosstalk by performing Ramsey experiments on A4 with and without the flux pump applied to B1. When the flux pump is turned on in the buffer B1, the coherence of the ancilla A4 is significantly degraded as illustrated by the orange curve (to be compared against the blue curve). This is due to the excitation of the readout resonator R4 which is caused by the flux pump on the buffer B1.   

In our experiment, ancilla readout and reset occur while the flux pumps are applied on the buffers.  As a result, for readout and especially reset to work properly we need to cancel out any crosstalk from flux pumps. We apply an additional compensation tones on the readout line at the same frequency as the buffer flux pump to cancel this crosstalk. The amplitudes and phases of the additional compensation tones need to be calibrated to cancel out the crosstalk from the flux pump.  In \cref{app_fig:crosstalk_calibration}(b) we show an example of calibrating the crosstalk cancellation phase for the crosstalk from B1 to A4's readout resonator R4. We determine the optimal cancellation phase by fitting the frequency shift of the ancilla A4 as a function of the cancellation phase. The green curve in \cref{app_fig:crosstalk_calibration}(a) demonstrates that the ancilla coherence can be successfully recovered by performing the crosstalk cancellation.   

In practice we need to apply two crosstalk cancellation tones to A4's readout resonator R4 due to crosstalk from the buffer flux pumps on B1 and B2. One important note is that we only calibrate crosstalk cancellation for the steady state.  The transients of the flux pump are not cancelled out and thus it is still important to conservatively pad the turning on of the buffer flux pump.  As discussed in \cref{app:distance_3_and_5_detection_comparison} nonideal padding of the buffer flux pump causes slightly elevated detection fractions on the stabilizer associated with A4 when running the full $d=5$ repetition code experiment.  Note that the reason we turn off the two-photon dissipation well before the coupler flux pulses for the $\text{CX}$ gates and even before the ancilla state preparation is to ensure that any excitations of the readout resonator due to crosstalk are fully dissipated away before these operations begin.    

\subsection{Repetition code X basis pulse sequences}
\label{app:X_basis_pulse_sequence}
\begin{figure*}[t!]
    \centering
    \includegraphics[width=1.9\columnwidth]{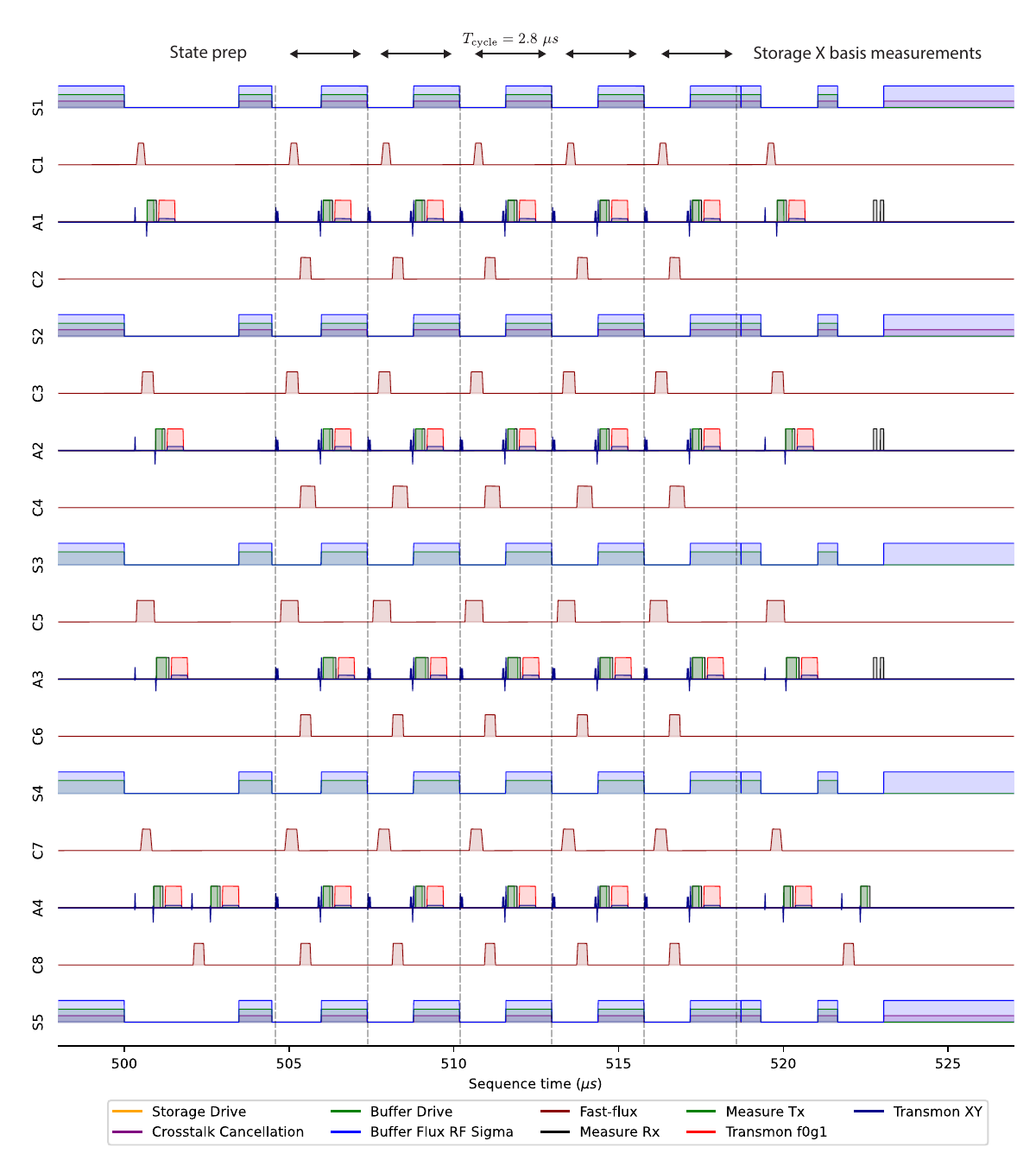}
    \caption{\textbf{Pulse sequence for characterizing the distance-5 repetition code logical $X$ lifetime.}  The storage modes are prepared into cat states, multiple rounds of syndrome extraction are applied, and finally the storage mode parity is readout.  Storage mode preparation into cat states is done by applying two-photon dissipation for $200~\mathrm{\mu s}$ and measuring parity. The sequence is simplified to not show phase offsets which are applied each cycle.  }
    \label{app_fig:X_basis_experiment}
\end{figure*}
In \cref{app_fig:X_basis_experiment} we show the pulse sequence for the characterization of the repetition code logical $X$ lifetime.  The sequence can be broken into three parts.  First the storage mode states are prepared into their initial states, next a variable number of cycles (in this case 5) of stabilizer measurements are applied, and finally the storage mode states are read out.  

For the logical $X$ lifetime experiment the storage modes are initialized into an even or odd cat state.  Before applying operations to the cat qubits we have a $300~\mathrm{\mu s}$ recycle delay between shots of the experiment.  Then, we apply two-photon dissipation on all the cat qubits for $200~\mathrm{\mu s}$ (not shown in its entirety) where all the storage modes start from the vacuum state. The duration of $200~\mathrm{\mu s}$ is chosen to be long enough compared to all of the storage $T_1$ lifetimes in the circuit so that the storage modes reach a steady-state parity.  The reason that we apply the two-photon dissipation rather than displacing the storage mode to a coherent state is that the steady-state parity at low photon number for the two-photon dissipation is different from the photon-number parity of a coherent state at the same photon number. After applying the two-photon dissipation we prepare the storage modes into either an even or odd cat state by non-destructively measuring the parity of each storage.  These parity measurements are performed with the ancilla transmons in the $|g\rangle / |e\rangle$ manifold.  The last ancilla A4 is used to measure the parity of the last two storage modes (S4 and S5), while all of the other ancillas (A1, A2, and A3) are used to measure the parity of only one storage mode (S1, S2, and S3). After preparing the storage mode states we briefly apply two-photon dissipation for $1~\mathrm{\mu s}$.  

Note that after the parity measurement, a tensor-product state of even and odd cat states (e.g. $|+\rangle|-\rangle|-\rangle|+\rangle|+\rangle$) is randomly prepared in the storage modes. For a distance-$d$ repetition code, there are $2^{d}$ possible combinations of such tensor-product states and the outcomes of the parity measurement can be used to determine which one of the $2^{d}$ states is prepared. With a sufficiently large $|\alpha|^{2}$ (e.g., $|\alpha|^{2} \gtrsim 1.5$), these $2^{d}$ states are drawn from a uniform distribution. However due to the asymmetric phase-flip rates between even and odd cat states, the distribution is skewed towards the even cat states in the small $|\alpha|^{2}$ regime (see \cref{app:distribution_of_initial_states_in_logical_X_lifetime_experiments}).

After the storage states are prepared we apply repeated rounds of stabilizer measurements with a conservative cycle time of $2.8~\mathrm{\mu s}$.  The stabilizer measurements begin with the preparation of the anclla into the state $|g\rangle + |f\rangle$ using a $ge$ $\pi/2$ pulse and two $ef$ $\pi/2$ pulses.  Next the CX gates are applied between ancillas and the storage modes.  We apply two layers of CX gates with a padding between.  Since the CX gates have different gate lengths the CX gates in the first layer are padded at front and the CX gates in the second layer are padded at the back to ensure the spacing padding between CX gates of the first and second layer are uniform across the device.  This serves to minimize the amount of time between the two CX gates since an ancilla decay to the $|g\rangle$ state between the CX gates induces a logical bit-flip error whereas an ancilla decay before or after the CX gates is unlikely to induce a logical bit-flip error.  After the CX gates the ancilla state is unprepared using two $ef$ $\pi/2$ pulses and a $ge$ $\pi/2$ pulse mapping the ancilla to the $|g\rangle$/$|e\rangle$ manifold in the absence of the ancilla decay. Note that if the ancilla decayed during the CX gates it ends up in the $|f\rangle$ state after the unpreparation pulses.  We subsequently also apply an $ef$ $\pi$ pulse.  This additional pulse allows us to detect ancilla $|f\rangle\rightarrow |e\rangle$ decay during the readout as an erasure error.  

Once the ancilla pulses have been applied, we read out and reset the ancilla tranmons. As mentioned in \cref{app:readout_and_reset}, the readout includes $80~\text{ns}$ of integration and we include an additional $64~\text{ns}$ of padding between readout and reset.  We apply the two-photon dissipation when the couplers are in the ``off'' position and in parallel to ancilla readout and reset.  Both the buffer drive and flux pump for realizing the two-photon dissipation use flat-top Gaussian waveforms.  In the steady-state of the two-photon dissipation a crosstalk cancellation tone is applied to some readout resonators as described in \cref{app:crosstalk}.  We do not turn on (or off) the two-photon dissipation directly after (or before) the CX gates to reduce the effects of the buffer-pump-induced crosstalk on the cat-qubit bit-flip rates.  For example, if the two-photon dissipation is turned off right before the CX gate, the transients of the flux pulse turning off (which are not cancelled by the crosstalk cancellation tone) can excite the readout resonator which can in turn excite the coupler as it passes through the readout resonator on its way to the ``on'' position.  This coupler excitation then increases the probability of a bit-flip error.  As will be discussed later the padding of the flux pulse could be further improved relative to the ancilla pulses to mitigate a slight increase in the detection probabilities of the A4 stabilizer in the repetition code experiment shown in \cref{app:distance_3_and_5_detection_comparison}.  Many of the conservative paddings in the circuit, such as that between the readout and reset or between readout and the state preparation for the next round, can be reduced to improve the performance of our repetition cat code.  The last cycle of the stabilizer measurements is slightly different because the two-photon dissipation stays on as we do not perform another ancilla state preparation.

After the last cycle of the syndrome measurements, we measure the individual photon-number parity of all the storage modes. Due to a data processing detail extra digitizations are present on some of the ancillas but these are not pulses played on the device.  At the end of the sequence we apply two-photon dissipation for $4~\mathrm{\mu s}$ to partially reset the storage modes by accelerating their return to the $|\hat{n}=0\rangle / |\hat{n}=1\rangle$ manifold. The subsequent recycle delay of $300~\mathrm{\mu s}$ resets the storage modes to the vacuum.

\subsection{Repetition code Z basis pulse sequences}
\label{app:Z_basis_pulse_sequence}
\begin{figure*}[t!]
    \centering
    \includegraphics[width=1.9\columnwidth]{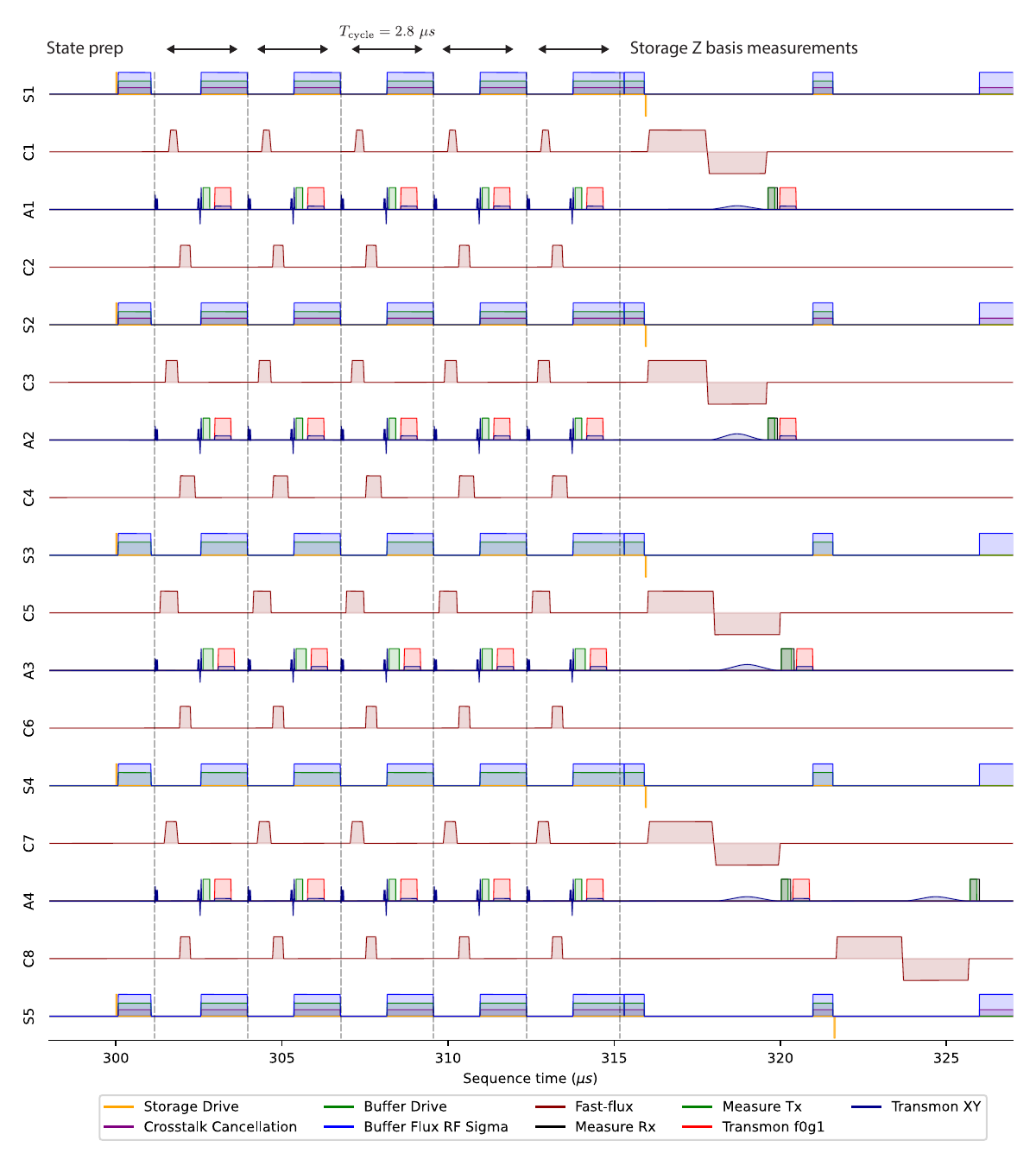}
    \caption{\textbf{Pulse sequence for characterizing the distance-5 repetition code logical $Z$ lifetime.}  The storage modes are prepared into coherent states, multiple rounds of syndrome extraction are applied, and finally the storage modes are readout using $Z$-basis measurement.  The sequence is simplified to not show phase offsets which are applied each cycle.  Photon number selective pulse amplitudes are increased to make them more visible.  }
    \label{app_fig:Z_basis_experiment}
\end{figure*}
The sequence for characterizing the logical $Z$ lifetime of the repetition code is shown in \cref{app_fig:Z_basis_experiment}.  The sequence follows the same steps of initial storage mode state preparation, multiple cycles of stabilizer measurement, and finally storage mode measurements.  

For state preparation in the logical $Z$ basis experiment we displace each for the storage modes to the state $|\alpha\rangle$.  We briefly apply two-photon dissipation for $1~\mathrm{\mu s}$ before starting the error correction cycles.  Though the states are identical up to an instrument phase we do also initialize the repetition code into the $|{-}\alpha\rangle^{\otimes d}$ state for completeness.

It is important that the central cycles of the logical $Z$ lifetime experiment apply identical pulses to the device as the sequence for characterizing the logical $X$ lifetime shown in \cref{app:X_basis_pulse_sequence}.  Since the measurement outcomes for the central rounds are not used for a logical $Z$ basis experiment we do not digitize the outcomes.  This does not affect the tones played on the device and is the only difference in the error correction cycle pulse sequence.  In \cref{app:X_and_Z_basis_detection_comparison} we perform a $Z$ basis experiment with digitization on to confirm the detection probabilities in the $Z$ basis experiment are commensurate with the $X$ basis experiment even with the different initialization.

After the error correction cycles each cat qubit in a storage mode is measured using the single-shot $Z$ basis measurement to distinguish $|\alpha\rangle$ from $|{-}\alpha\rangle$ as discussed in \cref{app:storage_z_measurement}.  The single-shot $Z$ basis measurement is symmetrized over all $2^d$ possible ways of $\pm\alpha$ displacements.  After the measurement, each storage mode is partially reset to the $|\hat{n}=0\rangle/|\hat{n}=1\rangle$ and a $300~\mathrm{\mu s}$ delay is applied in the same way as in the $X$ basis experiment.

\subsection{Repetition code characterization procedure}
\label{app:characterization_procedure}
To characterize the repetition code performance we perform a set of three interleaved calibrations and characterizations of the distance-3 and distance-5 sections.  More specifically we perform the following high level sequence.  

\begin{verbatim}
for i in [1,2,3]:
    for code in [
       distance_5, 
       first_distance_3, 
       second_distance_3
    ]:
        calibrate_ancilla_phases()
        if code == distance_5:
            storage_T1_with_dissipation_off()
            storage_T1_with_dissipation_on()
        calibrate_simultaneous_storage_
            Ramsey_with_dissipation_off(code)
        calibrate_simultaneous_storage_
            Ramsey_with_dissipation_on(code)
        calibrate_storage_phases(code)
        collect_Z_basis_logical_data(code)
        collect_X_basis_logical_data(code)
        
\end{verbatim}

The calibrations are kept identical between the different repetition code sections to not give an advantage to one section over the others.  We only characterize the storage lifetimes for the $d=5$ section since it is done purely for characterization purposes and not used to update the calibration. When we characterize the repetition code $Z$ logical lifetime we do so for the ancilla being in $|g\rangle+|f\rangle$ and $|g\rangle$. For the repetition code $X$ basis experiments the axis ordering used is [storage mean photon number, number of cycles].  For the $Z$ basis experiment the axis ordering is [storage mean photon number, ancilla state, storage initial state, number of cycles, readout symmetrization].  For the $d=5$ $Z$-basis experiments we use $50$ shots and for the $d=5$ $X$-basis experiment we use $5000$ shots.  For the $d=3$ $Z$-basis experiments we use $120$ shots and for the $d=3$ $X$-basis experiment we use $5000$ shots.  Note that for the $Z$-basis experiment more points are collected for $d=5$ compared to $d=3$ with the same number of shots due to the larger number of storage $Z$-basis readout symmetrizations as explained in \cref{app:storage_z_measurement}. To analyze the repetition code data we combine together the data from the interleaved runs into a large dataset which we perform the analysis on. The duration of the characterization procedure is roughly $6~\text{hours}$. 

\section{Repetition Code Data Analysis}

\subsection{Fitting procedure for the logical lifetimes}

To determine the logical lifetimes we fit the exponential decay of $\langle \hat{O}_L(0)\hat{O}_L(t) \rangle$ as a function of time (here $\hat{O}$ can be $\hat{X}$ or $\hat{Z}$).  

First, we describe how the value of  $\langle \hat{O}_L(0)\hat{O}_L(t) \rangle$ and its uncertainty are computed from the raw experimental data. The raw output from the experiment for one time step is a length-$N$ binary vector, where $N$ is the number of shots, and $i$th vector element is 0 (1) when the initial and final logical states agree (disagree).  From this vector we compute the sample mean $\mu_0$.  We assume the data comes from a binomial distribution which has the beta distribution as its conjugate prior. We take a uniform prior by starting from $\beta(1,1)$.  Specifically for the sample mean $\mu$ and the number of shots $N$, the conjugate prior beta distribution is given by
\begin{align}
    \beta(1+N\mu_0,1+N-N\mu_0).
\end{align}
We assign the mean ($\mu$) and standard deviation ($\sigma$) based on this distribution.  Then we compute $\langle \hat{O}_L(0)\hat{O}_L(t) \rangle=1-2\mu$ and $\sigma_{ \langle \hat{O}_L(0)\hat{O}_L(t)\rangle }=2\sigma$.

Once we have computed the value of $\langle \hat{O}_L(0) \hat{O}_L(t)\rangle$ and $\sigma_{ \langle \hat{O}_L(0)\hat{O}_L(t)\rangle}=2\sigma$ we fit to an exponential decay model.  In the model fitting the points are weighed by the inverse of their standard deviation.  For the fit to the $\hat{Z}_L$ observable we include no offset in the exponential fit.  For the fit to the $\hat{X}_L$ observable we allow for a constant offset in the exponential fit since the steady state parity of the individual storage modes is non-zero at low $|\alpha^2|$. In practice the constant offsets are small (magnitude less than $0.03$) even at $|\alpha|^2=1$ because even though the individual storage modes have significant parity offsets the offset of the logical observable is suppressed with distance. 

\subsection{Decoding without erasure information}
\label{app:decoding_without_erasure_information}
\begin{figure*}[t!]
    \centering
    \includegraphics[width=\textwidth]{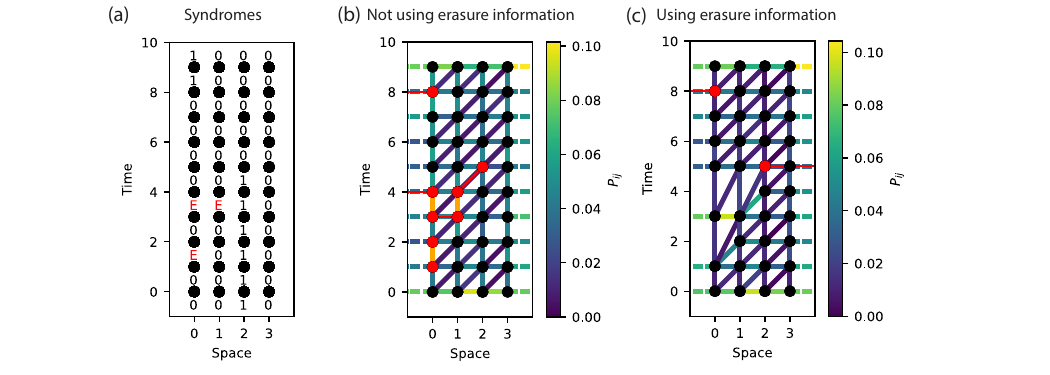}
    \caption{\textbf{Error correction graph.} (a) Syndrome outcomes for an example shot of a distance-5 repetition code run with $|\alpha|^2=1.5$.  Circles indicate detection locations and the numbers above and below denote the syndrome measurement outcomes from which the detection event is computed.  The red indicates syndromes which are classified as erasures. (b) Error correction graph not accounting for erasures.  The red vertices indicate triggered detection events. The red lines indicate the matching, which in this case indicates as a correction that the 2nd and 3rd data qubits should be flipped. The orange edges indicate syndromes where erasures occurred, but here this erasure information was not used to help inform the matching. (c) Modified error correction graph taking into account the erasure information.  The syndromes corresponding to erasure events are not trusted and we instead compute detection events only using non-erased syndromes.  Edge weights are combined together to account for the higher error probability between further separated detection events.  The new matching correctly disregards erasure information and suggests a different (and correct) correction compared to the decoder without erasure information. }
    \label{app_fig:error_correction_graph}
\end{figure*}

An error correction experiment returns syndrome outcomes $S_t^{(i)}$, where $i$ denotes the index of the ancilla measured and $t$ the time step.  The syndrome takes the value $S_t^{(i)}=0$ when the ancilla is measured to be in the state $|g\rangle$ and $S_t^{(i)}=1$ when the ancilla is measured to be in the state $|e\rangle$ or $|f\rangle$.  Detectors are defined by comparing syndrome outcomes from subsequent rounds of error correction
\begin{align}
    D_{t_1, t_2}^{(i)} = (S_{t_1}^{(i)}+S_{t_2}^{(i)})\text{ mod } 2.
\end{align}
We refer to detectors with value $0$ as trivial, and detectors with value $1$ as non-trivial.  A non-trivial detector indicates that an error occurred. Individual physical errors result in one or two nontrivial detectors.  

To infer what physical error occurred from the detectors we form a graph where each vertex corresponds to a detector.  Edges between detectors (or to a boundary) indicate physical errors which can trigger the connected detectors.  Each edge $e$ is assigned a weight $w_e$ related to the probability of the corresponding physical errors $p_e$ by $w_e = \log((1-p_e)/p_e)$.

Given a set of non-trivial detectors, we use minimum-weight perfect matching (MWPM) to infer the corresponding physical errors. In this procedure, each non-trivial detector is matched to another non-trivial detector or the boundary via the graph edges, and the set of matched detectors is referred to as a matching.  MWPM returns the matching which minimizes the total weight of the involved edges. Matching is performed using the PyMatching package~\cite{pymatchingv2}.  

The edge weights of the error correction graph are determined using the correlation based weighting procedure of~\cite{Chen2021}. We restrict allowed edges to only those that would arise in a simple Pauli error model: space-like edges for data qubit errors, time-like edges for syndrome measurement errors, and spacetimelike edges for data qubit errors occurring midway through an error correction cycle. Letting $P_{ij}$ denote the probability that detectors $i$ and $j$ are simultaneously nontrivial, the weight of an edge connecting detectors $i$ and $j$ is computed as $\log((1-P_{ij})/P_{ij})$. The $P_{ij}$ are computed from measured correlations between detectors using the first quarter of the experimental shots. 

In \cref{app_fig:error_correction_graph}(a) we show an example set of experimental syndrome measurement outcomes for the distance-5 code. When we do not account for erasure information, all syndromes with outcome $1$ and $2$ are treated as the same syndrome outcome. Detectors are indicated by black dots, and detection events are computed by taking the difference of the syndrome outcomes at subsequent rounds of the experiment.  In \cref{app_fig:error_correction_graph}(b) we show the error correction graph where vertices correspond to detection locations and edges correspond to errors which can trigger a pair of detection events.  We indicate in orange the edges which correspond to an erased syndrome though we do not yet use this information.  The edges selected by the MWPM are indicated in red.  In this case, the matching indicates that the second and third data qubits have each suffered a net phase-flip.  For this example, we can already see that not accounting for the erasure information can affect the decoding outcome: the matching indicates that the first data qubit did not suffer a net phase flip, since nontrivial detectors associated with an erasure at the third time step are interpreted as a phase flip.   

\subsection{Decoding with erasure information}
\label{app:decoding_with_erasure_information}

When accounting for erasures the ancilla measurement outcome $|e\rangle$ is no longer treated as $S_t^{(i)}=1$ but rather is classified as an erasure, $S_t^{(i)}=\text{E}$.  To incorporate this erasure information in decoding we employ the following three step procedure.  First, we generate a no-erasure baseline matching graph where the edge weights are computed excluding erasures.  Next, on a per-shot basis we compute new detectors that bypass the erased syndromes. Lastly, we update the error correction graph edges to account for the modified detectors.  We now describe each of these three steps in detail.  

When we account for erasure information, the probabilities of certain edges in the matching graph can change.  The main examples of this fact are the timelike edges which correspond to syndrome measurement errors.  In the simple method of \cref{app:decoding_without_erasure_information}, we always assign the erasure state ($|e\rangle$) to syndrome value 1.  Thus in this simple graph weighting an erasure event has a probability of roughly 50\% of causing a measurement error, and the timelike edges incorporate this probability.  Here instead, to construct a no-erasure baseline graph, we compute the edge probabilities conditioned on no erasures occurring.  For a given detector, we consider the neighborhood of detectors directly connected to it (i.e.~detectors one edge away).  We consider the shots of the experiment where none of the detectors in the neighborhood involve an erasure (a detector $D_{t_1, t_2}^{(i)}$ involves an erasure if $S_{t_1}^{(i)}=\text{E}$ or $S_{t_2}^{(i)}=\text{E}$).  Using only these shots, we apply the same weighting procedure of~\cite{Chen2021} to determine the weights of the edges involving the given detector.  This procedure is repeated for every detector in the graph.   Note that we do not allow for erasure events in the initial storage mode measurements for state preparation or the final storage mode parity measurements as both these measurements use the $|g\rangle/|e\rangle$ encoding of the ancilla.  

After forming the no-erasure baseline matching graph, we update the detectors on a per-shot basis such that detectors are formed only using non-erased syndromes.  Specifically when $S_t^{(i)}=\text{E}$, the detectors $D_{t-1, t}^{(i)}$  and $D_{t, t+1}^{(i)}$ are no longer meaningful.  We instead replace these two detectors with a single detector $D_{t-1,t+1}^{(i)}$.  This new detector compares the last non-erased syndrome before the erasure to the first non-erased syndrome after the erasure. Similarly, if multiple consecutive syndromes are erased $S_{t_1}^{(i)}\ldots S_{t_2}^{(i)}$ we replace them with a single detector $D_{t_1-1,t_2+1}^{(i)}$.  

Next we update the edges of the matching graph to account for the detectors that were removed.  As an example consider the case of an erasure at time $t$, the meaningful detector $D_{t-1,t+1}^{(i)}$ is now responsible for detecting data errors between times $t-1$ and $t+1$.  Since this time is twice as long as usual, we expect that the horizontal edges connecting to this detector should have roughly twice as large probability relative to that in the baseline no-erasure graph. Formalizing this intuition, we regard each edge in the baseline graph as representing an independent physical error mechanism that results in nontrivial values for the connected detectors. We construct new edges so that this same set of physical error mechanisms is still accurately accounted for after detectors have been deleted. In particular, we update the edges as follows starting from baseline no-erasure error correction graph:
\begin{algorithmic}[1] 
    \FOR{each cluster of consecutive erased detectors $C_{t_1, t_2}^{(i)} \equiv \{D_{t_1, t_1 +1}^{(i)}\ldots D_{t_2,t_2+1 }^{(i)}\}$}
        \STATE Add the non-erased detector $D_{t_1, t_2}^{(i)}$ to the graph (initially with no connecting edges)
        \FOR{each detector $D \not \in C_{t_1, t_2}^{(i)}$ with edges connecting to $C_{t_1, t_2}^{(i)}$}
            \STATE Enumerate the edge probabilities $p_e$ for each edge $e$ connecting $D$ to $C_{t_1, t_2}^{(i)}$
            \STATE Treating $p_e$ as probabilities of independent events, compute the total probability $p_{odd}$ of an odd number of these events occurring
            \STATE Add a new edge connecting $D$ to the non-erased detector $D_{t_1, t_2}^{(i)}$ with edge probability $p_{odd}$ 
        \ENDFOR
        \STATE Remove the detectors in $C_{t_1, t_2}^{(i)}$ and associated edges from the graph
    \ENDFOR
\end{algorithmic}
After this procedure, we perform MWPM on the resultant graph.

\begin{figure}
    \centering
    \includegraphics[width=0.8\linewidth]{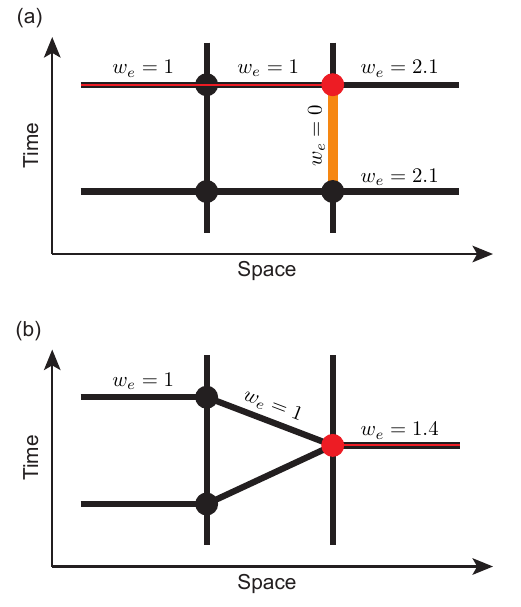}
    \caption{\textbf{Comparison of erasure decoding strategies.} (a) The ``naive'' strategy, where an edge corresponding to erased syndrome (orange) has its edge weight set to 0. Vertices indicate detectors, with the single nontrivial detector in red. The weights of relevant edges are specified, and the red line indicates the minimum-weight matching. (b) Our erasure decoding strategy, where detectors are constructed using only non-erased syndromes. The two $w_e = 2.1$ edges in (a), each corresponding to error probability $p_e \approx 0.11$, are effectively combined here into a single edge with a reduced weight $w_{e'} \approx 1.4$. This reduced weight corresponds to roughly twice the error probability, $p_{e'} = 2p_e (1-p_e) \approx 0.19$, since two time steps elapse between non-erased syndromes. The modification to edge weight changes the minimum-weight matching. }
    \label{app_fig:erasure_matching_comparison}
\end{figure}

In \cref{app_fig:error_correction_graph}(c) we show the matching graph updated to account for erasure information.  The erased syndromes indicated by the orange edges in (b) are no longer used and the adjacent detectors are removed.  Instead, each cluster of erased detectors is replaced by a single detector  involving only non-erased syndromes.  Notably, in this example accounting for erasures alters the logical outcome. Whereas the matching in \cref{app_fig:error_correction_graph}(b) incorrectly indicates that the 2nd and 3rd data qubits suffered net errors, the matching in \cref{app_fig:error_correction_graph}(c) correctly indicates that the 1st, 4th, and 5th data qubits suffered net errors. 

An alternate way to account for erased syndromes when decoding is to simply set the weight of the corresponding vertical edges to 0. This choice of weight corresponds to a physical error probability of $1/2$, reflecting the intuition that erased syndromes contain no useful information, just as would a syndrome measurement with error probability  $p=1/2$. While the simplicity of this alternate approach is appealing, it is not equivalent to our approach. In particular, setting the edge weight to 0 does not account for the fact that the probability of detecting a data error increases with the number of time steps between consecutive non-erased syndromes. 

\begin{figure*}[t!]
    \centering
    \includegraphics[width=1.8\columnwidth]{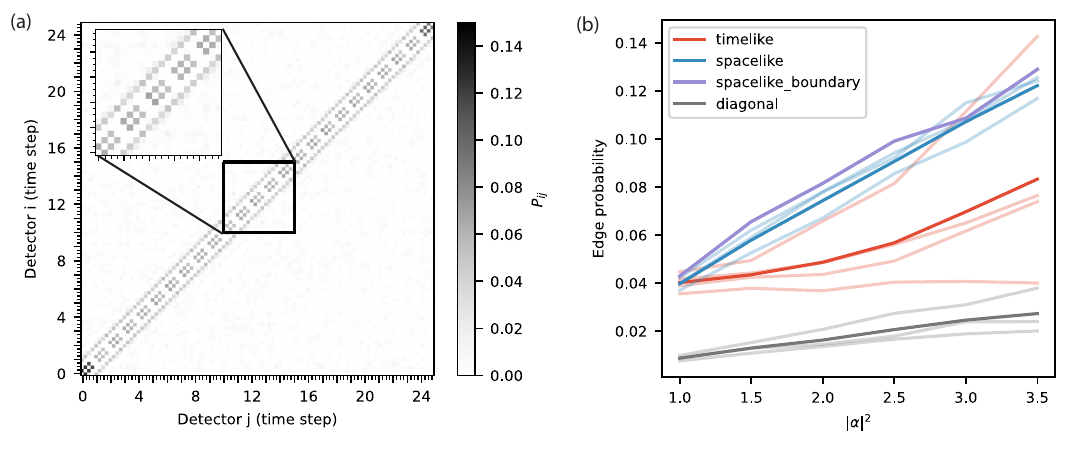}
    \caption{\textbf{Detection correlations and edge weights.} (a) Correlations between detection events for the distance-5 repetition code with $|\alpha|^2=1.5$.  The major ticks correspond to a time step and the minor ticks correspond to the spatial index within the time index. (b) Edge weights as a function of $|\alpha|^2$. The data is averaged across all of the bulk edges in a repetition code experiment with 135 cycles.  Faint lines indicate the traces for individual edge locations while the solid lines indicate the average.  As expected we observe an increase in spacelike and diagonal edge probabilities with $|\alpha|^2$ since these relate to storage mode phase-flips.  The timelike edges for some of the stabilizers increase with photon number (particularly on A4).  This is due to higher order nonlinarities shifting the optimal gate length which are particularly prevalent on interactions with small storage-ancilla detuning.  }
    \label{app_fig:edge_weight_details}
\end{figure*}

To illustrate this effect, \cref{app_fig:erasure_matching_comparison} presents a simple example where this difference alters the minimum-weight matching. In the example, an erasure has occurred, and there is a single adjacent nontrivial detector which must be matched with either the right or left boundary. When the erasure is accounted for by setting the corresponding timelike edge to weight to 0, the example edge weights are such that matching to the left boundary has lower weight. However, when the erasure is accounted for following our procedure, the weight of the edge connecting to the right boundary is reduced, since this single edge encompasses data errors over two time steps. With this modification, matching to the right boundary then has lower weight. While the example in \cref{app_fig:erasure_matching_comparison} is contrived, we do find that these two different erasure decoding methods result in different logical lifetimes when applied to our experimental data, with our approach yielding slightly longer lifetimes. The difference is primarily due to shots with multiple consecutive erasures.

\subsection{Correlations in detection events}
\label{app:detection_event_correlations}

In \cref{app_fig:edge_weight_details}(a) we show the complete correlation matrix between all of the detectors in the matching graph. Detectors are indexed by a single number, $i = x + (d-1)t$, where $x$ and $t$ are the detector's space and time coordinates.  Just off of the diagonal we observe the three correlations within each time index corresponding to correlations between the spatially neighboring detection events.  These correlations are caused by phase-flip errors on the data qubits S2,S3, and S4.  Next, off the diagonal we see sizable correlations between detections of the same spatial index but with time index offset by 1.  These detectors correspond to the timelike edges representing syndrome measurement errors.  We also observe fainter off-diaongal correlations that connect detectors at subsequent time and spatial indices, corresponding to diagonal edges of the error correction graph.  These correlations are caused by phase-flip errors that occur mid-cycle between CX gates.  These major detection correlations we observe are explained with a Pauli error model.  This motivates our use of the Pauli error model to downselect on which edges to include in the error correction graph.

\subsection{Syndrome measurement error rate versus $|\alpha|^2$}
\label{app:measurement_error_vs_n}

In \cref{app_fig:edge_weight_details}(b) we show the edge weights for the spacelike, timelike, and diagonal (spacetimelike) edges of the distance-5 repetition code as a function of $|\alpha|^2$ (computed from shots with 135 error correction cycles).  The numbers reported are the average of each edge type across the bulk time indices. The faint lines indicate the edge probabilities for each spatial index of the edge type.  The solid lines indicate the average edge probability for the edge type.  We observe an increase in the spacelike edge probabilities with $|\alpha|^2$ which is expected since these edges predominantly originate from data qubit phase-flip errors.  

\begin{figure*}[t!]
    \centering
    \includegraphics[width=1.8\columnwidth]{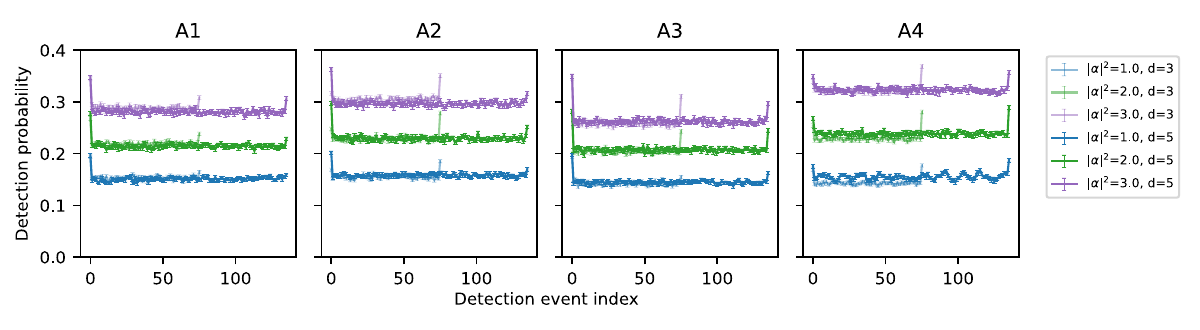}
    \caption{\textbf{Distance-3 and 5 detection probability comparison.}  We compare the detection probabilities for each stabilizer between the distance-3 and distance-5 sections (different number of cycles are used for the distance-3 and 5 experiment).  On the A4 stabilizer we observe that the detection probabilities for the distance-5 code are elevated relative to the distance-3 code and oscillate over time.  We attribute this to crosstalk from S1's flux pump.}
    \label{app_fig:distance_3_and_5_detection_comparison}
\end{figure*}

\begin{figure*}[t!]
    \centering
    \includegraphics[width=1.8\columnwidth]{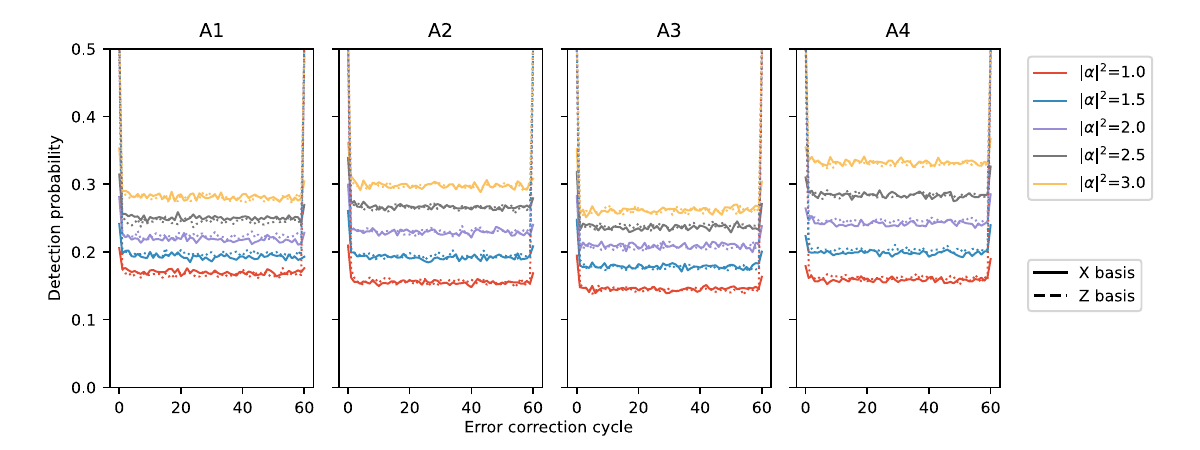}
    \caption{\textbf{Detection probability comparison between $X$ and $Z$ basis storage mode preparation.}  We perform a twice interleaved repetition code experiment with storage mode state preparation into the $X$ and $Z$ basis.  Detection probabilities are consistent between $X$ and $Z$ basis state preparation.  The first and last cycles are different as expected because for $Z$ basis state preparation the first and last comparison are random.}
    \label{app_fig:X_Z_basis_detection_experiment}
\end{figure*}

We also observe that the average timelike edge probability increases with $|\alpha|^2$.  The stabilizer which exhibits the most significant increase with $|\alpha|^2$ corresponds to A4 while the A3 stabilizer shows barely any increase with $|\alpha|^2$.  The increase in timelike edge probability (corresponding to syndrome extraction error) with photon number can be attributed to the higher order nonlinearities between the storage and ancilla of the form $K_{g} a^{\dagger 2} a^2 |g\rangle\langle g|$ and $K_{f} a^{\dagger 2} a^2 |f\rangle\langle f|$.  This Hamiltonian term corresponds to higher order dispersive shifts (or storage Kerr nonlinearities conditioned on the ancilla state).  These effects are strongest on the interactions with the smallest storage-ancilla detuning.  In our device the smallest detuning is on the A4 stabilizer while the largest is on A3.  This explains why the syndrome extraction error of A3 exhibits barely any increase while that of A4 exhibits significant increase with $|\alpha|^2$.  These terms lead to a shift in the optimal gate length as photon number increases.  Since we fix the gate length across $|\alpha|^2$ this leads to a degradation in the measurement error.  This effect could be mitigated by tuning up the gate length separately for each photon number.

\subsection{Comparison of detection probabilities between distance-3 and distance-5 experiments}
\label{app:distance_3_and_5_detection_comparison}

\begin{figure*}[t!]
    \centering
    \includegraphics[width=2\columnwidth]{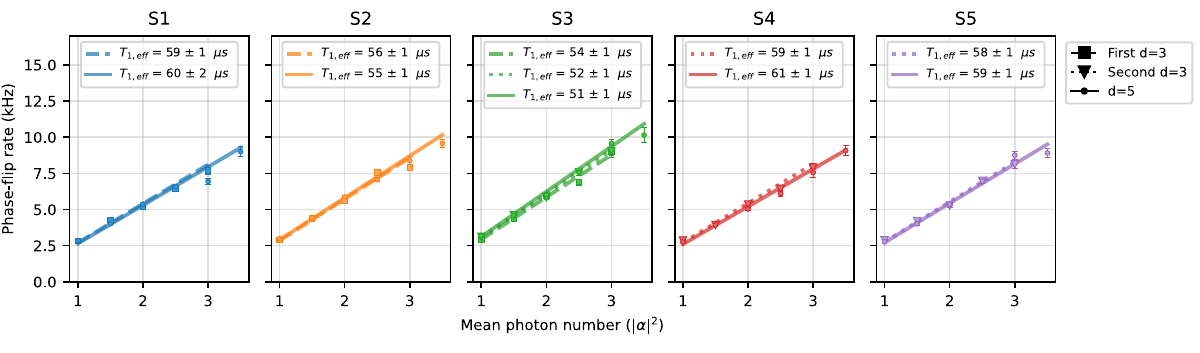}
    \caption{\textbf{Effective storage $T_1$ during the repetition code experiment.} Using the data from the $X$ basis repetition code experiment we extract the phase-flip decay of the individual storage modes during the repetition code experiments.  Specifically we fit $\langle X_i(0)X_i(t)\rangle$ for the individual storage modes and each photon number.  We find that the storage mode phase-flip rates are consistent between distance-3 and distance-5 experiments.  We fit the phase-flip rate as a function of the storage mean photon number to the functional form $\gamma_{\text{phase-flip}} = \kappa_{1,\text{eff}} |\alpha|^2$ to determine $T_{1,\text{eff}}=1/\kappa_{1,\text{eff}}$ during the repetition code experiment.  We observe the expected linear dependence between the storage mode phase-flip rates and $|\alpha|^2$.  }
    \label{app_fig:phase_flip_during_rep_code}
\end{figure*}

In \cref{app_fig:distance_3_and_5_detection_comparison} we compare the detection probabilities for distance-3 and distance-5 repetition code experiments.  Detection probabilities are consistent on the A1, A2, and A3 stabilizers between the distance-3 and distance-5 experiments.  On the A4 stabilizer we observe that the detection probabilities fluctuate and are slightly elevated by 1-2\% for the distance-5 section.  We attribute this increase to imperfect cancellation of the crosstalk from the S1 flux pump to the A4 readout resonator (see \cref{app:crosstalk}).  This minor increase could be mitigated by increasing the padding of the flux pump relative to the ancilla state preparation pulses.  In the dataset of \cref{app:X_and_Z_basis_detection_comparison} the padding was increased and these fluctuations were removed.  

\subsection{Comparison of detection probabilities between X and Z basis initialization}
\label{app:X_and_Z_basis_detection_comparison}

In this section we perform a test where we digitize the syndrome measurements for the $Z$ basis experiment to confirm the detection probabilities are consistent between $X$ and $Z$ basis state preparation.  We twice interleave repetition code experiments with $X$ and $Z$ basis state preparation.  In \cref{app_fig:X_Z_basis_detection_experiment} we show the detection probabilities for each of the stabilizers. The bulk detection probabilities are consistent between the $X$ and $Z$ basis initialization.  The first and final rounds detection probabilities are different between the $X$ and $Z$ basis experiments because the initial state of the stabilizer is non deterministic for the $Z$ basis experiment.  

Note that this data also used a different padding around the stabilization which mitigates the fluctuations observed in \cref{app:distance_3_and_5_detection_comparison}. 

\subsection{Effective storage $T_1$ during repetition code experiments}
\label{app:storage_t1_during_repetition_code}

In addition to studying the logical performance of the repetition code, we can also compare initial and final states to infer the phase-flip decay rates of the individual storage modes during the repetition code experiment.  In \cref{app_fig:phase_flip_during_rep_code} we show the individual storage mode phase-flip rates as a function of $|\alpha|^2$ during the repetition code experiment.  For each storage we fit the phase-flip rates vs. $|\alpha|^2$ to a linear model of $\gamma_{\text{phase-flip}} = \kappa_{1,\text{eff}} |\alpha|^2$ where $\kappa_{1,\text{eff}}=1/T_{1,\text{eff}}$.  The $T_{1,\text{eff}}$ values from different repetition codes segments are consistent to within 5\% of each other and all above $50~\mathrm{\mu s}$.  The $T_{1,\text{eff}}$ extracted from the repetition code experiment can be compared to the $T_1$ under dissipation in \cref{app:mode_frequencies_and_coherences} and the $T_{1,\text{eff}}$ extracted from the $\text{CX}^2$ phase-flip rate experiments.   

\subsection{Comparison of logical Z lifetime for the two different initial states and with ancilla in g}

\begin{figure}[b!]
    \centering
    \includegraphics[width=\columnwidth]{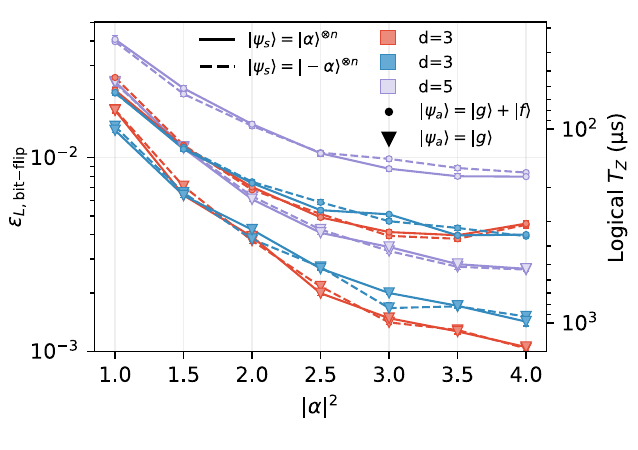}
    \caption{\textbf{Logical bit-flip probabilities for different ancilla and storage initial states.} Logical bit-flip probabilities for the different repetition code sections, ancilla states $|g\rangle$ and $|g\rangle+|f\rangle$, and storage initial states.  Dotted lines correspond to the storage initial state $|{-}\alpha\rangle^{\otimes n}$ and solid lines correspond to the storage initial state $|\alpha\rangle^{\otimes n}$. The colors enumerate the two distance-3 and the distance-5 sections.  Lastly the circles correspond to the ancilla being prepared in $|g\rangle+|f\rangle$ every cycle and the triangles to the ancilla being prepared in $|g\rangle$ every cycle.  }
    \label{app_fig:logical_Z_additional_states}
\end{figure}

For the $Z$ basis repetition code experiment we initialize the repetition code into both the $|\alpha\rangle^{\otimes d}$ and $|{-}\alpha\rangle^{\otimes d}$ states (see \cref{app:Z_basis_pulse_sequence}).  In practice the coherent states which makes up these repetition code states are identical up to a phase but we include both state preparations nonetheless for completeness.  In the main text we report the logical $Z$ lifetime averaged over these two state preparations.  In \cref{app_fig:logical_Z_additional_states}, we report the lifetime broken down into both of the initial states.  

\begin{figure*}[t!]
    \centering
    \includegraphics[width=1.8\columnwidth]{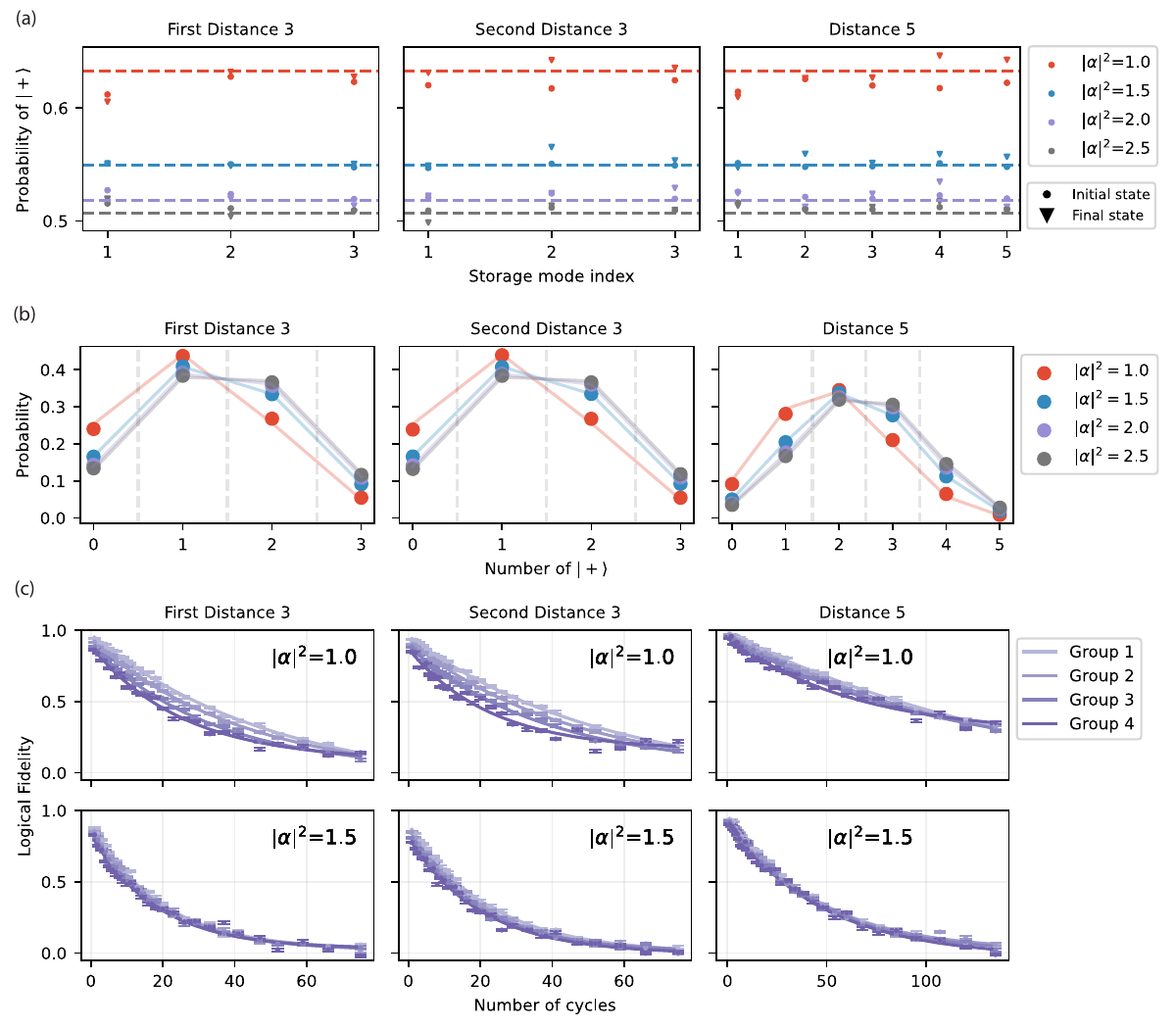}
    \caption{\textbf{Asymmetric phase-flip rates at low photon number.}  (a) Probability of individual storage modes being in $|+\rangle$ at the beginning and end of a repetition code logical $X$ experiment. 
 (b) Distribution of $|+\rangle$ in the initial repetition code state.  We report the number of storage modes measured to have even parity from the initial storage mode measurements of the $X$-basis repetition code experiment.  The markers correspond to the measured probabilities while the lines correspond to the expected probabilities from the binomial distribution and the analytic form for the steady state parity as a function of photon number.  The experimentally measured points are consistent with expectations.  At low photon numbers the bit-strings bias towards $|+\rangle$ consistent with the storage mode $|+\rangle$ state having a lower phase-flip decay rate than $|-\rangle$ at low storage mean photon number.  (c) Exponential fits for the distance-3 and distance-5 sections when grouped based on the number of $|+\rangle$ in the initial state.  At $|\alpha|^2=1$ we observe significant differences in error rates between groupings while for $|\alpha|^2=1.5$ the dispersion between groups is suppressed.}
    \label{app_fig:error_rate_asymmetry}
\end{figure*}

In \cref{app_fig:logical_Z_additional_states}, we also report the logical bit-flip probability for a control experiment where the ancilla is prepared to $|g\rangle$ every cycle.  This experiment is useful since it gives a baseline on the repetition code performance when errors relating to the ancilla being in $|f\rangle$, such as decay from $|f\rangle$, are removed.   With the ancilla in $|g\rangle$ a logical bit-flip error probability of $\sim3\times 10^{-3}$ is measured at $|\alpha|^2=4$ for the distance-5 section.  For the distance-3 sections logical bit-flip error probabilities under $2\times 10^{-3}$ are measured at $|\alpha|^2=4$.

\subsection{Distribution of initial states in logical X lifetime experiments}
\label{app:distribution_of_initial_states_in_logical_X_lifetime_experiments}

In the small $|\alpha|^{2}$ regime, where $e^{-2|\alpha|^{2}}$ is not negligible, typical physical error mechanisms like photon loss result in \emph{asymmetric} cat phase flips rates. That is, in the Lindblad dissipator $\gamma_{+\rightarrow -}\mathcal{D}[|-\rangle\langle +|] + \gamma_{-\rightarrow +}\mathcal{D}[|+\rangle\langle -|]$, the two phase-flip rates $\gamma_{+\rightarrow -}$ and $\gamma_{-\rightarrow +}$ are not identical to each other. For example under the single-photon loss channel $\kappa_{1}\mathcal{D}[\hat{a}]$, we have 
\begin{align}
    \gamma_{+\rightarrow -} &= \kappa_{1}|\alpha|^{2} 
    \tanh{|\alpha|^2}
    \nonumber\\
    \gamma_{-\rightarrow +} &= \kappa_{1}|\alpha|^{2} 
    \coth{|\alpha|^2}
\end{align}
Physically, this asymmetry can be understood as a consequence of the fact that the even- and odd-parity cat states have significantly different mean photon number at small $|\alpha|^2$. Indeed, these states defined such that they approach different Fock states, $|+\rangle \to |n=0\rangle$ and $|-\rangle \to |n=1\rangle$, in the limit $|\alpha|^2\to 0$. 
This asymmetry in the phase-flip rates leads to an imbalance in the populations of the even and odd cat states $|\pm\rangle$ in the long-time limit. In particular, the steady-state populations of the even and odd cat states are given by 
\begin{align}
    P_{t\rightarrow \infty}(|+\rangle) &= \frac{\gamma_{-\rightarrow +} }{\gamma_{+\rightarrow -} + \gamma_{-\rightarrow +}} = \frac{(1 + e^{-2|\alpha|^{2}})^{2}}{2 (1 + e^{-4|\alpha|^{2}})}, 
    \nonumber\\
    P_{t\rightarrow \infty}(|-\rangle) &= \frac{\gamma_{+\rightarrow -} }{\gamma_{+\rightarrow -} + \gamma_{-\rightarrow +}} = \frac{(1 - e^{-2|\alpha|^{2}})^{2}}{2 (1 + e^{-4|\alpha|^{2}})}. 
\end{align}
We show a comparison between these analytical predictions (horizontal lines) and experimentally measured steady parities for each of the cat qubit and code distance in \cref{app_fig:error_rate_asymmetry}(a).

These distributions in cat state population then affect the distribution of repetition-code initial states.  Rather than the initial states having a uniform distribution over the $2^d$ possible product cat states, at low $|\alpha|^2$ the distribution is biased towards states with even cats.  We show this bias in \cref{app_fig:error_rate_asymmetry}(b) where we plot for each distance of code the probability of the initial logical state containing a given number of individual $|+\rangle$ states.  At $|\alpha|^2=1$ the states are heavily biased towards those with higher numbers of $|+\rangle$ but already at $|\alpha|^2=1.5$ the distribution is more symmetric.

The asymmetry in error rates also means that the logical error rate per cycle depends on the state of the repetition code.  To show this effect we group the data by the number of $|+\rangle$ states in the logical state.  For the distance-3 sections, we make 4 groups based on 0, 1, 2, or 3 $|+\rangle$ states in the initial logical state.  For the distance-5 section, we make 4 groups based on 0/1, 2, 3, 4/5 $|+\rangle$ states in the initial logical state.  In \cref{app_fig:error_rate_asymmetry}(c) we show the decay of $\langle X_L(0) X_L(t)\rangle$  for each grouping.  At $|\alpha|^2=1$ the decay curves for the different groupings vary significantly.  Thanks to the exponential dependence of the steady state on the photon number, though, at $|\alpha|^2=1.5$ the curves are almost identical.  The decay times from these groupings are shown as faded points in \cref{fig:logical_X}.

When we group the data by initial state as above, $\langle X_L(0) X_L(t)\rangle$ will no longer necessarily exhibit a pure exponential decay.  To see this, first consider the case where the initial state of the repetition code is $|+\rangle^{\otimes d}$.  This initial state has a lower error rate than the steady state of the repetition code due to the error rate asymmetry of the underlying cat qubits.  This implies that at short times the logical error probability per cycle will be lower than at later times ($t\gg \kappa_1 |\alpha|^2$) when the cat qubits have reached their steady state distribution.  Importantly, this temporal dependence of the error rates means that the functional form of the decay for just one initial state at low $|\alpha|^2$ is not a pure exponential.  Nonetheless, we have found that exponential fits agree well enough with the data to allow us to indicate the spread due to the varying initial states.  We compute the overall decay rate by fitting an exponential decay to the data sampled from the steady state distribution, which ensures a pure exponential decay.  Because lower-error states like $|+\rangle^{\otimes d}$ are over represented in the steady state, bit-flip times at low $|\alpha|^2$ (especially at $|\alpha|^2=1$) are biased towards higher values. This effect benefits our distance-3 code, which has minimal logical error at $|\alpha|^2 \approx 1$, more than it does out distance-5 code, which has minimal logical error at $|\alpha|^2 \approx 1.5$.  By $|\alpha|^2=1.5$ the dispersion from this effect collapses and the data from different groupings agree.

We note that this asymmetry in the error rates taken to its most extreme limit is the case of a bit-flip repetition code with a $|n=0\rangle/|n=1\rangle$ Fock state encoding for the data qubits.  Subject to $T_1$ decay, the steady state is when all the data qubits are in the ground state and do not experience $T_1$ errors.  This is why the experiments~\cite{Kelly2015, Chen2021} apply $\pi$ pulses every round to make the steady state equally mixed between $|n=0\rangle$ and $|n=1\rangle$.

\subsection{Explanation of distance-3 and distance-5 $T_X$ scaling with $|\alpha|^2$}
\label{app:explanation_logical_phase_scaling}
As shown in \cref{fig:logical_X} of the main text, our repetition cat's logical phase-flip error per cycle scales as $(|\alpha|^2)^\gamma$, where we observe $\gamma = 1.63(4)$ and $\gamma = 1.86(3)$ for the two distance $d=3$ codes, and $\gamma = 2.31(2)$ for the $d=5$ code. These values of $\gamma$ are lower than the ideal values of $2$ and $3$, respectively, that one expects based on the scaling $p_L \propto (p/p_\text{th})^{(d+1)/2}$, where $p_L$ is the logical error and $p_\text{th}$ is the error correction threshold. However, corresponding simulations similarly predict the lower-than-ideal scaling (see the dashed line in \cref{fig:logical_X}(f), and a more detailed discussion in \cref{app:logical_error_budget}).

\begin{figure}[t!]
    \centering
    \includegraphics[width=\columnwidth]{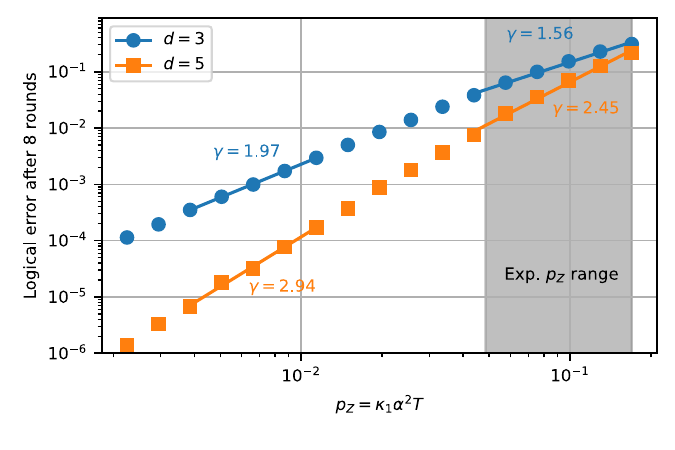}
    \caption{\textbf{Logical phase-flip error scaling in different physical error regimes.} We perform error correction simulations (see \cref{app:logical_error_budget}) of distance $d=3,5$ repetition codes over a range of data qubit phase-flip error probabilities $p_Z$, and with the syndrome measurement error rate artificially set to 0. Solid lines indicate fits of the form $\propto (|\alpha|^2)^\gamma$, with the values of the scaling exponent $\gamma$ indicated next to the corresponding fit line. Fit lines are drawn only over $p_Z$ values used in the respective fits. The gray rectangle indicates values of $p_Z$ that are accessible in our experiment by varying $|\alpha|^2$. }
    \label{app_fig:error_scaling_simulations}
\end{figure}

We perform additional error correction simulations to investigate why we do not observe the ideal scaling, and results are shown in \cref{app_fig:error_scaling_simulations}. In these simulations, we artificially set the syndrome measurement error rate to 0, and we vary the data qubit phase-flip probability $p_Z = \kappa_1 |\alpha|^2 T$ over a larger range of values than we are able to access experimentally. We plot the logical phase-flip probability after a fixed number of error correction rounds as a function of $p_Z$ for $d=3,5$ codes. We find that, even in the idealized case where our experiment had no syndrome measurement error, the data qubit error probabilities are nevertheless too high to observe the ideal scaling.  Indeed, the range of $p_Z$ that we probe experimentally by varying $|\alpha|^2$ is approximately $p_Z\in [5\%, 17\%]$, so good agreement with the leading-order $p_L \propto (p/p_\text{th})^{(d+1)/2}$ scaling should not be expected. As $p_Z$ is decreased, however, we do see the ideal scaling emerge. Note that because we have artificially set the syndrome measurement error rate to 0, we do not expect agreement between the simulated $\gamma$ values in \cref{app_fig:error_scaling_simulations} and the measured values in \cref{fig:logical_X} of the main text.

Non-zero syndrome measurement probability can also have an impact on the scaling exponent $\gamma$. As discussed in \cref{app:measurement_error_vs_n} and shown in \cref{app_fig:edge_weight_details}, we observe that some syndrome measurement error probabilities in our experiment increase with $|\alpha|^2$. Because the logical error increases monotonically with syndrome measurement error probability, this nontrivial $|\alpha|^2$ dependence serves to increase the measured value of $\gamma$ relative to the case of constant syndrome measurement error. Such an increase is not necessarily undesirable; indeed, in quantum error correction simulations one typically considers noise models where the syndrome measurement error is scaled in proportion to data qubit error. To properly mimic such noise models in experiment, it would generally be necessary to deliberately increase the syndrome measurement error in an $|\alpha|^2$-dependent way, so that $p_Z$ and syndrome error both scale proportionately with $|\alpha|^2$.

\section{Simulated logical performance and error budgets}
\label{app:error_budget}
In this section, we describe the theoretical models used to make performance projections for the logical phase- and bit-flip errors per cycle. These projections are shown in \cref{fig:logical_X} and \cref{fig:logical_Z} of the main text, respectively. Then, we describe how these models are used to produce a budget for the overall logical error per cycle, shown in \cref{fig:total_lifetime} of the main text.

\subsection{Simulated logical phase-flip performance}
To estimate the logical phase-flip performance of the repetition code, we perform Monte Carlo simulations of noisy syndrome measurement using a simplified circuit-level noise model. In contrast to usual circuit-level noise models that take as input full Pauli error channels for each operation in the syndrome measurement circuit, our model takes only a limited set of inputs: the probabilities of cat phase flips, the probabilities of syndrome measurement errors, and, optionally, the probability of ancilla erasure. It suffices to work with this simplified model because we are only interested in the logical phase-flip error of a repetition code. In this context, we can neglect bit-flip errors on the data and ancilla qubits, since they do not contribute to the logical phase-flip error. Moreover, ancilla phase-flip errors do not propagate through the CX gates to the data qubits, so all ancilla errors can be modelled as an effective increase in the syndrome measurement error or ancilla erasure. 

We determine the input parameters for this noise model as follows. To compute the cat phase-flip probabilities, we use effective storage-mode lifetimes $T_{1,\text{eff}}$ measured during the repetition code experiment as described in \cref{app:storage_t1_during_repetition_code}. We infer the effective syndrome measurement error probabilities from the average time-like edge weights in the matching graph. That is, we compute the associated probability for each edge by inverting the relation $w_e = \log((1-p_e)/p_e)$, where $w_e$ is the weight of edge $e$ and $p_e$ the corresponding probability, then we average these probabilities over all time steps. Erasure probabilities are determined by computing the fraction of erased syndrome measurement outcomes.

Noisy syndrome measurements generated from Monte Carlo sampling are decoded in exactly the same manner as the experiment, and the simulated logical phase-flip times $T_X$ are plotted alongside the measured data in \cref{fig:logical_X}. We observe reasonable agreement between the two, with both simulated and measured data exhibiting scaling exponents $\gamma$ less than the ideal value of $\gamma=3$ for a $d=5$ repetition code. As discussed in \cref{app:explanation_logical_phase_scaling}, we do not expect to observe this ideal value given the error rates in our system. 

\begin{figure}[t!]
    \centering
    \includegraphics[width=1.0\linewidth]{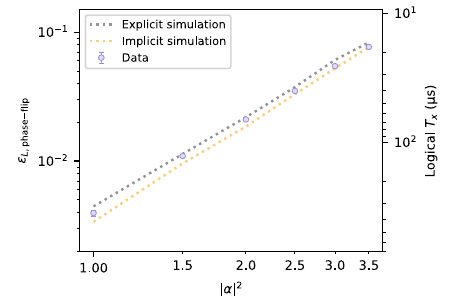}
    \caption{\textbf{Distance-5 erasure simulation method comparison.} Comparison of simulated logical phase-flip error using the explicit and implicit approaches to account for erasure detection described in the text. Error bars are smaller than the plot markers.}
    \label{app_fig:explicit_implicit_erasure_sim}
\end{figure}

Additionally, we note that rather than \emph{explicitly} modelling erasures and accounting for them when decoding, the impact of erasure detection can instead be approximately accounted for \emph{implicitly}, by only incorporating the effective reduction in syndrome measurement error probability (see \cref{fig:logical_X}(d)). That is, in the explicit approach syndrome measurements are incorrect with probability $p_\text{meas}$, erased with probability $p_\text{e}$, and correct with probability $1-p_\text{meas}-p_\text{e}$. In the implicit approach, syndrome measurement are incorrect with the same probability $p_\text{meas}$, correct with probability $1-p_\text{meas}$, and are never erased.
In the simulation results shown in \cref{fig:logical_X}(f), we explicitly account for erasures to obtain the best agreement with the experimental data, but in all other simulations we opt to implicitly account for erasure decoding for the sake of numerical efficiency. In \cref{app_fig:explicit_implicit_erasure_sim}, we confirm that the logical phase-flip errors for these two approaches do not significantly differ, indicating that the main impact of erasure detection is to effectively lower the syndrome error rate. The implicit approach performs marginally better due to the fact that, in the explicit case the occasional erased syndrome provides no useful information, whereas in the implicit case all syndromes provide nontrivial information.

\subsection{Simulated logical bit-flip performance}
\label{app:logical_bit_flip_projections}
To estimate the logical bit-flip error per cycle of the repetition code, we employ a simple phenomenological error model. This model contains contributions from two sources: $|\alpha|^2$-dependent idling bit-flip probabilities (per error correction cycle) for each cat, and additional $|\alpha|^2$-independent bit-flip probabilities associated with each CX. The values of these phenomenological probabilities are determined from independent experiments, as described below.

The phenomenological cat idling bit-flip probabilities are intended to capture cat bit-flip errors caused by mechanisms intrinsic to the stabilized cat, such as photon loss and dephasing. We assume these probabilities decrease exponentially with $|\alpha|^2$, so that the bit-flip probability of the $i$th cat is given by $p_i^\text{(idle)}\left(|\alpha|^2\right) = A e^{-B |\alpha|^2}$. The constants $A$ and $B$ are fit from measurements of the individual cats' bit-flip times, shown in \cref{fig:architecture}(d) of the main text. 

The phenomenological CX bit-flip probabilities are intended to capture additional cat bit-flip errors caused by the application of a CX gate, including bit-flip errors that may propagate from the ancilla transmon. We determine the values of these CX bit-flip probabilities using the CX$^2$ experiments described in \cref{app:cx_calibration}. We assume that the measured CX$^2$ bit-flip probability  is given by a sum of the cat idling bit-flip probability plus an additional additive contribution associated with the two CX gates,
\begin{equation}
    p_j^\text{(CX$_{|k\rangle}^2$)}\left(|\alpha|^2\right) = p_i^\text{(idle)}\left(|\alpha|^2\right) + 2 p_j^\text{(CX$_{|k\rangle}$)}.
\end{equation}
Here, $p_j^\text{(CX$_{|k\rangle}^2$)}\left(|\alpha|^2\right)$ is the measured CX$^2$ bit-flip probability of the $j$th CX gate (acting on cat $i$) for initial ancilla state $|k\rangle \in \{ |g\rangle, |f\rangle\}$, and the above equation defines the phenomenological CX bit-flip error probability $p_j^\text{(CX$_{|k\rangle}$)}$. We fit the measured values of $p_j^\text{(CX$_{|k\rangle}^2$)}\left(|\alpha|^2\right)$ using this expression to determine the value of the value of $p_j^\text{(CX$_{|k\rangle}$)}$. To improve accuracy, the fit is performed only over relatively large photon numbers, $|\alpha|^2 \geq 3$, where cat-idling errors are relatively small. 

We compute the logical bit-flip error per cycle from the phenomenological parameters by summing the idling bit-flip probabilities for each cat together with the probabilities for each CX, since a single bit-flip error at any cat or CX is equivalent to a logical bit-flip error. The logical error per cycle is thus given by
\begin{equation}
    \sum_i p_i^\text{(idle)}\left(|\alpha|^2\right) + \sum_j \frac{1}{2} \left[p_j^\text{(CX$_{|g\rangle}$)} + p_j^\text{(CX$_{|f\rangle}$)} \right],
\end{equation}
where we have averaged over the CX bit-flip probabilities for transmon states $|g\rangle$ and $|f\rangle$, since the transmon is prepared in an equal superposition of these two states. This expression is plotted alongside the measured logical bit-flip probabilities in \cref{fig:logical_Z} of the main text. 

\subsection{Logical error budget}
\label{app:logical_error_budget}

We use the simulation procedures described in the two preceding subsections to produce an error budget for the overall logical error per cycle of our $d=5$ repetition code. To do so, we follow the approach of Ref.~\cite{Chen2021}. We let $x$ denote a vector of physical error probabilities, and we regard the overall logical error per cycle as a function of these physical error probabilities, $\epsilon_{L}(x)$. The contribution of the $i$th physical error mechanism to $\epsilon_{L}$'s error budget is then defined as
\begin{equation}
    a_i \frac{\partial \epsilon_{L}(x)}{\partial x_i}\bigg|_{x=a/2},
\end{equation}
where $a_i$ denotes the nominal probability observed in the experiment of the $i$th error mechanism. 
While there are different ways one can conceive of defining an error budget, this definition has the appealing property that, if $\epsilon_{L}$ depends at most quadratically on the physical error probabilities $x$ and $\epsilon_{L}(0) = 0$, then the sum of the individual error budget contributions equals the total error,
\begin{equation}
\label{eq:error_budget_sum}
    \sum_i a_i \frac{\partial \epsilon_{L}(x)}{\partial x_i}\bigg|_{x=a/2} = \epsilon_{L}(a).
\end{equation}

We construct an error budget for $\epsilon_{L}$ of our $d=5$ repetition code using this approach, where we consider four different types of physical error mechanisms:  idling cat phase-flip probabilities, idling cat bit-flip probabilities, CX bit-flip probabilities, and syndrome measurement error probabilities. It is perhaps surprising that above definition of an error budget works well---meaning that the individual contributions sum to close to the overall error as in \cref{eq:error_budget_sum}---when applied to this situation. In particular, the cat phase-flip probability would be expected to contribute beyond quadratically to $\epsilon_{L}$, since ideally the $d=5$ code would achieve a cubic suppression of these errors. \cref{eq:error_budget_sum} is thus not generally expected to hold in this situation. However, in practice we find that \cref{eq:error_budget_sum} holds to a good approximation, which we attribute to the fact that our experiment does not actually exhibit the ideal cubic scaling due to its relatively high physical error rates (see \cref{app:explanation_logical_phase_scaling}). Separately, we note that because the bit-flip mechanisms contribute linearly to $\epsilon_{L}$, it is not necessary to actually compute derivatives with respect to these errors. That is, for these linear contributions we have simply that $a_i \frac{\partial \epsilon_{L}(x)}{\partial x_i}\big|_{x=a/2} = a_i$.

The resulting error budget for $\epsilon_{L}$ is shown in \cref{fig:total_lifetime}(b) of the main text. For the cat phase- and bit-flip probabilities, as well as the CX bit-flip probabilities, we opt to compute individual contributions from each component (i.e.~individual contributions for each cat and each CX), in order to give a sense of the relative contributions of these individual components.
For syndrome measurement, we opt to compute only an aggregate contribution, since this type of mechanism contributes least to the overall budget.

\subsection{Logical performance estimates assuming idealized coherence-limited, bias-preserving gates}
\label{app:optimistic_overhead_estimates}
In the conclusion of the main text, we present rough estimates for the logical memory overhead under the idealized assumption of coherence-limited, bias-preserving cat-cat gates. In this section we provide justification for these estimates. 

The overall logical error per cycle can be roughly estimated as follows. Starting with the logical phase-flip error per cycle, under the assumption coherence-limited gates the cat phase-flip probability per error correction cycle is simply given by $p_Z = |\alpha|^2 T_\text{cycle}/T_1$. We assume the logical phase-flip error per cycle is then $\epsilon_{L,\text{phase-flip}} = A(p_Z/p_\text{th})^{(d+1)/2}$, where the threshold error per cycle is $p_\text{th} \approx 0.1$ ~\cite{Dennis2002}, and the constant $A\approx 0.1$~\cite{Chen2021,Acharya2024}.  Separately, we estimate the repetition code logical bit-flip error per cycle as $\epsilon_{L,\text{bit-flip}} = d T_\text{cycle}/2T_\text{Z}$. 

We discuss two examples in the main text, both of which assume $d=11$ and $T_\text{cycle} = 1~\mu s$. In the first example we assume $T_1 = 300~\mu s$, and $T_Z = 1~s$ at $|\alpha|^2 = 5$. This yields $\epsilon_{L,\text{phase-flip}} = 2.1 \times 10^{-6}$, $\epsilon_{L,\text{bit-flip}} = 5.5\times 10^{-6}$ and hence $\epsilon_L = 3.8\times 10^{-6}$. In the second example we assume $T_1 = 1~ms$, and $T_Z = 100~s$ at $|\alpha|^2 = 7$, where these choices for $T_Z$ and $|\alpha|^2$ are determined by extrapolating the exponential scaling measured in~\cite{singlecat2024}. This yields $\epsilon_{L,\text{phase-flip}} = 1.1 \times 10^{-8}$, $\epsilon_{L,\text{bit-flip}} = 5.5\times 10^{-8}$ and hence $\epsilon_L = 3.3\times 10^{-8}$.

\section{Simulated CX performance and error budgets}

While the phenomenological model for logical bit-flip error per cycle in \cref{app:logical_bit_flip_projections} enables us to quantify how much additional cat bit-flip error is associated with each CX, it does not provide any insight into the physical mechanisms responsible for this error. In this section, we thus simulate individual CX gates in order to reveal which physical mechanisms contribute most to their bit-flip errors. We begin by first describing the simulations and how simulation parameters are determined. We pay particular attention to the ancilla $T_1$, as we find an effective enhancement of ancilla decay during gate operation that degrades CX performance. Then, we use the simulations to quantify the tolerance of CX gates to $\chi$ mismatches and to construct error budgets for individual gates.

\subsection{CX simulations and parameters}
\label{app:cx_sims_and_params}

We perform master equation simulations of the CX$^2$ experiments described in \cref{app:cx_calibration}. These simulations include the storage mode, buffer mode, tunable coupler, and ancilla transmon. Whenever possible, we directly use representative measured values for Hamiltonian parameters. For example, we use measured values for the storage-ancilla dispersive couplings $\chi_{ge}$ and $\chi_{gf}$ and $g_2$, $\kappa_b$ are consistent with the storage decay to the $|\hat{n}=0\rangle/|\hat{n}=1\rangle$ manifold under two-photon dissipation. In some cases, we do not directly measure relevant parameters, and so we rely instead on numerical predictions (see, e.g., \cref{app:buffer_model,app:CX_circuit_quantization_model}). For example, we find that it is important to incorporate the joint ancilla- and coupler-state dependence of the storage dispersive shift and self Kerr. We estimate the values of these parameters by numerically diagonalizing the Hamiltonian of a lumped-element circuit model incorporating the storage, ancilla, and coupler. The underlying circuit parameters of the model are chosen to reproduce a set of experimentally measured parameters such as the mode frequencies, $\chi_{ge}$, and $\chi_{gf}$.

\begin{figure*}[t!]
    \centering
    \includegraphics[width=\textwidth]{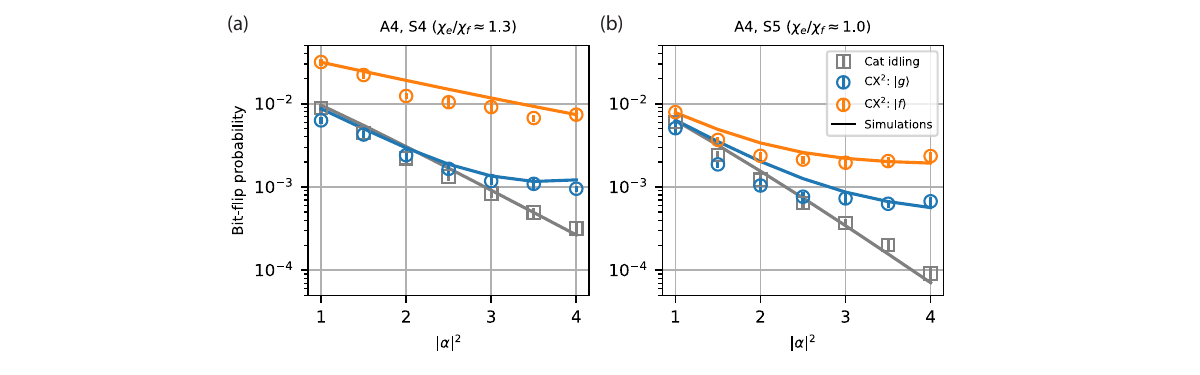}
    \caption{\textbf{Comparison of simulated and measured CX$^2$ bit-flip probabilities.}
    Measured bit-flip values (markers) are plotted alongside simulated values (lines) for the A4, S4 (a) and A4, S5 (b) gates. The former gate is poorly $\chi$ matched ($\chi{ge}/\chi_{gf}\approx 1.3$), while the latter is well matched ($\chi{ge}/\chi_{gf}\approx 1.0$).  For reference, bit-flip probabilities of stabilized cat idling are also shown in gray. 
    }
    \label{app_fig:cx_simuations_vs_n}
\end{figure*}

Other simulation parameters are determined by fitting against experimental cat bit-flip data. In particular, as described below, we sequentially determine the storage dephasing rate, coupler heating rate, and ancilla decay rates this way. While one could instead use directly measured values of these quantities as inputs to the simulations, we find that tuning is generally required to obtain good agreement with measured bit-flip data. Moreover, we find that comparing the independently-measured and tuned values provides useful physical insight, especially in the case of ancilla decay (see \cref{app:cx_ancilla_T1}).

To determine the storage dephasing rate, we compare bit-flip times of a stabilized cat against measured values, and tune the simulated storage dephasing rate such that we see good agreement between the two across a range of $|\alpha|^2$. While we could instead use the directly measured value of the storage mode's dephasing rate, the noise spectrum of the experimental dephasing noise may not match the white noise used in our simulations. The cat bit-flip probabilities can be sensitive to this spectrum, so using the measured dephasing time to determine the simulated white-noise dephasing rate would not necessarily produce good agreement~\cite{singlecat2024}.

Next, we tune the coupler heating rate by comparing simulated and measured values of $p^\text{(CX$_{|g\rangle}^2$)}$, the cat bit-flip probability during CX$^2$ experiments for initial ancilla state $|g\rangle$. We assume that at sufficiently large $|\alpha|^2$ these probabilities are dominated by heating of the coupler. We thus tune the simulated coupler heating rate so that the simulated bit-flip probability due to this heating agrees well with the measured value at a specified large value of $|\alpha|^2$. We choose $|\alpha|^2 = 4$, since this is the value we will consider when constructing our CX error budgets in order to highlight the contributions of $|\alpha|^2$-independent error mechanisms. Note that ancilla heating also generally contributes to $p^\text{(CX$_{|g\rangle}^2$)}$. However, since we can not easily separate out the impacts of ancilla and coupler heating on cat bit flips, we opt to model all heating in simulation solely as coupler heating. 

Finally, we tune the ancilla transmon decay rates by comparing simulated and measured values of $p^\text{(CX$_{|f\rangle}^2$)}$, analogous to how the coupler heating rate was tuned. We discuss the tuned values in detail in the next section.

Additionally, the simulations assume that with 1\% probability the initial ancilla state $|f\rangle$ is mistakenly prepared as $|e\rangle$. As discussed below, this state preparation error can be a significant contributor to the budget in cases where $\chi$ matching is poor, and so in such cases it is important include in simulations where the ancilla is initialized in $|f\rangle$. The choice of 1\% for the state preparation error is an approximate choice based on the amount of state preparation error of $|f\rangle$ as $|g\rangle$ we observe. State preparation error of $|f\rangle$ as $|g\rangle$ is quantified by characterizing the bit-flip rate when applying cycles with only one $\text{CX}$ gate and initial ancilla state $|f\rangle$.  With only one CX gate the bit-flip error probability is directly sensitive to state preparation error.  Across the interactions in the circuit we find an average state preparation error of $|f\rangle$ as $|g\rangle$ of $\sim 0.7\%$.   Based on this we roughly assume a 1\% state preparation error into $|e\rangle$ given the lower coherence in the $|e\rangle$, $|f\rangle$ manifold.

With the simulation parameters determined, we compare simulated bit-flip probabilities against measured values in \cref{app_fig:cx_simuations_vs_n}. There is general qualitative agreement, and for some values of $|\alpha|^2$ the results agree quantiatively. Recall that the coupler heating and ancilla decay rates were tuned to ensure reasonable agreement between simulation and data only at $|\alpha|^2 = 4$. The fact that there is reasonable agreement at other values of $|\alpha|^2$ as well is thus a nontrivial indication that the tuned simulation parameters are reasonable.

\subsection{Ancilla $T_1$}
\label{app:cx_ancilla_T1}

\begin{figure}
    \centering
    \includegraphics[width=1\linewidth]{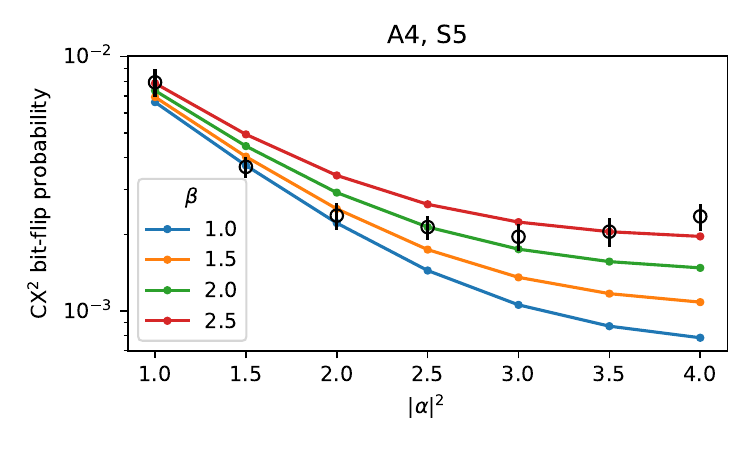}
    \caption{\textbf{Comparison of measured CX$^2$ bit-flip probability against simulations with reduced ancilla $T_1$.} Black markers indicated measured CX$^2$ bit-flip probability for the A4,S5 gate with initial ancilla state $|f\rangle$, and the different curves indicate corresponding simulations where independently-measured ancilla $T_1$ values (separately measured at both the coupler on and off positions) are scaled down by a factor $\beta$. At large $|\alpha|^2$, where the impacts of ancilla decay are most significant, reduced ancilla $T_1$ values ($\beta>1$) are required to explain the measured data.}
    \label{app_fig:CX2_vary_ancilla_loss}
\end{figure}

As previously mentioned, the ancilla $T_1$ values used in simulation are tuned by comparing against measured bit-flip data.  Specifically, this tuning is performed by taking the independently measured $T_1$ values at the coupler-off (idling) and coupler-on positions for the $|e\rangle$ and $|f\rangle$ states, then dividing these values by some factor $\beta$. As shown in \cref{app_fig:CX2_vary_ancilla_loss}, we generally find that values $\beta > 1$ are required to obtain good agreement. This finding suggests that the independently measured $T_1$ values are not representative of the effective ancilla decay rates experienced \emph{during gate operations}. 

\begin{figure*}[t!]
    \centering
    \includegraphics[width=\textwidth]{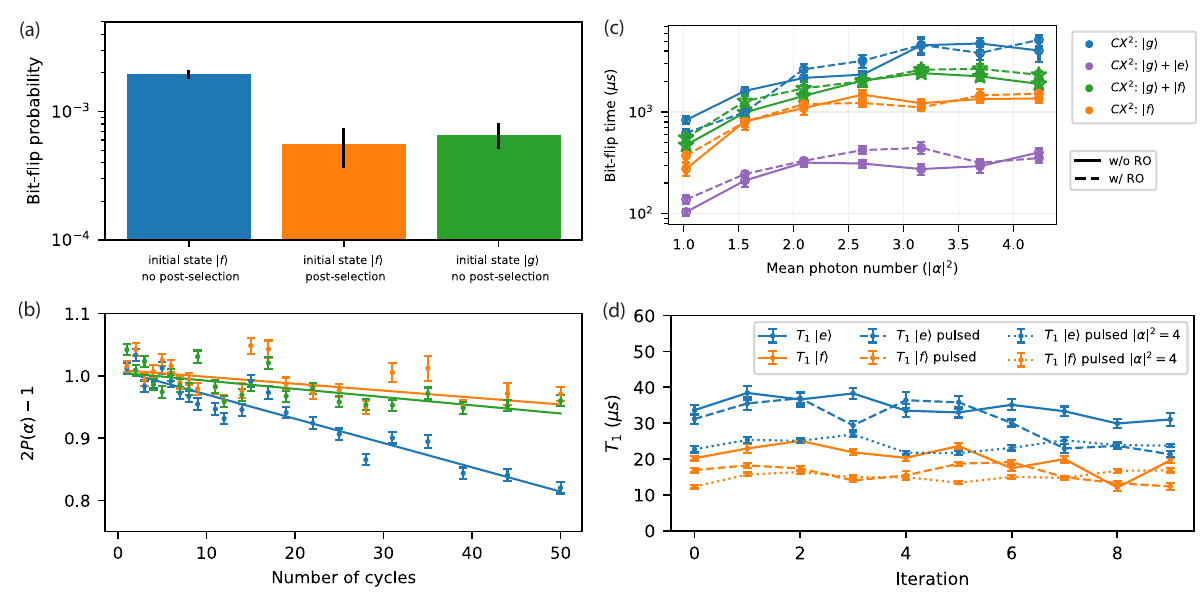}
    \caption{\textbf{CX Error Rate Details.} All these measurements are performed on the $A4 \leftrightarrow S5$ interaction (a) $\text{CX}^2$ error rates at $|\alpha|^2=3.5$ in a few different situations.  In all these experiments readout is applied every cycle.  In blue we show the bit-flip probability per cycle when the ancilla initialized to state $|f\rangle$.  In orange we show the same experiment except that we reject all shots of the experiment in which the ancilla decayed to $|g\rangle$.  The error rate in orange is comparable to the green which is an experiment where the ancilla is initialized to $|g\rangle$.  (b) The short-time linear fits corresponding to the data in (a).  The markers and curves in (b) use the same color scheme as in the bars in (a).  As a point of comparison a long-time exponential fit when the ancilla is prepared to $|g\rangle$ yields a bit-flip probability of $(2.5\pm 0.2) \times 10^{-3}$ (c) $\text{CX}^2$ bit-flip times as a function of the cat-qubit mean photon number $|\alpha|^2$ with and without readout.  The ancilla state is not unprepared so that the ancilla is in a superposition of $|g\rangle$ and $|f\rangle$ during readout. (d) Repeated measurements ancilla $T_1$ of $|e\rangle$ and $|f\rangle$ under three different conditions.  In the solid curves we vary the length of one flux pulse that takes the coupler ``on'' position to measure the ancilla $T_1$ with the coupler on.  In the dashed curve we measure the effective ancilla $T_1$ when repeatedly pulsing the coupler to the ``on`` position. Specifically, during cycles of length $1.6~\mathrm{\mu s}$, a coupler flux pulse with a length of the $\text{CX}^2$ gate is applied. In the dotted curve the coupler is also repeatedly pulsed to the on position but the storage is additionally prepared into a state with mean photon number $|\alpha|^2=4$. The measurements are repeated to show temporal variations.  }
    \label{app_fig:cx_deep_dive}
\end{figure*}

In \cref{app_fig:cx_deep_dive}, we present further experimental investigation of the CX error mechanisms. First, in \cref{app_fig:cx_deep_dive}(a), we verify that the large CX$^2$ bit-flip probability for initial ancilla state $|f\rangle$ in comparison to $|g\rangle$ is due to effective decay from $|f\rangle$ to $|g\rangle$ over the course of the gate. In particular, we perform a characterization of the $\text{CX}^2$ error rate with initial states $|g\rangle$ and $|f\rangle$.  In these experiments we do not apply the pulses to undo the preparation of the ancilla state, so ideally the ancilla should remain in its initial state after the gate and before the readout.  This means the measurements each round allow us to determine the ancilla state after the gate.  This gives us the capability to post-select against decay to $|g\rangle$ for the case when we start with the initial state $|f\rangle$.  Without post-selection we find that the CX$^2$ bit-flip probability is significantly higher for initial ancilla state $|f\rangle$ compared to $|g\rangle$.  When we postselect and discard shots where initial ancilla state $|f\rangle$ decayed to $|g\rangle$ in any round, the bit-flip probabilities are comparable between initial state $|g\rangle$ and $|f\rangle$. These results indicate that the sizable difference between the bit-flip probabilities for initial state $|g\rangle$ and $|f\rangle$ is attributable to effective ancilla decay and not some other mechanism associated with the ancilla being in $|f\rangle$. 

Next, in \cref{app_fig:cx_deep_dive}(c), we investigate whether application of the readout drive could be responsible for enhanced ancilla decay~\cite{Sank2016}. We find that the readout drive is not the cause: comparing measured CX$^2$ bit-flip probabilities with and without the readout drive applied, we find that the probabilities do not change significantly regardless of whether the readout drive is applied.  Note that unlike \cref{app:cx_calibration}, in this experiment, the ancilla unpreparation is not applied each round, so that the readout is applied with the ancilla in both $|g\rangle$ and $|f\rangle$.  

Finally, in \cref{app_fig:cx_deep_dive}(d), we investigate whether the flux pulses for the CX gates may lead to an enhancement of the ancilla decay.  We compare the results from three experiments that measure the ancilla $T_1$ in different conditions.  In the first experiment (solid lines), we measure the ancilla $T_1$ while the coupler is continuously biased to the on position for the duration of the measurement.  In the second experiment (dashed lines), we measure the ancilla $T_1$ while the coupler is pulsed to the on position for the duration of a CX$^2$ gate repeatedly, once every $1.6\mu s$.  Two-photon dissipation is pulsed on while the gate is not being applied.  The shorter $1.6\mu s$ cycle duration serves to make the time in the on position is a larger part of the sequence.  The lifetime from this pulsed experiment can not be directly interpreted as the ancilla $T_1$ in the on position since it combines together contributions of when the coupler is on and off and from transients. Nonetheless it is a valuable signal to indicate mechanisms for $T_1$ reduction coming from the application of the flux pulse.  We observe noticeable degradation of the $|f\rangle$ $T_1$ relative to the first experiment.  The third experiment (dotted lines) is the same as the second except that the storage is excited to $|\alpha|^2=4$ at the beginning of the experiment.  We observe degradation of the ancilla $|e\rangle$ $T_1$ relative to the other experiments.  A hypothesis is that these $T_1$ reductions are due to the ancilla passing through or being biased near TLS (see \cref{app:noise_mechanisms} for an example of TLS in our system).   Furthermore storage excitations may shift the ancilla frequency close to that of a TLS due to the dispersive coupling (similar to~\cite{Thorbeck2024}).  For example, the $T_1$ of the $|e\rangle$ state may be reduced when adding the storage excitation because corresponding to certain Fock states the ancilla is shifted into a lossy frequency band.  Note that in all cases we observe temporal $T_1$ fluctuations~\cite{Klimov2018} which are also not currently being taken into account in our modeling of $\text{CX}$ error rates.

Bearing in mind the aforementioned caveats about interpreting the pulsed $T_1$, we can ask how the measured pulsed $T_1$ values in \cref{app_fig:cx_deep_dive}(d) compare against the values that are needed in simulations to obtain reasonable agreement with measured bit-flip data. The simulations assume $T_1$ values reduced by a factor $\beta  \approx 2.5$, relative to the measured values without pulsing, at $|\alpha|^2=4$. For the data in \cref{app_fig:cx_deep_dive}(d), comparing the values without pulsing to those with pulsing with $|\alpha|^2=4$, the average reduction is roughly $\beta \approx 1.5$. 
The reduced discrepancy suggests that the measured $T_1$ reduction in \cref{app_fig:cx_deep_dive}(d) is likely a contributor to the $\text{CX}^2$ gates underperforming the ideal simulations. However, temporal fluctuations, caveats about interpreting the pulsed $T_1$ values, and TLS effects would need to be considered more carefully to properly quantify the discrepancy. While current repetition code performance was not limited by ancilla decay to $|g\rangle$, improving the error modeling to accurately predict $\text{CX}$ error rates based on independently measured quantities is an important future direction.  Other error mechanisms, such as direct double decay from $|f\rangle$ to $|g\rangle$, which can be mediated by TLS (see \cref{app:noise_mechanisms}), and the storage-photon-number-dependent ancilla $T_{1}$ are also important areas of future investigations.

As an additional consistency check, we verify that the enhanced ancilla decay rates indicated by simulations and the results in \cref{app_fig:cx_deep_dive}
are consistent with the measured ancilla erasure probabilities. In particular, the ancilla erasure probability is equal to the probability of a decay from $|f\rangle \to |e\rangle$ during the time span from ancilla state preparation to approximately halfway through the readout pulse. This duration is approximately $1.25~\mathrm{\mu s}$ for the CX in \cref{app_fig:cx_deep_dive}. The ancilla erasure probability is $p_e \approx 5\%$, implying an $|f\rangle\to|e\rangle$ decay time of $(1.25~\mathrm{\mu s}/0.05)/2 = 12.5~\mathrm{\mu s}$, where the factor of 2 is included because the ancilla is initialized in the equal superposition $(|g\rangle + |f\rangle)/\sqrt{2}$. This value is roughly consistent with the measurements in \cref{app_fig:cx_deep_dive}(d). Moreover, the CX$^2$ bit-flip probability induced by ancilla double decay can also be crudely estimated from the erasure probability by computing the probability of an ancilla double decay $|f\rangle \to |e\rangle \to |g\rangle$ over the duration of a single CX. For initial ancilla state $|f\rangle$ this probability is given by $(2p_e T_\text{CX}/1.25~\mathrm{\mu s})^2 \approx  10^{-3}$, where $T_\text{CX}\approx 400~\mathrm{n s}$ is the duration of the CX gate. This estimate is roughly consistent with the measured CX$^2$ bit-flip probability in \cref{app_fig:cx_deep_dive}(a). 

\subsection{CX $\chi$-matching tolerance}
\label{app:chi_matching_requirements}

\begin{figure}
    \centering
    \includegraphics[width=\columnwidth]{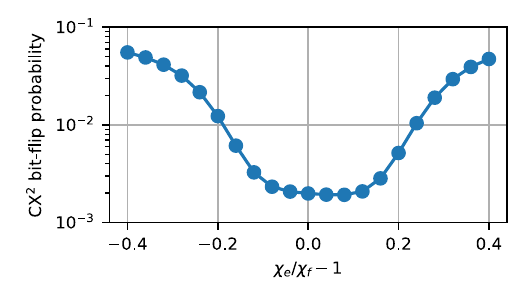}
    \caption{\textbf{$\chi$-matching requirements.} Simulated $\text{CX}^2$ bit-flip probabilities as a function of the $\chi$ mismatch (other parameters are held fixed at the values used for the A4, S5 gate).  Note that the tolerance to $\chi$ mismatch is lower for a $\text{CX}^2$ gate compared to a $\text{CX}$ gate, as discussed in the text.  }
    \label{app_fig:chi_match_requirements}
\end{figure}

We use simulations to investigate the robustness of CX gates to $\chi$ mismatches. In \cref{app_fig:chi_match_requirements}, we plot simulated CX$^2$ bit-flip probabilities for the A4, S5 gate as a function of the mismatch, $\chi_{ge}/\chi_{gf} - 1$. The bit-flip probability remains relatively constant for mismatches $|\chi_{ge}/\chi_{gf} - 1| \lesssim 0.1$. This robustness to small mismatches is due to the application of engineered dissipation after the gates are applied in each cycle. In particular, even though any non-zero mismatch will lead to ancilla-decay-induced over- or under-rotations of the storage mode, these small rotations are corrected by the engineered dissipation.

In comparison to CX$^2$ cycles considered in \cref{app_fig:chi_match_requirements}, repetition code error correction cycles exhibit approximately twice as much robustness to $\chi$ mismatch. That is, mismatches $|\chi_{ge}/\chi_{gf} - 1| \lesssim 0.2$ do not lead to significant additional logical bit-flip error during repetition code operation. The factor of two difference is simply due to the fact that a CX$^2$ gate is twice the length of a CX gate used in repetition code operaiton. As a result, ancilla-decay-induced over- or under-rotations of the storage mode are twice as large in CX$^2$ gates. 

\subsection{CX bit-flip error budgets}
\label{app:cx_error_budgets}

Finally, we produce error budgets for the average CX$^2$ bit-flip error, $(p^\text{(CX$_{|g\rangle}^2$)} + p^\text{(CX$_{|f\rangle}^2$)})/2$, at $|\alpha|^2 = 4$. The budgets are produced in the same manner as the total logical error rate budget presented in the main text, i.e.~following the approach of Ref.~\cite{Chen2021}. Notably, as discussed in \cref{app:logical_error_budget} this approach allows for a well-defined error budget---where the individual contributions to the budget sum to the total---so long as the physical error mechanisms contribute at most quadratically to the total error. We exploit this property here because certain logical bit-flip mechanisms, like ancilla double decays from $|f\rangle\to|e\rangle\to|g\rangle$ or $|f\rangle \to |e\rangle$ state-preparation errors followed by single decay from $|e\rangle \to |g\rangle$, are expected to scale quadratically in underlying physical error probabilities.

\begin{figure}
    \centering
    \includegraphics[width=\columnwidth]{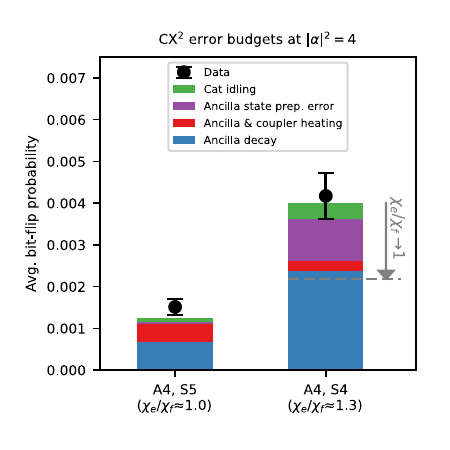}
    \caption{ \textbf{CX$^2$ bit-flip error budgets.} Bit-flip probabilities are averaged over initial ancilla states $|g\rangle$ and $|f\rangle$, and are evaluated at $|\alpha|^2 = 4$. For the poorly $\chi$-matched A4, S4 CX gate, the gray dashed line indicates the predicted bit-flip probability when the $\chi$ mismatch is artificially removed in simulation.}
    \label{app_fig:CX_budgets}
\end{figure}

In \cref{app_fig:CX_budgets}, we present budgets for the two different CX gates involving the ancilla A4 and storage modes S4 and S5. The budgets for these two CX gates serve as illustrative examples because the A4, S5 CX gate is very well $\chi$ matched while the A4, S4 CX gate is not. In both cases, we see that ancilla transmon decay is the dominant contributor to the error budget (see \cref{app:cx_ancilla_T1}), but the magnitude of this contribution and the total error are quite different between the two CX gates. We attribute these differences to difference in $\chi$ matching in the two cases. For example, in the poorly $\chi$ matched A4, S4 CX gate, we expect that a single $|f\rangle \to |e\rangle$ ancilla decay or state preparation error has a non-negligible probability of inducing a bit-flip error, consistent with the relatively large contributions associated with these error mechanisms. Indeed, we separately simulate the A4, S4 CX gate with the $\chi$ mismatch artificially set to 0, and we find a significant reduction in the expected bit-flip error as indicated by the gray arrow and dashed line in the figure. 

\section{Additional error correction experiment data}
\begin{figure*}[t!]
    \centering
    \includegraphics[width=1.8\columnwidth]{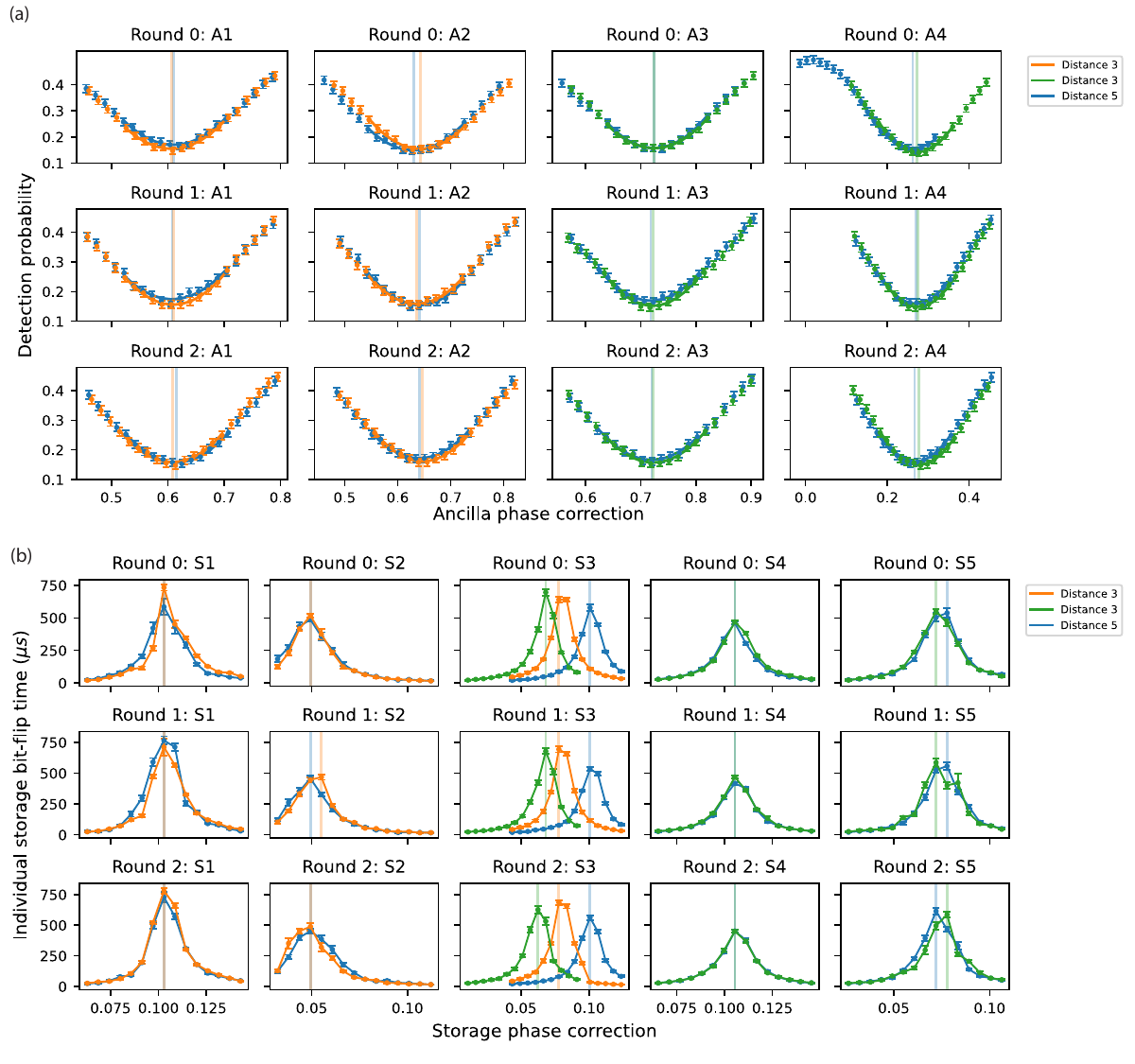}
    \caption{\textbf{Ancilla and storage phase calibrations for the repetition code experiment.} We report data for each of the three interleaved runs of the repetition code experiment as discussed in \cref{app:characterization_procedure}.  (a) Ancilla phase calibrations for the reptition code experiment.  The ancilla phase is optimized to minimize the detection probability as discussed in \cref{app:insitu_calibration}. Vertical lines indicate the fitted optimal value. (b) Storage phase calibrations for the repetition code experiment.  The storage phase is optimized to maximize the individual storage bit-flip times while running a repetition code style pulse sequence with the ancilla prepared to $|g\rangle$ each round (as discussed in \cref{app:insitu_calibration}).  The vertical lines correspond to the selected value of the phase which maximized the storage bit-flip time.}
    \label{app_fig:all_phase_optimizations}
\end{figure*}

\begin{figure*}[t!]
    \centering
    \includegraphics[width=1.9\columnwidth]{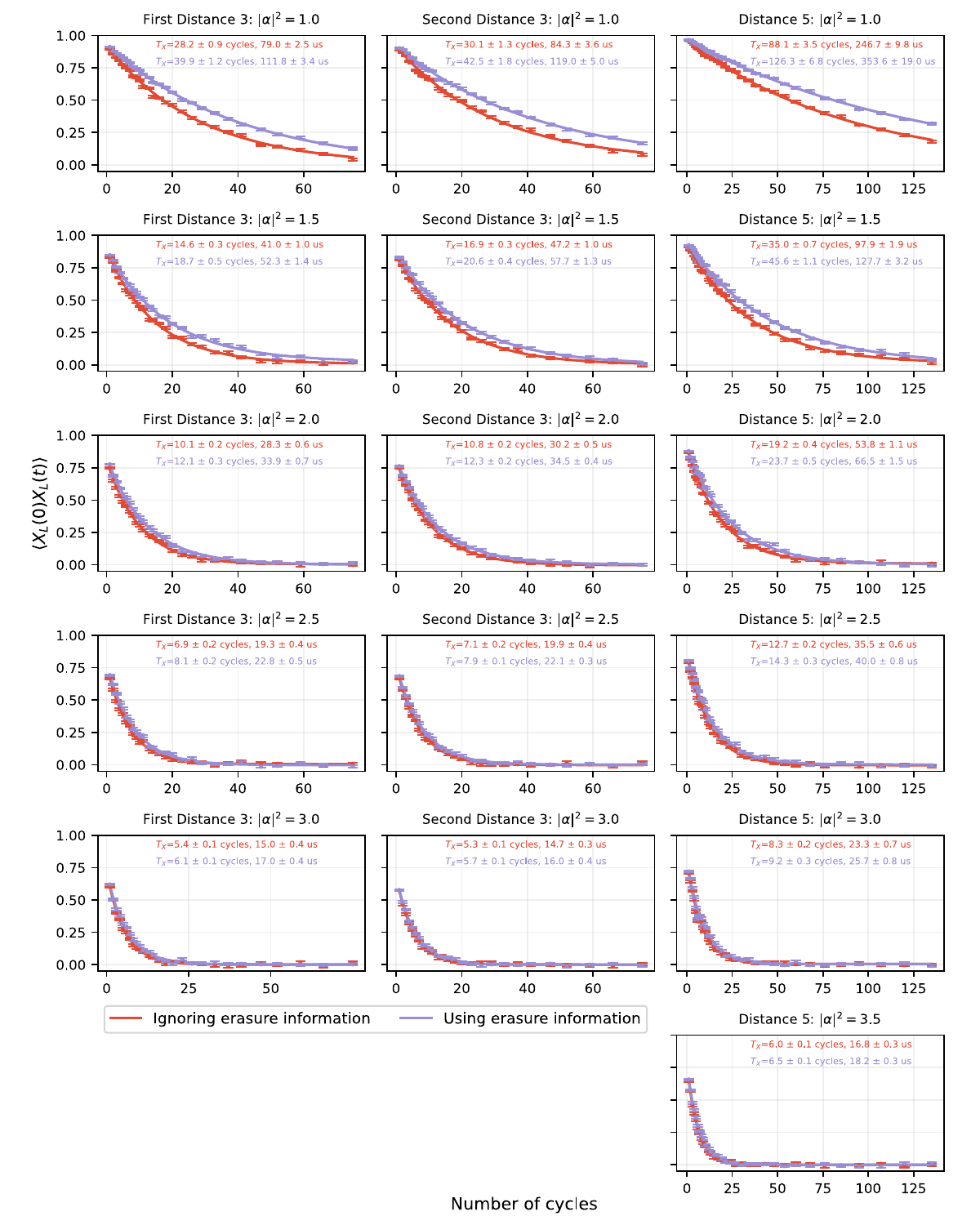}
    \caption{\textbf{Fits of logical $T_X$.} The fits here correspond to the data reported in \cref{fig:logical_X}.  The red curves are data where we do not use erasure information and the purple curves are data where we do use erasure information.}
    \label{app_fig:all_logical_X_fits}
\end{figure*}

\begin{figure*}[t!]
    \centering
    \includegraphics[width=1.9\columnwidth]{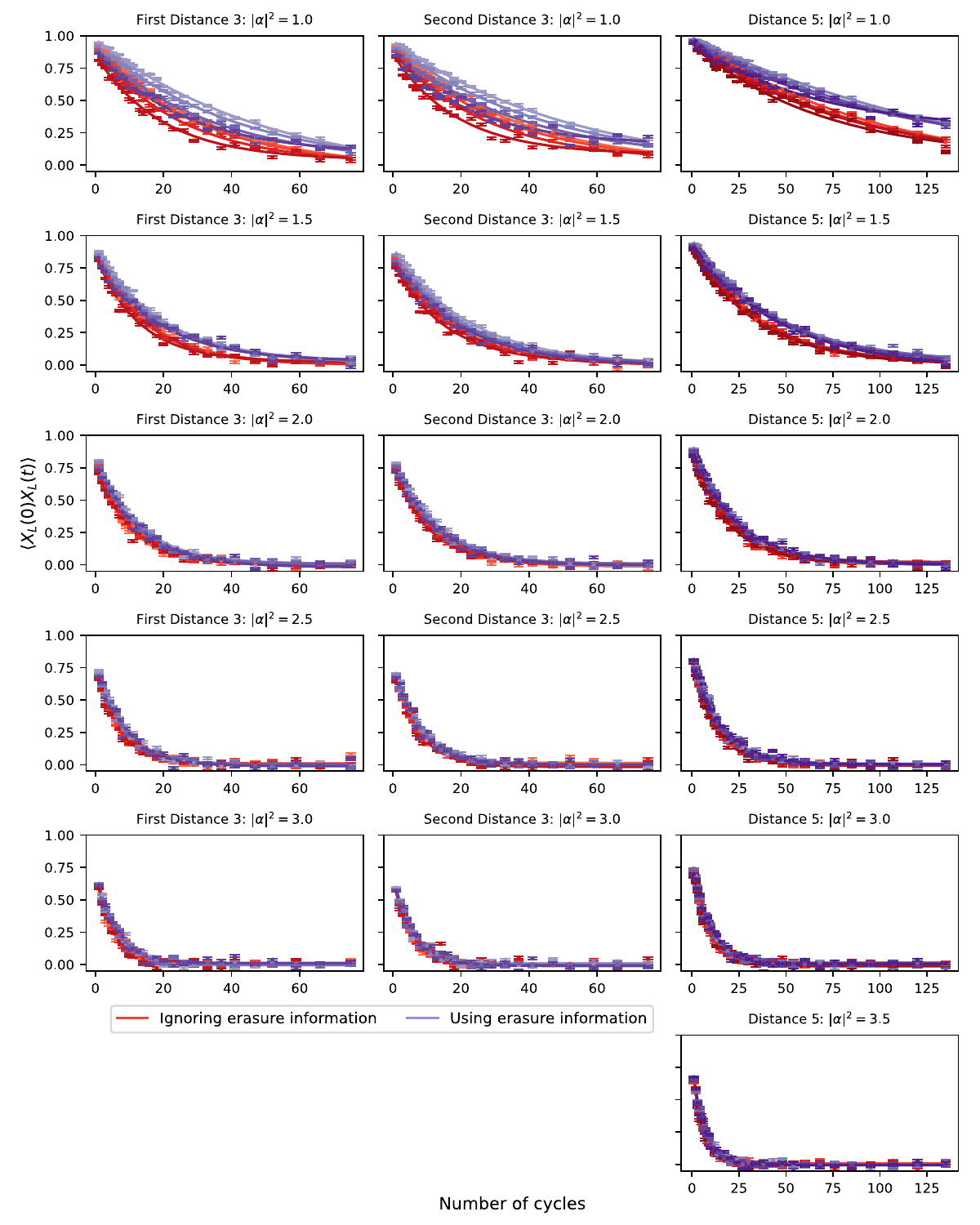}
    \caption{\textbf{Fits of logical $T_X$ binned by the number of $|+\rangle$ in the initial state.}  The red curves are data where we do not use erasure information and the purple curves are data where we do use erasure information.  Different shades of each color correspond to a different bin for the number of $|+\rangle$ in the initial state.  The bins are given in \cref{app:distribution_of_initial_states_in_logical_X_lifetime_experiments}.  As discussed in \cref{app:distribution_of_initial_states_in_logical_X_lifetime_experiments} we expect to see differences in error rates between different groupings due to the asymmetric phase-flip probabilities at low $|\alpha|^2$. }
    \label{app_fig:all_logical_X_fits_binned}
\end{figure*}

\begin{figure*}[t!]
    \centering
    \includegraphics[width=1.8\columnwidth]{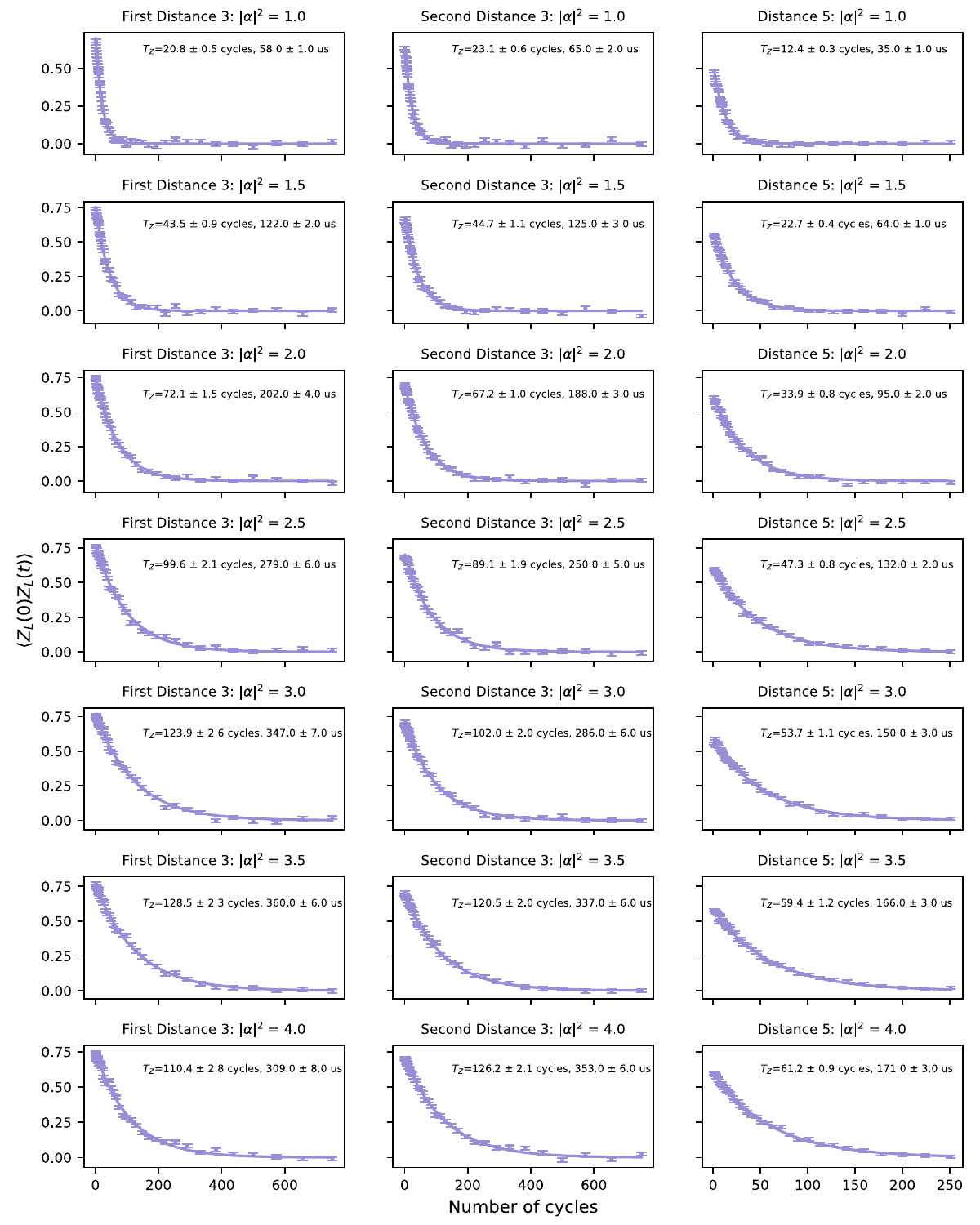}
    \caption{\textbf{Fits to determine logical $T_Z$.} The fits here correspond to the data reported in \cref{fig:logical_Z}.  }
    \label{app_fig:all_logical_Z_fits}
\end{figure*}

We present some additional information about the repetition code run reported in the main text.  In \cref{app_fig:all_phase_optimizations} we report the ancilla and storage phase calibrations (see \cref{app:insitu_calibration}) for all of the ancilla and storage modes for each of the three loops of the experiment (see \cref{app:characterization_procedure}).  In \cref{app_fig:all_logical_X_fits} we show the data and fits of the logical X basis experiment.  In \cref{app_fig:all_logical_X_fits_binned} we report the data and fits from the logical X basis experiment when they are grouped by the number of $|+\rangle$ cat states in the initial state (see \cref{app:distribution_of_initial_states_in_logical_X_lifetime_experiments}).  Lastly, in \cref{app_fig:all_logical_Z_fits} we report the data and fits from the logical Z basis experiments.  

\begin{figure*}[t!]
    \centering
    \includegraphics[width=\textwidth]{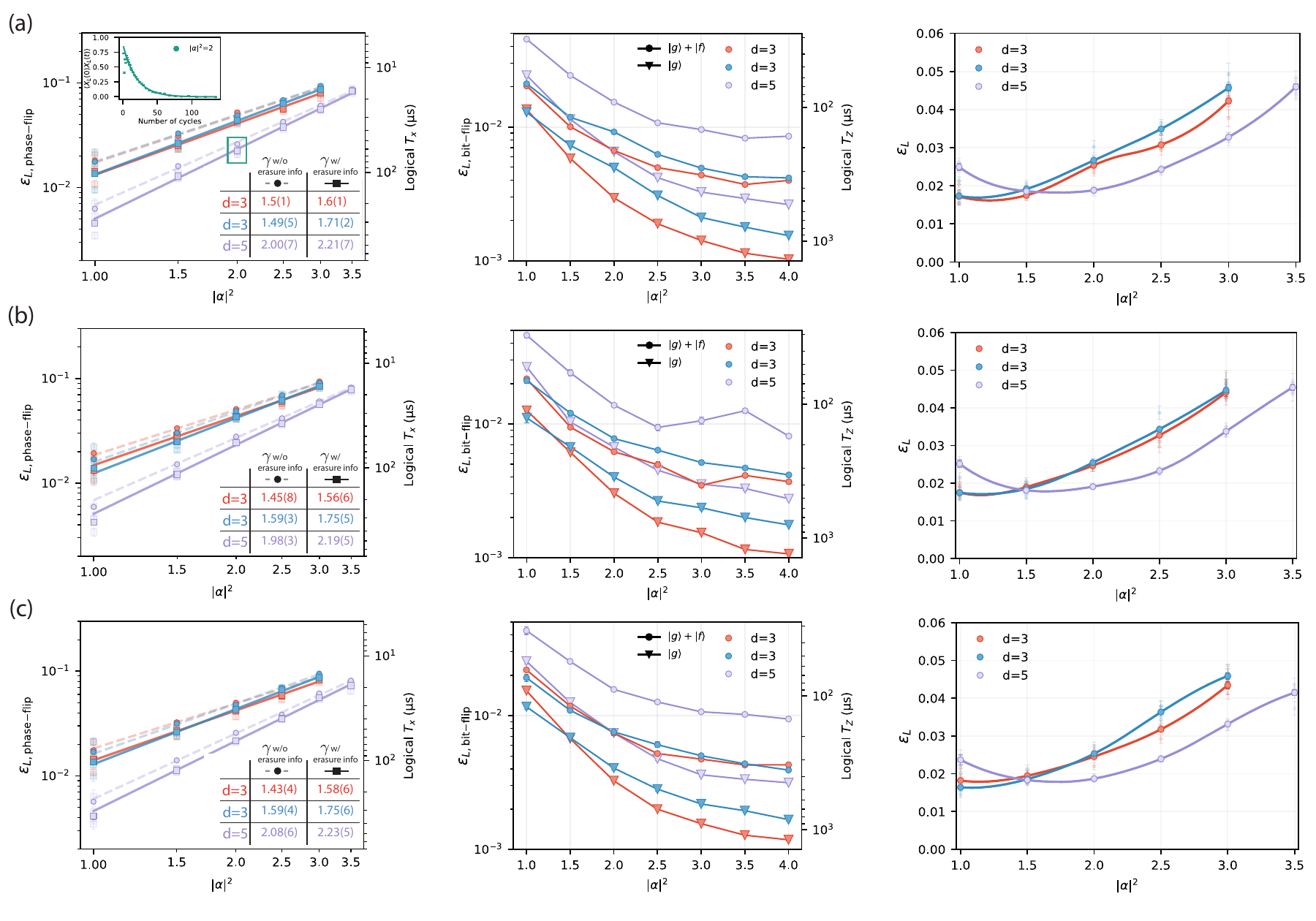}
    \caption{\textbf{Example data from other experiment runs.} We show the data from three other experiment runs (a,b,c) from before and after the repetition code run reported in the main text.  These experiments use different pulse sequence parameters compared to those discussed in \cref{app:X_basis_pulse_sequence} and \cref{app:Z_basis_pulse_sequence}.  Specifically these experiments do not use an $ef$ $\pi$ pulse before the readout and in some cases different padding around the stabilization.  In the logical phase-flip data for (a) there is a temporal fluctuation for a few time steps at $|\alpha|^2=2$.  The reported lifetime for $|\alpha|^2=2$ in (a) excludes these points in the fit as they yield an artificially high lifetime. When performing erasure matching in these datasets, $p_{\mathrm{odd}}$ (see \cref{app:decoding_with_erasure_information}) for a list of edges over 20 is assigned a probability near $50~\%$ . }
    \label{app_fig:data_from_other_runs}
\end{figure*}

In \cref{app_fig:data_from_other_runs} we provide logical memory performance from three other characterizations taken before and after the run reported in the main text. These runs use slightly different parameters including no $ef$ $\pi$ pulse before the readout and in some cases different paddings around the stabilization. However, each run follows the same structure of the interleaved experiments as described in \cref{app:characterization_procedure}.  We observe below threshold operation of the phase-flip correcting repetition code in all the datasets.  Additionally in all datasets we observe the distance-5 repetition code reaching logical bit-flip probabilities of $\sim 1\%$ or less.  In each dataset the minimum measured logical error per cycle for the distance-3 and distance-5 sections are comparable.  Owing to the increased protection from phase-flip errors and noise bias the distance-5 code outperforms the distance-3 when $|\alpha|^2\gtrsim 1.5$ for all the datasets.

\end{document}